\documentclass{aa}
\pdfoutput=1
\usepackage{amssymb}
\usepackage{amsmath}
\usepackage{natbib}
\bibpunct{(}{)}{;}{a}{}{,} 
\usepackage{graphicx}
\usepackage{subfig}     
\usepackage{txfonts}
\usepackage{color}
\usepackage{epstopdf}
\usepackage{wasysym}
\usepackage[utf8]{inputenc}
\usepackage{gensymb}
\DeclareUnicodeCharacter{2032}{-}
\usepackage{hyperref}
\definecolor{darkblue}{rgb}{0.,0.,1.0}
\hypersetup{colorlinks=true,breaklinks=true,
           linkcolor=darkblue,urlcolor=darkblue,citecolor=darkblue}

\begin{document}

\title{Gas and stellar dynamics in Stephan's Quintet \thanks{This paper uses data taken with the MODS spectrographs built with funding from NSF grant AST-9987045 and the NSF Telescope System Instrumentation Program (TSIP), with additional funds from the Ohio Board of Regents and the Ohio State University Office of Research. The LBT is an international collaboration among institutions in the United States, Italy and Germany. LBT Corporation partners are: LBT Beteiligungsgesellschaft, Germany, representing the Max-Planck Society, The Leibniz Institute for Astrophysics Potsdam, and Heidelberg University; The University of Arizona on behalf of the Arizona Board of Regents; Istituto Nazionale di Astrofisica, Italy; The Ohio State University, and The Research Corporation, on behalf of The University of Notre Dame, University of Minnesota and University of Virginia.}}

   \subtitle{Mapping the kinematics in a closely interacting compact galaxy group}

   \author{M. Yttergren
          \inst{1,2,4}\fnmsep\thanks{Member of the International Max Planck Research School (IMPRS) for Astronomy and Astrophysics at the Universities of Bonn and Cologne. \newline \email{yttergren@ph1.uni-koeln.de}}
        \and
          P. Misquitta\inst{2}
        \and
          \'A. S\'anchez-Monge\inst{2}
        \and
          M. Valencia-S\inst{2,3}
        \and
          A. Eckart\inst{1,2}
        \and
          A. Zensus\inst{1}
        \and
          T. Peitl-Thiesen\inst{2}
          }

   \institute{Max-Planck Institut für Radioastronomie (MPIfR), Auf dem Hügel 69, D-53121 Bonn, Germany
         \and
             I.Physikalisches Institut, Universität zu Köln, Z\"ulpicherstrasse 77, D-50939 Köln, Germany
         \and
            Regionales Rechenzentrum (RRZK), Universität zu Köln, Weyertal 121, D-50931 Köln, Germany 
         \and 
            Department of Space, Earth and Environment, Chalmers University of Technology, SE-412 96 Gothenburg, Sweden}

\authorrunning{M. Yttergren et al.}

\titlerunning{Gas and stellar dynamics in Stephan's Quintet}

\date{Received 21 December 2020 / Accepted 08 September 2021}

  \abstract
   {
    In nearby compact galaxy groups we can study the complex processes of galaxy interactions at high resolution and obtain a window into a time in the history of the Universe when the galaxies were closely spaced and the intergalactic medium was awash with gas. 
   Stephan's Quintet is a nearby compact galaxy group and a perfect laboratory for studying the process of galaxy evolution through galaxy harassment and interaction. 
   By analysing the kinematics of Stephan's Quintet we aim to provide an increased understanding of the group, the history of the interactions, their cause and effect, and the details regarding the physical processes occurring as galaxies interact. 
   Ionised gas and stellar kinematics have been studied using data from the Large Binocular Telescope, while the molecular gas kinematics have been obtained from CO observations using the IRAM 30m telescope. 
   Large areas of the group have been mapped and analysed. 
   We obtain a total ionised gas mass in the regions chosen for closer analysis of $20.1\pm0.2 \cdot 10^{10} M_{\odot}$ and a total H$_2$ gas mass of $21 \pm 2 \cdot 10^{9} M_{\odot}$ in the observed area (spectra integrated over the velocity range covering Stephan's Quintet), while the star-forming clouds show an impressive complexity, with gas congregations at multiple velocities at many locations throughout the group. 
   We map the large-scale nuclear wind in NGC7319 and its decoupled gas and stellar disk. 
   With our high resolution data we can, for the first time, reveal the Seyfert 1 nature of NGC7319 and fit the narrow-line region and broad-line region of the H$\alpha$ line. 
   While the $^{12}\text{CO~}(1-0)$ map shows significant emission in the area in or near NGC7319, the bridge, and the star-forming ridge, the $^{12}\text{CO~}(2-1)$ emission shows a prevalence to the star-forming ridge, an area south of the NGC7318 pair, and shows an extension towards NGC7317 -- connecting NGC7317 to the centre of the group, indicating a previous interaction. 
   NGC7317 may also be a prime candidate for studies of the process of galaxy harassment. 
   Furthermore, we connect the kinematical structures in Stephan's Quintet to the history of the group and the ongoing interaction with NGC7318B. 
   Through our extensive observations of Stephan's Quintet we trace and present the kinematics and evolution of the complex processes and structures occurring in this nearby interactive group. 
   }


\maketitle
\section{Introduction}
In the early Universe, the close proximity between galaxies is expected to have enabled a high rate of galaxy interactions and mergers \citep{rod2014}, which were vital in driving galaxy evolution. 
At lower redshifts we can see similar dynamics in compact galaxy groups. 
Compact galaxy groups exhibit high amounts of intergalactic gas, a high proximity between the galaxies, and frequent galaxy interactions and mergers, and nearby compact galaxy groups can provide a high resolution view into the processes and morpho-kinematical structures. 
Compact galaxy groups are therefore perfect laboratories for the study of galaxy evolution through galaxy harassment and interaction, and they can reveal key information regarding the connections between galaxy evolution and the environment. 
Furthermore, the physical processes occurring in the environment of galaxy groups play a fundamental role in determining the star formation history of the Universe (e.g. \citet{natale2010}). 

Galaxy-galaxy and galaxy-intergalactic medium (IGM) interactions should effectively remove interstellar gas from the group's galaxies, leading to a quenching of star formation in the galaxies. 
However, this process simultaneously causes the chemical enrichment of the IGM, which in turn can cool the IGM and facilitate accretion onto the galaxies and new star-forming (SF) systems. 
The interplay of these processes, and their overall effect on galaxy evolution and star formation activity, is yet to be determined. 

There is a multitude of additional interesting questions that arise from the study of compact galaxy groups \citep{hickson1997}, such as: what the end products of the evolution of these groups are; whether the end products have properties consistent with any known population of objects; where compact groups fit into the overall clustering hierarchy; what the connection between compact groups and morphological segregation is; and what the role of compact groups in the evolution of galaxies is throughout cosmic time. 

\begin{figure}[h]
    \centering
    \includegraphics[width=0.48\textwidth]{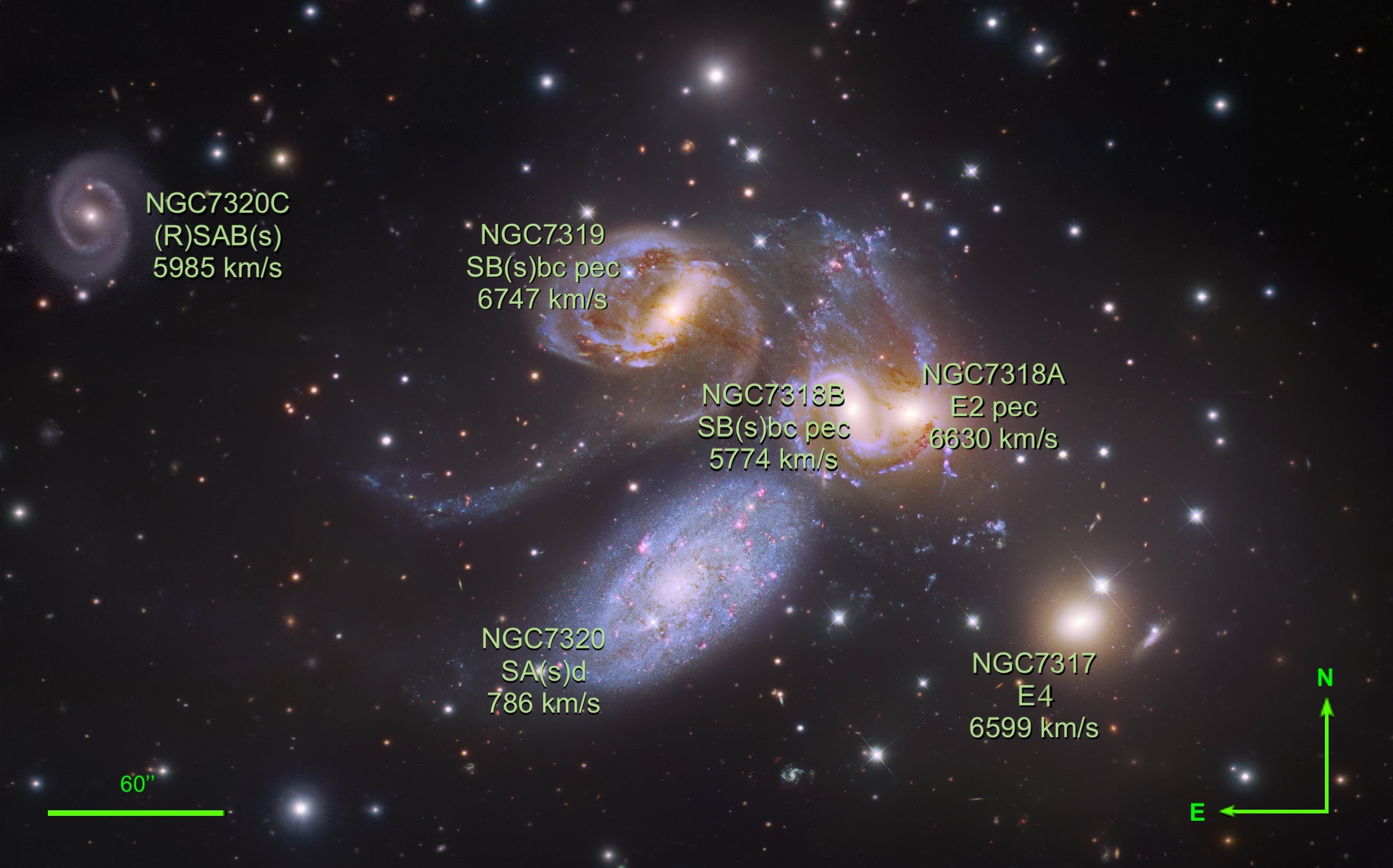}
    \caption{Luminance, Red, Green and Blue (LRGB) composite colour image of SQ obtained with the Subaru telescope and Hubble Space Telescope WFC3 (based on the image processed by Robert Gendler and Judy Schmidt). The velocities and morphologies stated are provided by the NASA/IPAC Extragalactic Database (NED), and the average distance to SQ is 88.6 Mpc (\citet{duarte2019}), where 1$''$ corresponds to $\sim$435 pc.}
    \label{fig:hst} 
\end{figure}

\begin{figure}[h]
    \centering
    \includegraphics[width=0.48\textwidth]{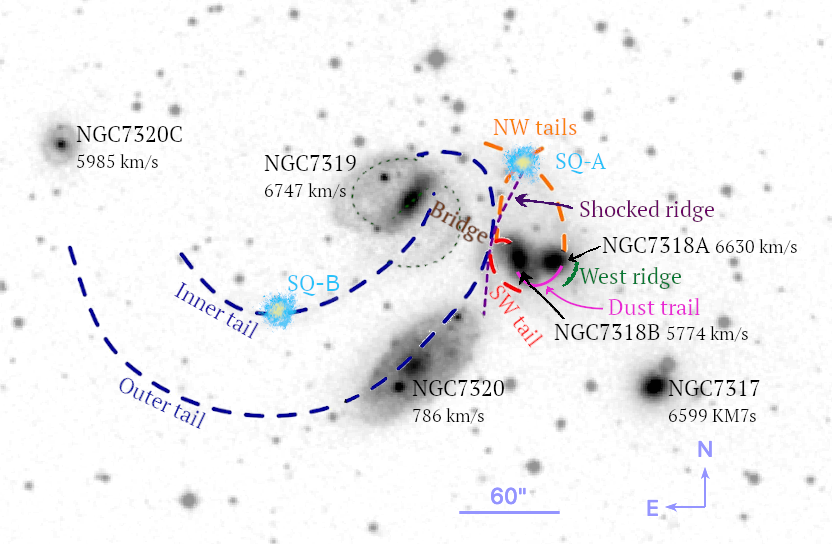}
    \caption{{Adjusted Digitized Sky Survey 2 (DSS2) R-band image of SQ overlaid with markings stating the naming convention of the different important tidal and interaction structures in the group. The figure design is inspired by \citet{renaud2010}.} }
    \label{fig:structures}
\end{figure}

The compact galaxy group \object{Stephan's Quintet} (SQ), shown in Fig. \ref{fig:hst}, was discovered in the 19th century \citep{stephan1877} and has since been studied across the electromagnetic spectrum. 
Stephan's Quintet consists of five galaxies, \object{NGC7317}, \object{NGC7318A}, \object{NGC7318B}, \object{NGC7319}, and NGC7320C (NGC7320 is a foreground galaxy at a much lower redshift; \citealt{burbidge1961}), and exhibits impressive tidal structures that reveal a rich history of past interactions (e.g. \citet{allen1972,allen1980,shostak1984,moles1997, moles1998,fedotov2011}). 
In the past, NGC7320C passed through the centre of the group, west of NGC7319 \citep{moles1997,lisenfeld2004}, depositing most of its interstellar medium (ISM) into the IGM and creating large tidal tail(s) (as marked in Fig. \ref{fig:structures}). 
Whether both the inner and outer tidal tail were created in the interaction of NGC7319 with NGC7320C or whether an interaction between NGC7319 and NGC7318A was involved as well is still unclear \citep{renaud2010,hwang2012}. 
NGC7319 is expected to have been involved in nearly all of the past interactions and is classified as a Seyfert 2 galaxy with a large-scale outflow \citep{aoki1996}. 
Furthermore, it is yet to be determined whether NGC7317 passed through the group in the past \citep{sulentic2001,rod2014}, though a diffuse stellar halo extending towards the galaxy indicates that an interaction has occurred \citep{duc2018}. 

There is galaxy-wide shocked ridge located between the NGC7318 pair and NGC7319 (as marked in Fig. \ref{fig:structures} together with additional important structures in SQ), flanked by star formation towards NGC7318B and in the ends of the radio shock \citep{cluver2010}. 
This structure is created by the collision of the IGM and the intruder galaxy NGC7318B, which is entering the group from behind at a relative line-of-sight velocity of $\sim900$ km/s \citep{xu2003}. 
The star formation near the shocked ridge towards NGC7318B, in the south-west (SW) tail and the north-west (NW) tail, is what is called the SF ridge. 
The SF ridge shows widespread star formation and high velocity dispersion (e.g. \citet{gao2000,sulentic2001,xu2003,xu2005,osullivan2009,iglesiasparamo2012,konstantopoulos2014,duarte2019}). 
Stephan's Quintet  exhibits an interesting distribution of HI and cool molecular gas, where most of the HI is located outside of the optical emission \citep{allen1980,shostak1984,williams1999,williams2002,renaud2010}, and the cool molecular gas is displaced from the galaxies and appears in isolated clumps in the IGM and near NGC7319 \citep{gao2000,appleton2017}.

The state of SQ fits into the scenario described by \citet{verdesmontenegro2001}, where compact groups evolve from HI rich to HI poor. 
Young compact groups are expected to be HI rich and mainly contain spirals, whereas older groups would contain mainly ellipticals and be HI poor. 
As the galaxies in the younger group interact, the HI is displaced from the galaxies and deposited into the IGM. 
As SQ is HI poor and most of its gas is outside of the galaxies themselves, SQ is thereby expected to be an older group. 
The HI deficiency in SQ may be accounted for by gas having been heated up in the shock in the SF ridge, as shown by the high X-ray emission present in that region \citet{trinchieri2003,trinchieri2005}. 
The emergence of NGC7318B (similar to the past interaction with NGC7320C) in SQ is expected to play an important role in increasing the energy and gas content in the group, ensuring the longevity of the group.

Stephan’s Quintet is a well-studied object and has been observed across the electromagnetic spectrum by multiple scientists, such as: in radio by \citet{gao2000}, \citet{smith2001}, \citet{lisenfeld2002}, \citet{pepitas2005}, and \citet{nikiel2013}; 
in infrared by \citet{appleton2006}, \citet{cluver2010}, \citet{guillard2010}, \citet{natale2010}, and \citet{appleton2013}; 
in optical by \citet{iglesiasparamo2001}, \citet{gallagher2001}, \citet{trancho2012}, \citet{duc2018}, and \citet{duarte2019}; 
in ultraviolet by \citet{xu2005} and \citet{demello2012}; 
and in X-ray by \citet{sulentic1995}, \citet{pietsch1997}, \citet{trinchieri2003}, \citet{trinchieri2005}, and \citet{osullivan2009}. 
The interactions have also been simulated by scientists such as \citet{renaud2010} and \citet{hwang2012}. 
Every time the group has been observed in a new wavelength regime or with higher resolution or sensitivity, new information and fascinating details have emerged. 

However, the structures have primarily been observed individually and often using single pointings. 
To understand this kind of complex interconnected interactive group, we need the benefit of studying the structure as a whole and not as multiple separate smaller structures. 
Despite the extensive observations of the group, there is still a clear lack of available information regarding the group's kinematics and regarding the content and distribution of the molecular gas. 
Therefore, we have carried out a study of the kinematics of the atomic gas, molecular gas, and stars in large areas covering SQ. 

Section \ref{obs} begins by detailing how we used the Large Binocular Telescope (LBT) in Tucson, Arizona, to analyse the atomic gas kinematics, the stellar kinematics, and the gas excitation mechanisms. 
The section continues by presenting the CO data obtained using the IRAM 30m telescope in Sierra Nevada, Spain. 
Section \ref{obs} also includes details regarding the data reduction and analysis procedures. 
In Sect. \ref{res} we present our results, for each area separately, and discuss their implications on the individual structures as well as on the whole group. 
Finally, in Sect. \ref{sum} we conclude our paper and summarise the presented analysis.

\section{Observations and data reduction}
\label{obs}  
In this section we present the data acquisition process and the primary data reduction and analysis procedures. 
We begin with the optical data and continue with the CO data. Since there is a difference between the optical and radio convention for velocity (see \citet{greisen2006}), whenever we state a velocity (in text and tables) we always clarify whether it is from the stellar, ionised gas, or molecular gas component. 

\subsection{Atomic gas and stars}
Optical long-slit spectroscopy observations were carried out using the Multi-Object Double Spectrograph (MODS) at the LBT in Tucson, Arizona, USA, on 13 and 15 November 2017. 
The observations were performed in single mode with MODS1 only, since MODS2 was out of order at this time. 
The spectra obtained cover the wavelength range $3200-10000$ Å  with the 1.0 arcsec wide and 5 arcmin long slit, and the Differential Image Motion Monitor (DIMM) seeing averaged on $0.8''$ throughout the observation period (thus the emission can be expected to have been contained within the slit). 

\begin{figure}[h]
    \centering
    \includegraphics[width=0.48\textwidth]{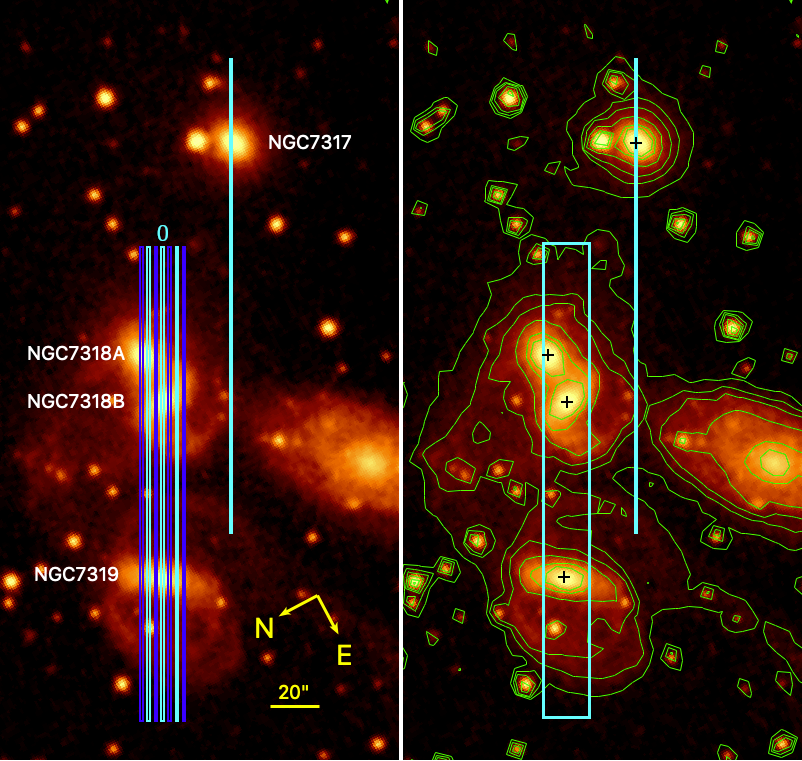}
    \caption{\textit{Left:} Eight observed slit positions marked on a DSS9 red image. Seven slits cover NGC7319 and the NGC7318 pair, alternately coloured in light and dark blue, while the eighth slit covers NGC7317. \textit{Right:} Observed region marked on the DSS9 red image, which has been overlaid with arbitrary contours to show the extension of the group's emission. The seven slits construct the rectangular area, whereas the crosses mark the galaxies, from top to bottom: NGC7317, NGC7318A, NGC7318B, and NGC7319.}
    \label{fig:slitregion}
\end{figure}

The core region of SQ, NGC7319, the NGC7318 pair, and the intergalactic area between them were covered with several slit positions (as marked in Fig. \ref{fig:slitregion}). 
The slit positions were carefully chosen to cover the nucleus of each of the galaxies in the observed area, whereas the extended regions were under-sampled to increase the area observed and focus our study on the global properties of SQ. 
Adopting a telescope position angle of 242 degrees, the first slit was centred on the nucleus of NGC7318B and NGC7319 -- this position is marked with a 0 in Fig. \ref{fig:slitregion}. 
Thereafter, several horizontal ($dx$) shifts of 3 arcsec each were carried out to either side of the 0 position, covering the region as shown in Fig. \ref{fig:slitregion} with seven slit positions. 
It should be noted that positive $dx$-shifts shift the slit to the left of the central position in the image, so position $dx=9''$ covers the NGC7318A nucleus. 
Also illustrated in Fig. \ref{fig:slitregion} is the additional slit that was positioned to cover NGC7317. 

The exposure times for each position are stated in Table \ref{table:exps}. 
At each slit position, three exposures of 300s each were obtained, and after a 10$''$ dithering along the slit, another three exposures of 300s each. 
Due to the weather conditions, NGC7317 and slit position $dx=9''$ were observed for a slightly shorter period, as indicated in Table \ref{table:exps}. 

\begin{table}[]
    \centering
    \caption{Exposure time at the different slit positions. }
    \label{table:exps}
    \begin{tabular}{cc} \hline \hline
        Shift & Exposure time \\ 
        (arcsec) & (seconds) \\\hline
        $9$ & 4x300\\
        $6$ & 6x300\\
        $3$ & 6x300\\
        $0$ & 6x300\\
        $-3$ & 6x300\\
        $-6$ & 6x300\\
        $-9$ & 6x300\\
        NGC7317 & 4x300\\ \hline
    \end{tabular}
\tablefoot{
The position angle of the slit is 242 degrees.}
\end{table}

To enable the subsequent flux calibrations the spectrophotometric standard stars G191-B2B (DA0 type) and BD+33 2642 (B2IV type) were observed using the $5''$ slit. 
The data were reduced using the MODS Basic CCD Reduction package (provided by the LBT Observatory) and the Image Reduction and Analysis Facility (IRAF); performing bias subtraction, flat field correction, background subtraction, wavelength calibration, flux calibration, and for the redder part of the spectra also telluric corrections. 
The background subtraction was a particularly delicate procedure due to the large amounts of extended gas, 
but could, using the upper and lower fifths of the slit and a low-emission central part, be corrected by using a fourth order Legendre polynomial baseline and careful visual inspection. 

The 1D spectra were extracted, prior to flux calibration, using a $1''$ aperture along the spatial extension of the slits, and adjusted to the average redshift of SQ, $z=0.0215$, to facilitate the analysis procedure. 
The spectra covering the rectangular area marked in Fig.\ref{fig:slitregion} were combined into a map with linear interpolation covering the under-sampled area between the slits. 

The stellar continuum was analysed using Penalized Pixel-Fitting method \citep[pPXF;][]{cappellari2004,cappellari2017} 
with the MILES stellar template library \citep{sanchez2006,vazdekis2010,falcon2011}, which contains 985 well-calibrated stars in the wavelength range $3525-7500$ Å at a spectral resolution of 2.51 Å. 
The gaseous emission lines were fitted using manually defined multiple-Gaussian functions and Python's SciPy curvefit optimisation module. 
Manually defined multiple-Gaussian functions allow for a choice of the number of components, expected line centres and expected amplitudes, thereby sufficient constraints on the resultant fits of the complex kinematics of this group could be achieved. 
Using the velocity dispersion values obtained from the Gaussian fitting of the emission lines, masses of HII regions could be estimated using the virial theorem, following the discussion in \citet{rozas2006}. 
As the extent of the clumpiness of the gas is unclear, it is also not clear whether case A or B recombination should be assumed. Although shocked gas is more likely to be optically thin (i.e. case A).

\begin{table}[t!]
\caption{IRAM\,30m observational parameters.}\label{table:30mobservations}
\centering
\begin{tabular}{c c c c c c c}
\hline\hline
\noalign{\smallskip}

& Frequency
& HPBW
& $F_\mathrm{eff}$
& $B_\mathrm{eff}$
& rms$^{(a)}$
\\
Transition
& (MHz)
& (arcsec)
& (\%)
& (\%)
& (mK)
\\
\hline
$^\mathrm{12}$CO\,(1--0) & 112835.58203 & 23.5 & 94 & 78 & 3.0 \\
$^\mathrm{13}$CO\,(1--0) & 107872.85774 & 23.5 & 94 & 78 & 1.8 \\
$^\mathrm{12}$CO\,(2--1) & 225666.85350 & 10.5 & 92 & 59 & 3.7 \\
\hline
\end{tabular}
\tablefoot{
\tablefoottext{a}{Root mean square (rms) noise level per channel for the final data cubes smoothed to a half-power beam width (HPBW) of 50\arcsec\  and to a velocity resolution of 40~km~s$^{-1}$.}}
\end{table}

\subsection{Molecular gas}
\begin{figure}[t!]
    \centering
    \includegraphics[width=0.9\columnwidth]{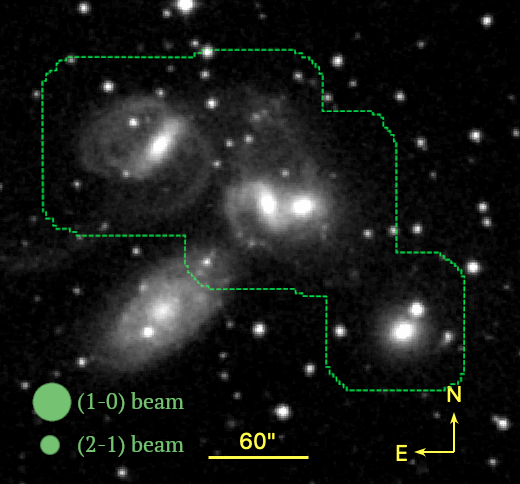}
    \caption{Area of SQ observed using EMIR and FTS200 at the IRAM 30m telescope. The background is a DSS2 red image, and the circles illustrate the beam sizes.}
    \label{fig:iram_boxes}
\end{figure}

The CO distribution and kinematics in SQ were observed using the IRAM\,30m telescope in Sierra Nevada, Spain. The observations (project ID 077--18) were conducted on 14--18 September 2018 under regular weather conditions, with high levels of precipitable water vapour (pwv) between 7 and 15~mm and opacities\footnote{The atmospheric opacity $\tau$ at 225~GHz is calculated from the expression $\tau(225)=0.058\times\mathrm{pwv}+0.04.$} between 0.5 and 1.0. We used the on-the-fly (OTF) mapping technique to cover a field of view of 5.67~arcmin$^2$ at 3~mm and 1~mm in dual polarisation mode using the EMIR receivers \citep{Carter2012}, with the fast Fourier transform spectrometer (FTS) at 200~kHz of resolution \citep{Klein2012} as the back end. The observed area is indicated in Fig.~\ref{fig:iram_boxes}. We used position switching mode, with the reference position being an empty region of the sky with coordinates RA(J2000)=22$^\mathrm{h}$35$^\mathrm{m}$32$.\!^\mathrm{s}$439, Dec.(J2000)=+34$^\circ$00$\arcmin$22$\farcs$600). The area was sampled using a mapping step of 4~arcsec (2.75 times smaller than the smallest beam; see Table~\ref{table:30mobservations}) and a mapping speed of 4\arcsec/s, resulting in Nyquist sampling conditions. We mapped the emission of three different transitions of the CO molecular species: $^{12}$CO\,(1--0), $^{13}$CO\,(1--0) and $^{12}$CO\,(2--1). More details are provided in Table~\ref{table:30mobservations}. During the observations, the pointing was corrected by observing the strong nearby quasars 2251+158 and 2201+315 every 1--2~hours, and the focus by observing the planet Saturn. Pointing and focus corrections were stable throughout all the runs. 

The data were reduced with a standard procedure using the CLASS/GILDAS package \citep{Pety2005}. For each one of the three observed transitions, we created individual data cubes centred at the red-shifted frequencies listed in Table~\ref{table:30mobservations}, and spanning a velocity range of $\pm$3\,000~km~s$^{-1}$. In order to perform a proper comparison of the line profiles of every transition, we smoothed the spectral resolution of each cube to a common value of 40~km~s$^{-1}$. A third-order polynomial baseline was applied for baseline subtraction. The rms noise level of each spectrum, as determined from the baseline fit, was used to filter out the noisier scans, usually associated with large opacity values. We used a threshold of 0.5~K for the $^{12}$CO and $^{13}$CO\,(1--0) observations, and 1.5~K for the $^{12}$CO\,(2--1) observations, which allowed us to exclude bad spectra improving the quality of the final maps. The data were also corrected for platforming effects as well as for the presence of spikes (bad channels). We used the main beam brightness temperature ($T_\mathrm{MB}$) as intensity scale for the different spectra and cubes. For this, we converted the antenna temperature ($T_\mathrm{A}^{*}$) by applying the factor $F_\mathrm{eff}/B_\mathrm{eff}$, where $F_\mathrm{eff}$ is the forward efficiency and $B_\mathrm{eff}$ is the beam efficiency, both listed in Table~\ref{table:30mobservations}. Each cube was created with a pixel size of 2\arcsec$\times$2\arcsec, and the maps were smoothed to a final half power beam width (HPBW) of 50\arcsec, in order to improve the signal-to-noise ratio. The final rms of each data cube is about 2.8~mK per 40-km~s$^{-1}$ channel (see Table~\ref{table:30mobservations}). 

The individual CO emission lines were fitted using manually defined multiple-Gaussian functions (1-4, as needed) and Python's SciPy curvefit optimisation module at a signal-to-noise ratio (S/N) >1.9. 
The line-of-sight velocity and velocity dispersion were derived directly from the fitted Gaussian while the integrated intensity also has been derived as
\begin{equation}\label{eq:Ico}
    I_{\rm{CO}}\left[{\rm K~km~s^{-1}}\right] = \int_{\langle v\rangle-v_{min}}^{v_{max}-\langle v\rangle}\tau_{mb} dv
\end{equation}
to better account for potential asymmetry in the line profile for the non-nested lines. 
$\tau_{mb}$ is the CO main beam temperature over the velocity range $v_{min}$ to $v_{max}$, and $\langle v\rangle$ is the first moment mean intensity-weighted velocity \citep{guillard2012}. 
Due to the nested nature of the lines present in SQ, we present the fluxes obtained from the Gaussian fits. Comparing the flux obtained from integrating over the individual lines using Eq. \ref{eq:Ico} to the flux obtained from the Gaussian fit, shows that the flux of the Gaussian fit is on average overestimated by $0.6\sigma$ with an added uncertainty of $\sim0.95\sigma$. Therefore,  we accept a $2\sigma$ uncertainty in the tables presented in the subsequent sections of this paper. 

Thereafter, the $^{12}\text{CO~}(1- 0)$ flux together with the Galactic conversion factor, 
\begin{equation}
   \frac{N(\text{H}_2)}{I_{CO}} = 2\times 10^{20} cm^{-2} ~\left[{\rm K~km~s^{-1}}\right],
\end{equation}
provides the molecular hydrogen column density, $N(\text{H}_2)$ \citep{guillard2009}, from which the H$_2$ gas mass is calculated as
\begin{equation} \label{eq:H2_mass}
    M_{\rm{H_2}} = 75 I_{\rm{CO}} D^2 \Omega ~\left[\rm{M}_{\odot}\right].
\end{equation}
Here $I_{CO}$ is the velocity integrated $^{12}\text{CO~}(1- 0)$ line intensity [${\rm K~km~s^{-1}}$], $D$ is the distance (in Mpc), and $\Omega$ is the area covered (in $\rm{arcsec}^2$) \citep{braine2001,lisenfeld2002}, where $\Omega = 1.13\Theta^2$ for a single pointing with a Gaussian beam of full width at half maximum (FWHM) $\Theta$.

\section{Results, analysis, and discussion}
\label{res}
\begin{figure}[h!]
    \centering
    \includegraphics[width=0.4\textwidth]{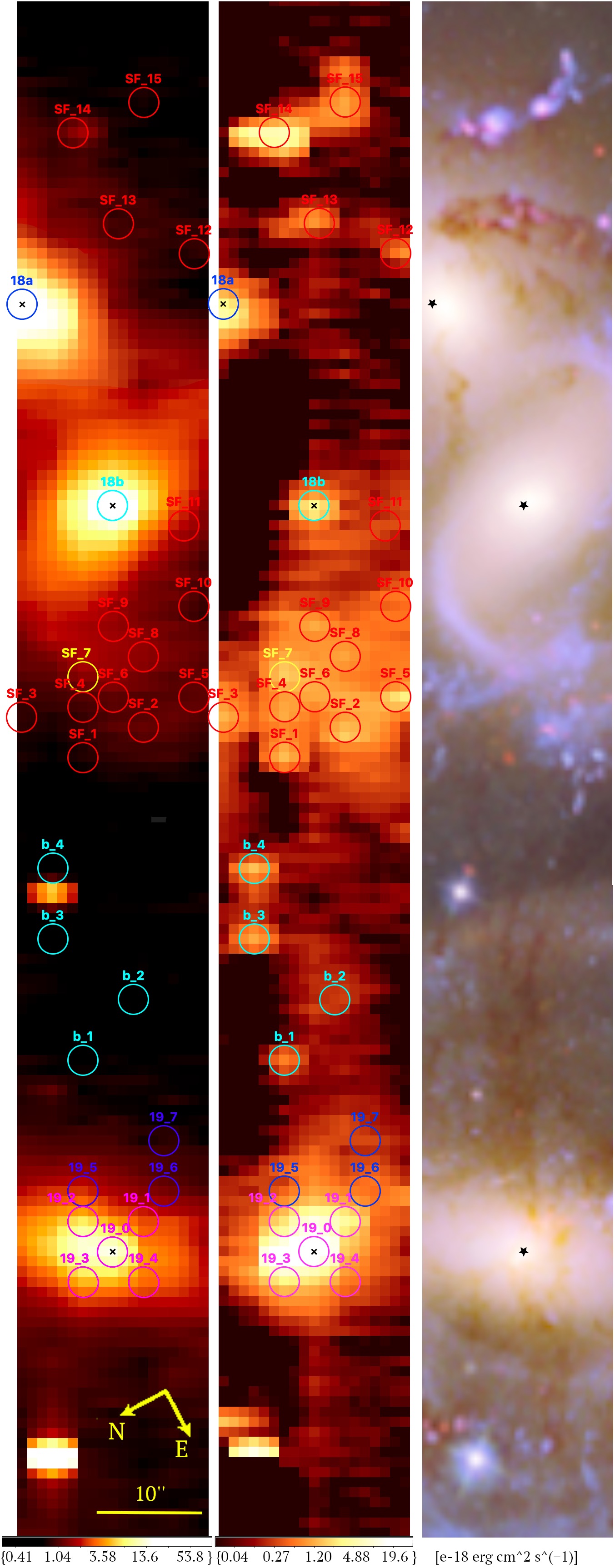}
    \caption{\textit{Left:} Stellar continuum image constructed from the LBT data with the regions marked for closer analysis. The right ascension (RA) and declination (Dec.) of each region is listed in Appendix \ref{app:regions_ra_dec}. \textit{Middle:} Emission of the H$\alpha\_[\ion{N}{II}]$ sub-band observed with the LBT, which provides the basis of the choice of regions. \textit{Right:} Observed area as imaged by a Subaru telescope and a Hubble Space Telescope WFC3 composite colour image (based on the image processed by Robert Gendler and Judy Schmidt). } \label{fig:regions} 
\end{figure}

\begin{figure}[h]
    \centering
    \includegraphics[width=0.48\textwidth]{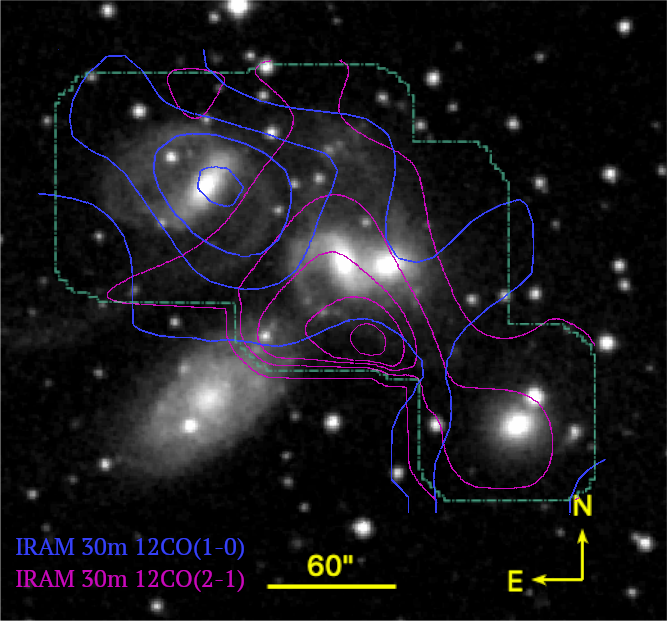}
    \caption{Contours of the CO emission summed over $5100-7500$ km/s per spaxel, as displayed separately in Appendix \ref{app:CO_maps}, for the $^{12}\text{CO~}(1- 0)$ line (contours at 4, 35, 50, and 70 mK) and the $^{12}\text{CO~}(2- 1)$ line (contours at 4, 80, 120, 140, and 160 mK), overlaid on a DSS2 R-band image. The green outline shows the edge of the observed area (the extension of the contours beyond this outline is artificial).} \label{fig:co} 
\end{figure}

The distribution of the stellar continuum emission (in the observed rectangular area as marked in Fig. \ref{fig:slitregion}) is displayed in Fig. \ref{fig:regions}, together with the flux of the H$\alpha\_[\ion{N}{II}]$ sub-band of the spectra, next to the Subaru and Hubble Space Telescope composite image. 
These maps show that while the stellar continuum predominantly originates from the galaxies the ionised gas is concentrated in the SF ridge, the west ridge, the bridge and in or around NGC7319. 
The molecular gas also exhibits a preference for NGC7319 and the SF ridge, as can be seen in Fig. \ref{fig:co}, which displays the extent of the $^{12}\text{CO~}(1- 0)$ and $^{12}\text{CO~}(2- 1)$ emission, although the $^{12}\text{CO~}(2- 1)$ indicates an extension towards NGC7317. 

\begin{figure}[h]
    \centering
    \includegraphics[width=0.48\textwidth]{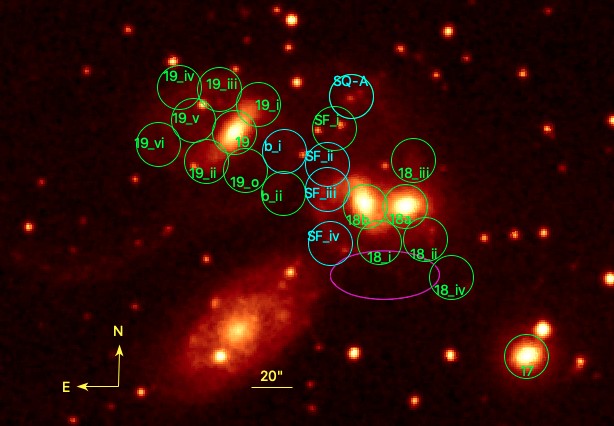}
    \caption{Our 22 regions, with an 11$''$ radius, chosen for closer CO analysis on a DSS2 red image, marked by green and cyan circles. The magenta ellipse is discussed in Sect. \ref{ch:7317}. The cyan regions coincide with the single beam IRAM 30m observations carried out by \citet{guillard2012}. The naming convention of the regions is within each circle, and the RA Dec. of each region is listed in Appendix \ref{app:regions_ra_dec}.}
    \label{fig:radio_regions}
\end{figure}

For further investigation of the data, we chose 29 regions in the optical data, as marked in Fig. \ref{fig:regions}, and 22 regions in the CO data, as shown in Fig.\ref{fig:radio_regions}; the RA Dec. positions for these regions are listed in Appendix \ref{app:regions_ra_dec} together with their HII and H$_2$ gas masses. 
The optical regions have radii of $1.5''$, whereas the CO regions have radii of $11''$. 
We present the spectra of a selection of regions in Appendix \ref{app:spec} and \ref{app:CO_spec}. 
We estimate a total HII region mass of $20.1\pm0.2\cdot10^{10} M_{\odot}$ in the regions chosen for closer analysis. However, this should be considered to be an upper limit estimate as it includes the mass of the HII regions associated with stars. 
In the subsequent sections we will see that SQ contains significant amounts of shocked gas, it is unclear how far the ionised gas mass estimates in shocked regions deviate from those in virial clouds and what the fraction of C- and J-shocks amongst each other and regions of virial clouds is \citep{flower2003,guillet2009,guillet2011}. 

In the area mapped with the IRAM 30m telescope, as marked in Fig.\ref{fig:iram_boxes}, we obtain a total H$_2$ gas mass of $21 \pm 2 \cdot 10^{9} M_{\odot}$ via Eq.\ref{eq:H2_mass} by summing the $^{12}\text{CO~}(1- 0)$ spectra over the velocity range $5100-7500$ km/s, supporting the \citet{gao2000} mass of $\gtrsim 4.7\cdot 10^9 M_{\odot}$. 
In the regions chosen for closer analysis, Fig.\ref{fig:radio_regions}, we obtain a total H$_2$ gas mass of $10 \pm 0.1 \cdot 10^{9} M_{\odot}$ by summing the spectra over the velocity range of $5100-7500$ km/s and $9.8 \pm 2 \cdot 10^{9} M_{\odot}$ from the Gaussian fits. Increasing the S/N threshold from 1.9 to 3 on the Gaussian fits decreases the total H$_2$ gas mass by 24\% to $7.5 \pm 0.4 \cdot 10^{9} M_{\odot}$, whereas it does not affect the masses obtained via summation over the spectra as these have a higher S/N. 
A couple of the CO spectra are subjected to baseline ripples, these have not been corrected for as we aim to give a homogeneous treatment of the entire mapped area, the subsequent discussion does not depend on these spectra, but they have been included in the mass estimates. Furthermore, where the emission line is weak we were restricted to setting an upper limit for the flux obtained via the line fitting procedure, as seen in the tables in Appendix \ref{app:tables} - as the uncertainty includes deviations from the Gaussian line shape, we used a 2$\sigma$ limit on the line flux. 

\citet{rod2014} present optical spectra of the galaxies NGC7319, NGC7318A and B, which our data, detailed in the sections below, show agreement with.\ Our data retain an additional increased extension of the wavelength range covered and a higher resolution, which allows for a kinematical analysis of the area. Our data also solidify the indication made by \citet{natale2010}, that no active star formation occurs near the galaxy centres, as discussed further below using optical diagnostic diagrams when suitable. 
When it comes to previously published CO data of SQ, there are discrepancies. For example, \citet{lisenfeld2002} and \citet{guillard2012} both used an IRAM 30m single pointing of region SQ-A (the position of SQ-A is marked in Fig.\ref{fig:structures}) to obtain the $^{12}\text{CO~}(1- 0)$ spectrum; though they have emission at similar line-of-sight velocities, \citet{guillard2012} show an amplitude a factor of $\sim5$ larger than \citet{lisenfeld2002}. \citet{smith2001} used the National Radio Astronomy Observatory 12 m telescope and obtain an amplitude a factor of $\sim2$ larger than \citet{guillard2012} for the same region, SQ-A. Our data, which in general can be regarded as more consistent across the group than previous observations due to the on-the-fly mapping observing technique adapted (compared to observations and calibrations of single pointings), show a ratio between the lines present at $\sim 5800-6000$ km/s and $\sim 6500-6700$ km/s similar to the line ratio of the \citet{smith2001} lines and an amplitude a factor of two to three lower than that of \citet{guillard2012}. Furthermore, our $^{12}\text{CO~}(1- 0)$ spectrum for SQ-A does not show the velocity component at $\sim 6900$ km/s detected by \citet{guillard2012} as our spectrum shows better agreement with that of \citet{smith2001}. Our $^{12}\text{CO~}(2- 1)$ spectrum does not show a velocity component at $\sim 6500-6700$ km/s either, as observed by \citet{guillard2012}, this is unexpected, as we still detect the lower velocity component. \citet{lisenfeld2002} and \citet{guillard2012} centre the contribution of the intruder galaxy, NGC7318B, to SQ-A at $\sim6000$ km/s, whereas our data state $\sim5800$ km/s, closer to the line-of-sight velocity of the galaxy itself. Further and deeper observations of the molecular gas content and kinematics may prove highly rewarding. 

In this section we present and discuss our data and the kinematics in the regions as a part of an area of SQ, starting with the regions in or near NGC7319 and working our way SW through the bridge, the SF ridge, the NGC7318 pair, the west ridge, and NGC7317.

\subsection{The active galaxy NGC7319}
Classified as a Seyfert 2 galaxy, NGC7319 is the only confirmed active galaxy in the group. 
The extensive emission in this galaxy allowed for a spaxel-by-spaxel fitting of the ionised gas emission lines and the stellar continuum. 
Although there are parts of NGC7319 that display dual gas emission lines, the higher signal-to-noise of the $1.5''$ region spectra is required for a proper fit of the dual lines; therefore, the continuum-subtracted line-emission maps are created with an allowance for a single broader line. 
When studying the multiple velocity components in or near NGC7319, we find that they can often be related to the line-of-sight velocities of not only the central velocity of NGC7319 but also NGC7320C (likely in the creation of the inner tidal tail). 

From the pPXF fit of the stellar kinematics we obtain the mean stellar line-of-sight velocity and velocity dispersion in the central $1.5''$ of NGC7319 as $6814\pm12$ km/s and $159.4\pm13.0$ km/s, respectively. 
The velocity dispersion maps of the gaseous emission show an inclination of a lower velocity dispersion in the centre surrounded by a ring-like structure of higher velocity dispersion. 
We also note that the velocity dispersion measured is a combination of the actual velocity dispersion and the change in the velocity field in that spaxel. 

\begin{figure*}[ht!]
   \subfloat[\label{fig:7319_stellar_flux}]{%
      \includegraphics[width=0.3\textwidth]{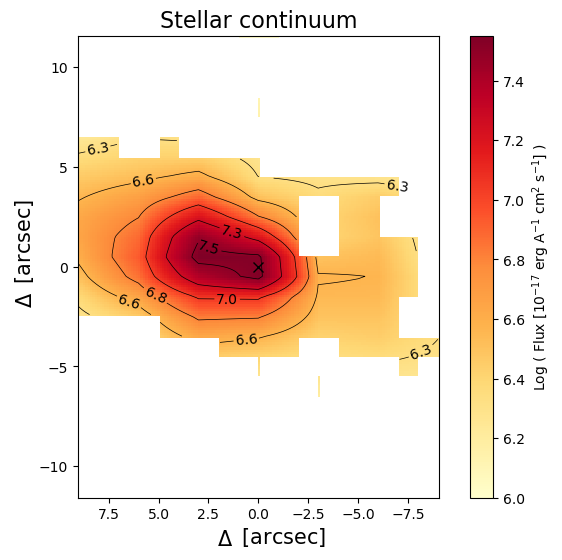}}
\hspace{\fill}
   \subfloat[\label{fig:7319_stellar_vlos} ]{%
      \includegraphics[width=0.3\textwidth]{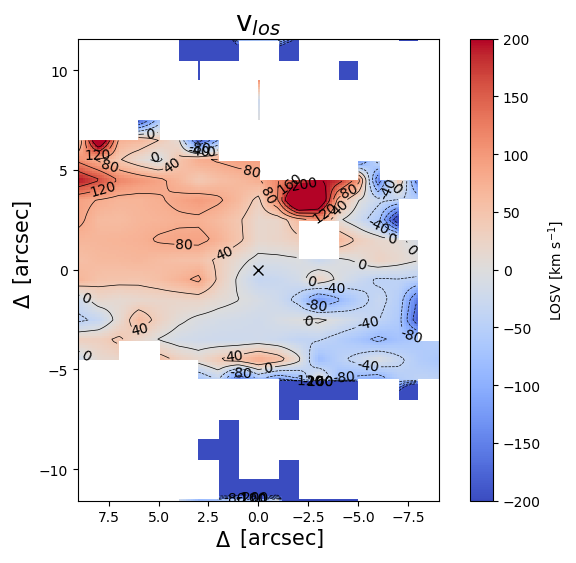}}
\hspace{\fill}
   \subfloat[\label{fig:7319_stellar_sigma}]{%
      \includegraphics[width=0.3\textwidth]{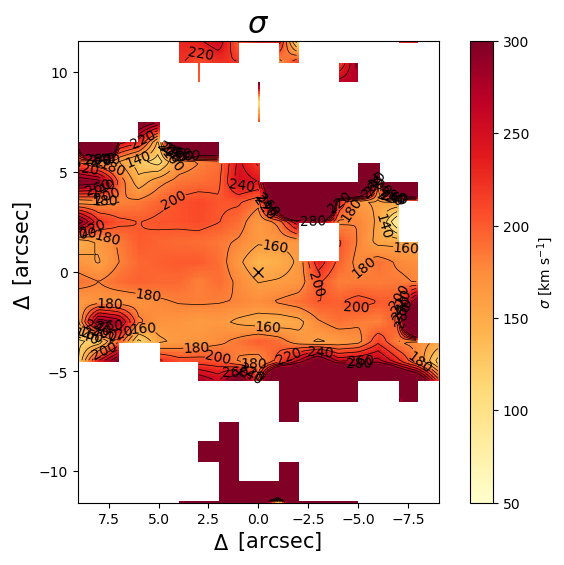}}
    \caption{\label{fig:7319_stellar}{Stellar continuum in NGC7319; map orientation as in Fig. \ref{fig:regions}. The cross marks the stellar continuum peak emission, set as the centre of the galaxy. The 0 line-of-sight velocity is that of the central 1.5$''$. (a) Integrated stellar continuum emission. (b) Line-of-sight velocity. (c) Velocity dispersion.}}
\end{figure*}

\begin{figure*}[ht!]
        \subfloat[\label{fig:7319_OIII_flux}]{%
            \includegraphics[width=0.3\textwidth]{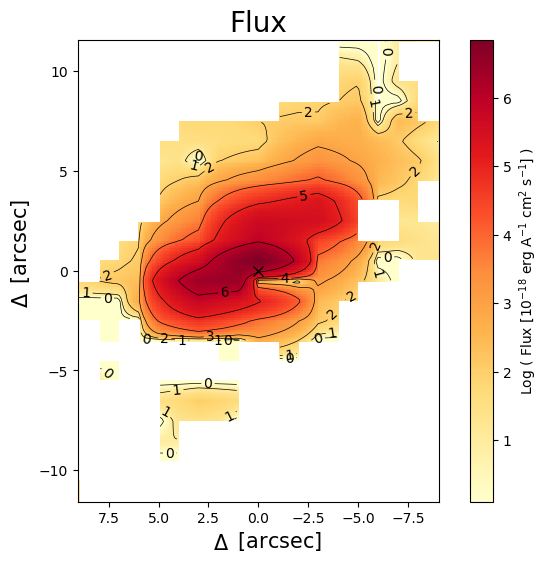}}
            \hspace{\fill}
        \subfloat[\label{fig:7319_OIII_vlos} ]{%
            \includegraphics[width=0.3\textwidth]{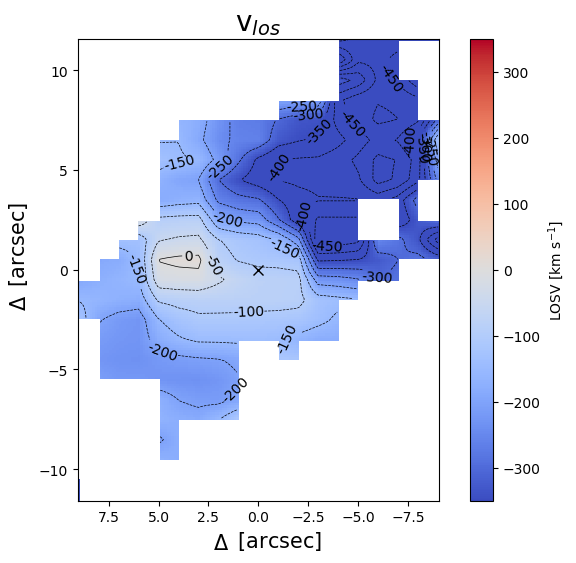}}
            \hspace{\fill}
        \subfloat[\label{fig:7319_OIII_sigma}]{%
            \includegraphics[width=0.3\textwidth]{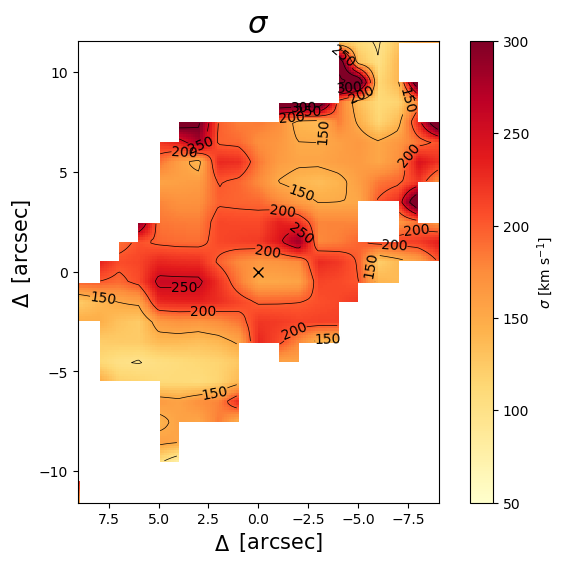}}
        \caption{\label{fig:7319_OIII}[OIII]$\lambda$5008 emission in NGC7319; map orientation as in Fig. \ref{fig:regions}. The cross marks the stellar continuum peak emission, set as the centre of the galaxy. The 0 line-of-sight velocity is that of the central 1.5$''$ of the stellar velocity field. (a) Flux. (b) Line-of-sight velocity. (c) Velocity dispersion.}
    \end{figure*}

\begin{figure}[ht!]
    \centering
      \includegraphics[width=0.3\textwidth]{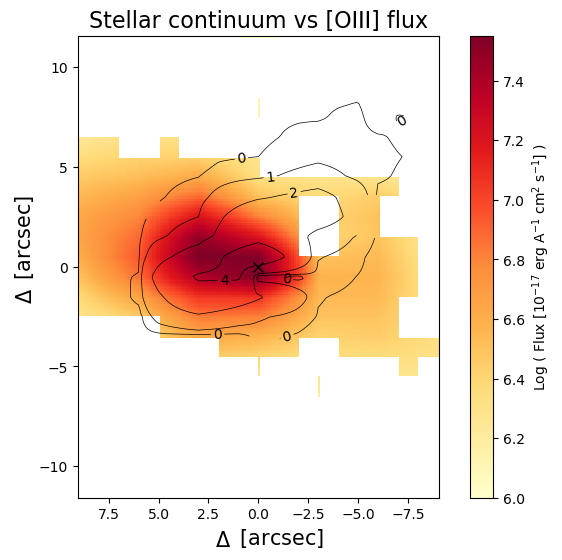}
    \caption{\label{fig:7319stellarOIII_flux} Discrepancy between the extension of the stellar and [OIII]$\lambda5008$ emission in NGC7319. The integrated stellar continuum emission is overlaid with the integrated [OIII] emission line flux contours.}
\end{figure}

\begin{figure}[ht!]
    \centering
      \includegraphics[width=0.3\textwidth]{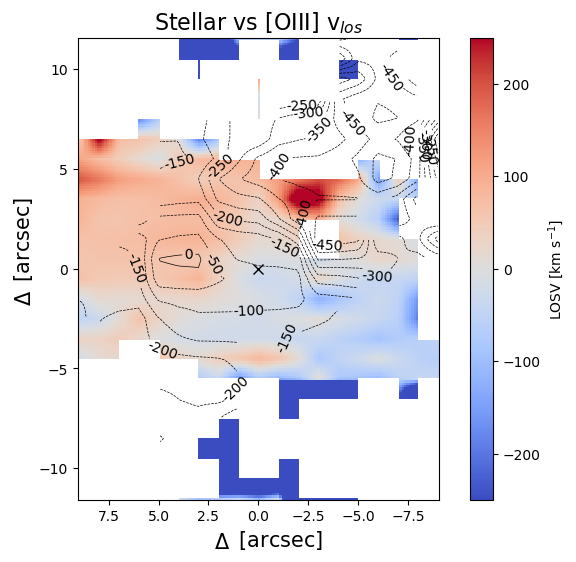}
    \caption{\label{fig:7319stellarOIII_vlos} Perpendicular nature of the stellar and [OIII]$\lambda5008$ velocity fields in NGC7319. The stellar line-of-sight velocity in a blue-to-red colour map is overlaid with the [OIII] line-of-sight velocity contours.}
\end{figure}

The stellar disk and gas disk in NGC7319 are close to perpendicular, as can be seen in Figs. \ref{fig:7319_stellar} and \ref{fig:7319_OIII}, which display the stellar continuum and [OIII]$\lambda5008$Å flux, line-of-sight velocity, $v_{\rm LOS} = v-v_{\rm centre}$ with $v_{\rm centre}$ the stellar line-of-sight velocity measured at the position of the central peak emission, and the velocity dispersion. 
Maps for the emission lines of [OII]$\lambda$3727, H$\beta$, [OI]$\lambda$6302, H$\alpha$, [NII]$\lambda$6585 and [SII]$\lambda$6716,6732 are displayed in Appendix \ref{app:7319_maps}. 
All gas emission maps show a similar extension as the [OIII] map, apart from the [OII] emission, which is more coupled to the bar. 
Figures \ref{fig:7319stellarOIII_flux} and \ref{fig:7319stellarOIII_vlos} clarify the perpendicular nature of the stellar and gas disk, by overlaying the contours of the [OIII] flux and line-of-sight velocity on the stellar counterparts. 

In addition, the centres of the stellar and gas velocity fields appear offset from each other. 
The gas velocity field is shifted by approximately $3.2\Delta''$ to the SW (up and to the right in the figures). 
This velocity field offset may be due to an extinction of the stellar continuum or a displacement of the gas by the previous interactions in the group. 
The outflow may have an important role in the discrepancy between the stellar and gas disks as well. 

The offset and the perpendicular nature of the gas and stellar velocity fields may be indicative of a highly disturbed galaxy, where the gas disk was pulled out of the stellar rotation plane during previous interactions. 
However, it is likely that a large-scale tidal disturbance would have severely affected the bar and stellar disk. 
Although the decoupled gas and stellar disk indicate that a Population I spiral pattern cannot be sustained, the bar and stellar disk remain intact and what we are looking at here is most likely ionised gas in the form of a nuclear or stellar wind. 
In the next section we present our investigation of the ionisation mechanisms that favour the nuclear wind scenario.

\subsubsection{The outflow}
NGC7319 lack radio jets at 22 GHz \citep{baek2019}, but the outflow is clear in optical wavelengths \citep{aoki1996,boschetti2003,dimille2008}. 
Our data support the claim, and south-south-west of the nucleus we find a mean blueshifted line-of-sight velocity of $476.6\pm13.8$ km/s in the [OIII] emission. 
Fortunately our higher resolution data can further map the outflow. 

We study the excitation mechanisms along the outflow using the optical diagnostic diagrams \citep{baldwin1981,veilleux1987}. 
The optical diagnostic diagrams utilises the emission lines [OIII]$\lambda$5008Å, H$\beta$, [NII]$\lambda$6585Å, H$\alpha$, [OI]$\lambda$6302Å, [SII]$\lambda$6718Å and [SII]$\lambda$6732Å to separate active galactic nucleus (AGN) excitation mechanisms from SF mechanisms using six equations:

    \begin{equation} \label{eq:bpt_SII}
        \log\left(\frac{\left[\ion{O}{III}\right]}{\text{H}\beta}\right) = \frac{0.72}{\log(\left[\ion{S}{II}\right]/\text{H}\alpha)-0.32}+1.30,
    \end{equation}
    \begin{equation} \label{eq:bpt_OI}
        \log\left(\frac{\left[\ion{O}{III}\right]}{\text{H}\beta}\right) = \frac{0.73}{\log(\left[\ion{O}{I}\right]/\text{H}\alpha)+0.59}+1.33,
    \end{equation}
        \begin{equation} \label{eq:bpt_NII_0}
        \log\left(\frac{\left[\ion{O}{III}\right]}{\text{H}\beta}\right) = \frac{0.61}{\log(\left[\ion{N}{II}\right]/\text{H}\alpha)-0.47}+1.19,
    \end{equation}
    \begin{equation} \label{eq:bpt_NII_kauffmann}
        \log\left(\frac{\left[\ion{O}{III}\right]}{\text{H}\beta}\right) = \frac{0.61}{\log(\left[\ion{N}{II}\right]/\text{H}\alpha)-0.05}+1.30.
    \end{equation}
     \begin{equation}
        \log\left(\frac{\left[\ion{O}{III}\right]}{\text{H}\beta}\right) = 1.89\log\left(\frac{\left[\ion{S}{II}\right]}{\text{H}\alpha}\right)+0.76,
    \end{equation}
    \begin{equation}\label{eq:bpt_last}
        \log\left(\frac{\left[\ion{O}{III}\right]}{\text{H}\beta}\right) = 1.18\log\left(\frac{\left[\ion{O}{I}\right]}{\text{H}\alpha}\right)+1.30.
    \end{equation}   
    While the first four equations were introduced and expanded upon by \citet{baldwin1981} and \citet{veilleux1987}, Eq. \ref{eq:bpt_NII_kauffmann} is the pure SF line and is based on observations \citep{kauffmann2003}; Eq. \ref{eq:bpt_NII_0} is based on theoretical modelling and is called the maximum starburst line \citep{kewley2001}. 

\begin{figure}
        \centering
        \includegraphics[width=0.48\textwidth]{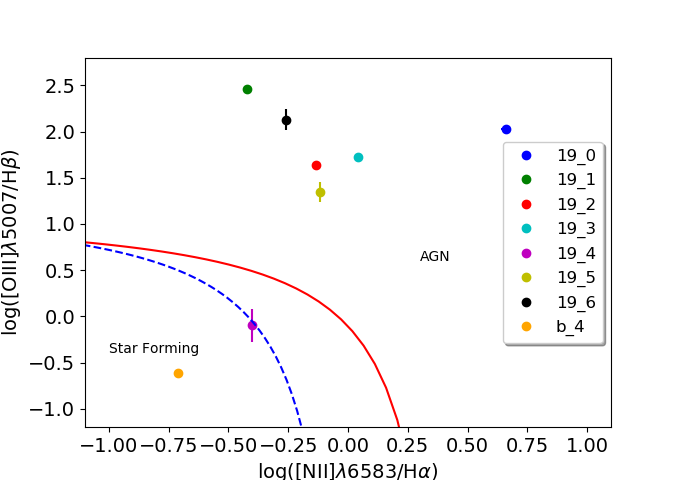}
        \caption{Optical diagnostic diagram, $\log($[NII]$\lambda 6585/$H$\alpha)$ vs. $\log($[OIII]$\lambda 5008/$H$\beta)$, for the regions in or near NGC7319, region 19\_0-6, and the region in the bridge with sufficient emission, b\_4.}
        \label{fig:BPTn}
    \end{figure}

\begin{figure}
        \centering
        \includegraphics[width=0.48\textwidth]{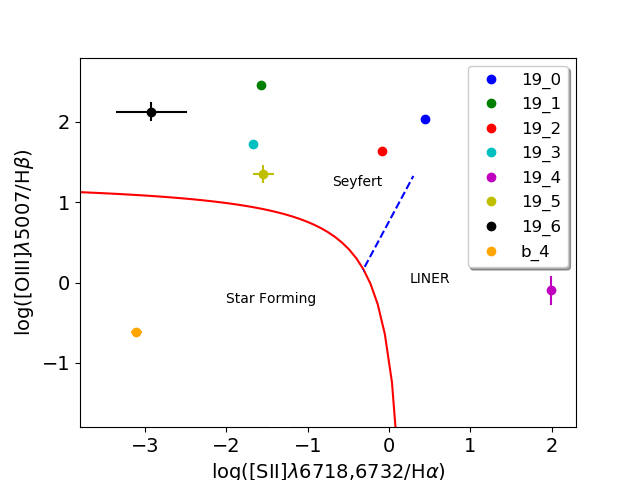}
        \caption{Optical diagnostic diagram, $\log($[SII]$\lambda6718,6732/$H$\alpha)$ vs. $\log($[OIII]$\lambda 5008/$H$\beta)$, for the regions in or near NGC7319, region 19\_0-6, and the region in the bridge with sufficient emission, b\_4.}
        \label{fig:BPTs}
    \end{figure}

\begin{figure}
        \centering
        \includegraphics[width=0.48\textwidth]{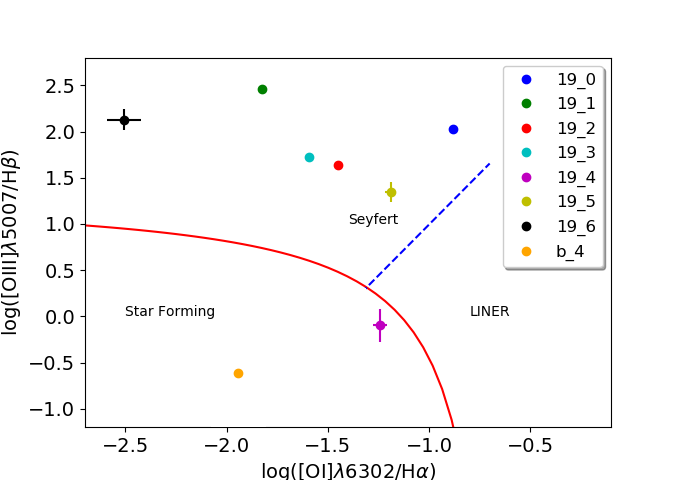}
        \caption{Optical diagnostic diagram, $\log($[OI]$\lambda6302/$H$\alpha)$ vs. $\log($[OIII]$\lambda5008/$H$\beta)$, for the regions in or near NGC7319, region 19\_0-6, and the region in the bridge with sufficient emission, b\_4.}
        \label{fig:BPTo}
    \end{figure}

Studying the individual 1.5$''$ radii regions as marked in Fig. \ref{fig:regions}, we place these regions in the optical diagnostic diagrams and reveal that the gas ionisation throughout the disk is due to nuclear processes, as Figs. \ref{fig:BPTn}-\ref{fig:BPTo} illustrate. 
Region 19\_4 deviates from this claim but the emission in this region is too weak to allow for a definitive statement. 
The results from the fitting of the spectra of all regions in or near NGC7319 are displayed in Appendix \ref{app:tables_7319}. 
    
\begin{figure}[ht!]
        \centering
            \includegraphics[width=0.3\textwidth]{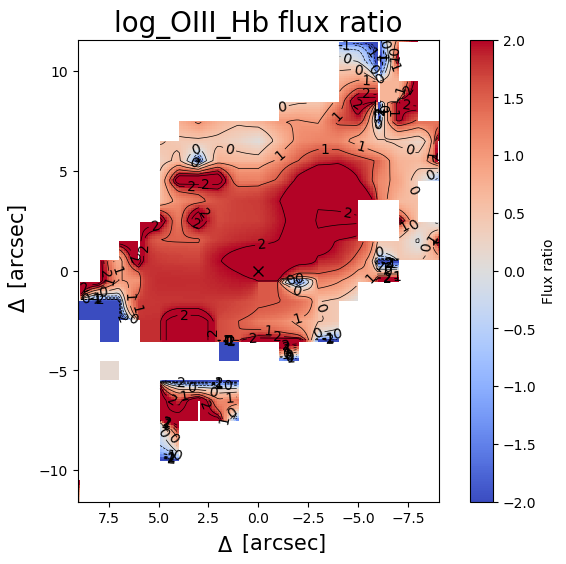}
        \caption{\label{fig:7319_bpt} NGC7319 line ratio map of log([OIII]/H$\beta$).}
    \end{figure}

Furthermore, temperature sensitive line ratios such as [OIII]/H$\beta$ are good indicators of shocked regions since shocked regions have the ability to retain higher temperatures. 
The [OIII]/H$\beta$ ratio, mapped for NGC7319 in Fig.\ref{fig:7319_bpt}, illustrates how the higher ratio values trace the outflowing wind. 

\begin{figure*}[ht!]
        \subfloat[\label{fig:7319_2vel_regions}]{%
            \includegraphics[width=0.3\textwidth]{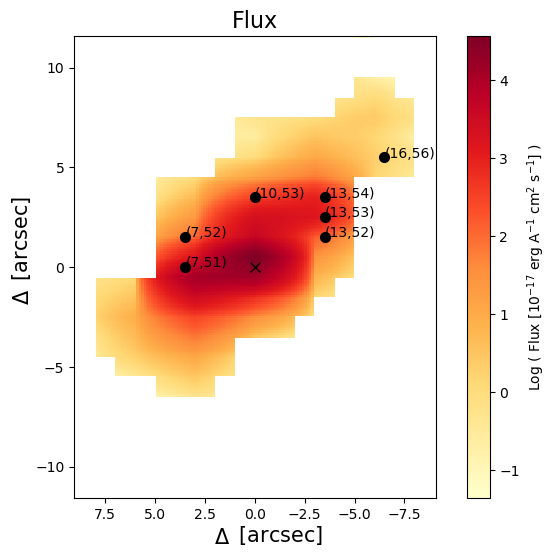}}
        \hspace{\fill}
        \subfloat[\label{fig:7319_2vel_high} ]{%
            \includegraphics[width=0.3\textwidth]{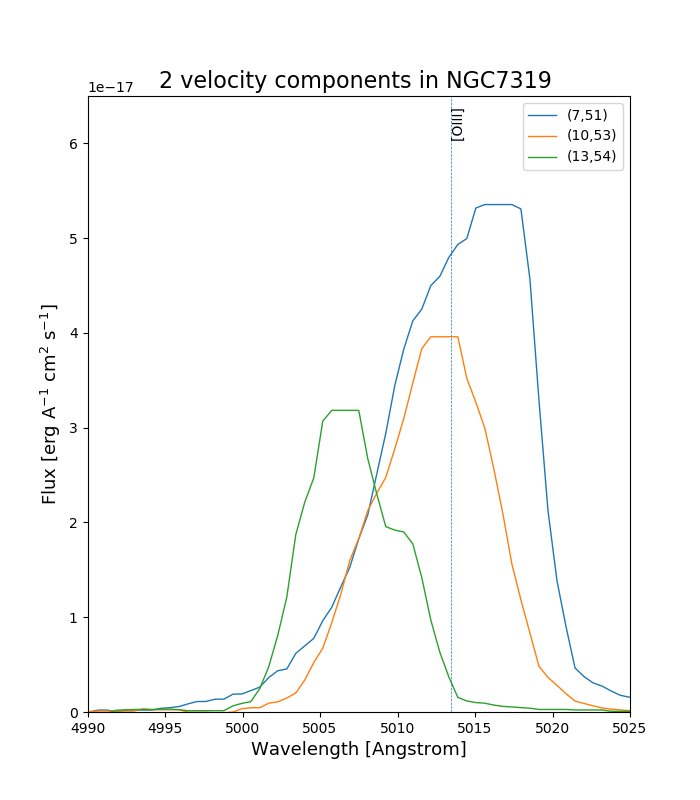}}
        \hspace{\fill}
        \subfloat[\label{fig:7319_2vel_low}]{%
            \includegraphics[width=0.3\textwidth]{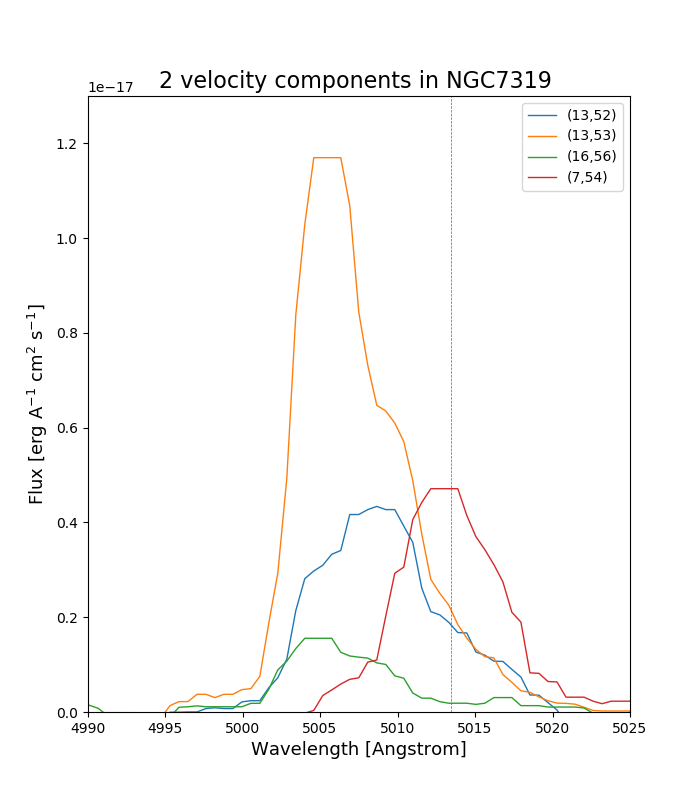}}
        \caption{\label{fig:7319_2vel} Multiple velocity components present in the $\text{[OIII]}\lambda5008$ line in NGC7319 at different positions. (a) Selection of positions in which the spectra show broadening and multiple velocity components in the $\text{[OIII]}$ line.\ Each point represents the $1''\times1''$ spaxel, from which the spectra in (b) and (c) have been extracted (for the reader's reference: the centre of NGC7319 is located at (10,51)). (b) and (c) $\text{[OIII]}$ line shape in the positions marked in (a), separated into two graphs due to differences in peak flux.}
    \end{figure*}

As a gaseous rotating disk without a stellar population is unlikely and as the bar remains undisturbed, it is possible that the brunt of the gas observed in NGC7319 is in the form of a nuclear wind, although a significant stellar contribution cannot be excluded. 
We find duality in the velocity component in the [OIII] line, as illustrated in Fig.\ref{fig:7319_2vel}, which indicates that the gas is on more than one congregation (supported by the fits of the regions, values presented in the table in Appendix \ref{app:tables_7319}). 

\subsubsection{Revealing the Seyfert 1 nature}
\begin{figure}[h]
    \centering
    \includegraphics[width=0.4\textwidth]{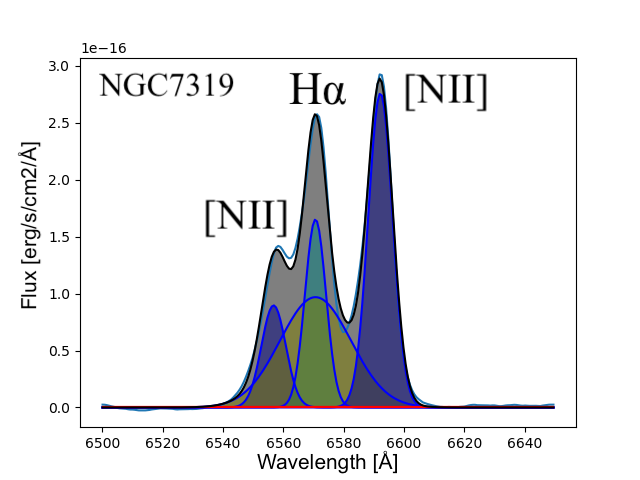}
    \caption{Gaussian fit of the emission in the H$\alpha\_$[NII] sub-band in the central 1.5$''$ of NGC7319.}
    \label{fig:7319_HalphaNII_0}
\end{figure}

Although this galaxy has been classified as a Seyfert 2, our higher resolution spectrum of the inner 1.5$''$ of NGC7319, presented in Fig. \ref{fig:19_0}, shows a broad-line region (BLR). 
The requirement of a BLR for an accurate fit is clear in Fig.\ref{fig:7319_HalphaNII_0}, which displays the fit of the H$\alpha\_$[NII] sub-band of the spectrum and provides the FWHM of the narrow-line region (NLR) and the BLR as $374\pm10$ km/s and  $1265\pm43$ km/s, respectively. 

Furthermore, we can estimate the black hole mass of NGC7319 to $\sim 6\cdot 10^7 M_{\odot}$ using the $M_{\rm{BH}}-\sigma$ relation \citep{mcconnell2011}, 
\begin{equation}
    \frac{M_{\rm{BH}}}{10^8M_{\odot}}\approx 1.9\left( \frac{\sigma}{200~\text{km/s}} \right)^{5.1}.
\end{equation}
The relation used to calculate the black hole mass has intrinsic scatter, and so we refrain from quoting any error bars. The black hole mass should be taken to be an order of magnitude estimate. 

\subsubsection{The molecular gas content}
The results of the fits of the regions in or near NGC7319 are listed in the tables in Appendix \ref{app:tables_7319}, including the estimation of the H$_2$ gas mass calculated using Eq. \ref{eq:H2_mass}. 
Of all of the regions analysed, marked in Fig. \ref{fig:radio_regions}, $56$\% of the molecular gas mass is located in or near NGC7319, as calculated using the masses estimated from the Gaussian fits, and in the central 11$''$ of NGC7319 we observe an H$_2$ gas mass of $10\pm1\cdot 10^8 M_{\odot}$ (obtained from the Gaussian fits). 
In region 19, 19\_i-vi and 19\_o we obtain a total H$_2$ gas mass of $55\pm3 \cdot 10^8 M_{\odot}$ from the Gaussian fits, supporting \citet{gao2000} estimate of $\gtrsim 36 \cdot 10^8 M_{\odot}$, and by summing the spectral emission over the velocity range $5100-7500$ km/s we obtain a total H$_2$ gas mass in or near NGC7319 of $53\pm1 \cdot 10^8 M_{\odot}$. 
The spectrum of the central region is presented in Appendix \ref{app:CO_spec} together with a selection of additional regions' spectra. 

The central gas deposit may be important in the feeding of the AGN or the outflow. 
Furthermore, the brunt of the CO gas in or near NGC7319 lie in the velocity range of $6400-6600$ km/s, although there is also a significant higher velocity component at $\sim6800$ km/s. 
These velocities are in agreement with the velocity range of $6400-6800$ km/s found in the NGC7319 region by \citet{gao2000}. 
In addition, we see CO emission throughout the bar and extending to the north-east and towards the ridge (emission that does not appear in the interferometric maps of \citet{gao2000} as it was likely filtered out due to the extended nature of the emission).

\subsection{The bridge}
\begin{figure}[ht!]
    \centering
      \includegraphics[width=0.4\textwidth]{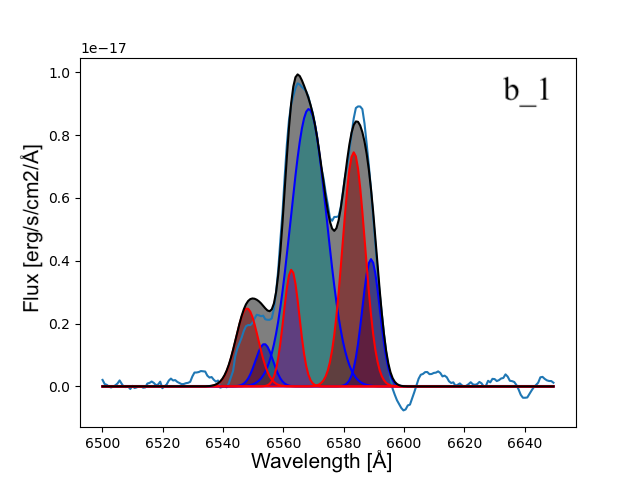}
\caption{\label{fig:b1_Ha} Fit of the emission in the H$\alpha\_$[NII] sub-band in the bridge region b\_1. The blue lines denote one velocity component and the red ones the other the two velocity components, which are separated by $261\pm14$ km/s (see the table in Appendix \ref{app:tables_bridge_ridge}). The black line is the resultant best fit.}
\end{figure}

The bridge (marked in Fig. \ref{fig:structures}) is a fairly newly detected structure \citep{gao2000}, and we confirm its presence in both optical and radio, as shown in Figs. \ref{fig:regions} and \ref{fig:radio_regions}, and detail its kinematics. 
The velocity components range $5650-7000$ km/s and in several cases show split lines, as seen Fig. \ref{fig:b1_Ha}, which shows the fit of the H$\alpha\_$[NII] sub-band for region b\_1. 
To enable the dual fit of the complex mixed emission lines a pre-condition for the fit must be set. 
This condition fixes the ratio of the [NII] lines to [NII]$\lambda6585.26/6549.86=3$ and the ratio [SII]$\lambda6732.67/6718.29=1.3$. 
These ratio values are the averages as provided by \citet{osterbrockbible} and enable a sufficient fit, which allows us to focus on the kinematics and velocities in the group without a significant decrease in the effect and robustness of the kinematical analysis. 

Closer to NGC7319, the bridge's velocities are higher, whereas closer to the SF ridge they are lower and show a tendency to agree with those of the SF ridge. 
As we move closer to the SF ridge, the amount of ionised gas also increases, as seen in the tables containing the values of the fits of the emission lines presented in Appendix \ref{app:tables_bridge_ridge}. 
In CO the bridge also shows a proclivity to similarities of the SF ridge, while leaning towards higher line-of-sight velocities, as expected due to its proximity to the high velocity NGC7319. 

The only region with sufficient signal-to-noise in the emission lines required for the optical diagnostic diagram is region b\_4, and this region has been placed in the diagrams presented in Figs. \ref{fig:BPTn}-\ref{fig:BPTo}. 
As can be seen region b\_4 is located close to the ridge and is excited by stellar processes.  
Understanding the bridge may give vital clues to the current interaction with NGC7318B and the group. 

\subsection{The star-forming ridge}
\begin{figure}[h]
\centering
      \includegraphics[width=0.4\textwidth]{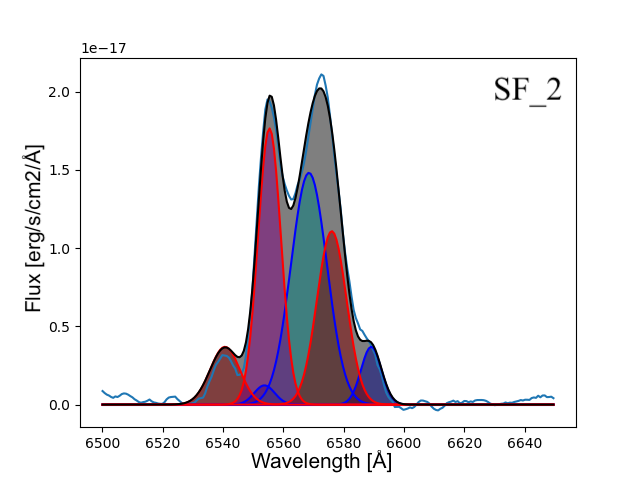}
\caption{\label{fig:SF_2_Ha} {Illustrating the dual velocity components in the H$\alpha\_$[NII] emission in the SF ridge, region SF\_2. The black line is the best fit, the lines that belong to one velocity component are in red, and those belonging to the other component are in blue. The two velocity components are separated by $598\pm10$ km/s (table in Appendix \ref{app:tables_bridge_ridge}).}}
\end{figure}

\begin{figure}[h]
\centering
      \includegraphics[width=0.4\textwidth]{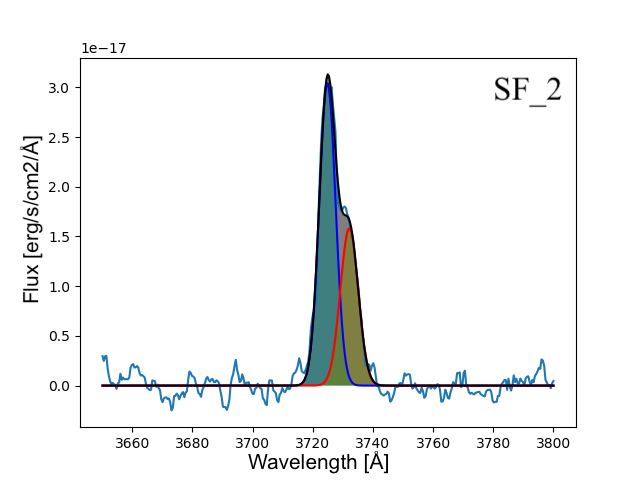}
\caption{\label{fig:SF_2_OII} Illustrating the dual velocity components in the [OII]$\lambda$3727 line in the SF ridge, region SF\_2. The black line is the best fit, and the two different colours denote the different velocity components. The two velocity components are separated by $583\pm21$ km/s (table in Appendix \ref{app:tables_bridge_ridge}).}
\end{figure}

Gas stripped from the participants of the past interactions has been deposited in the IGM, facilitating the creation of the galaxy-wide shock-induced SF ridge in the collision of the IGM and NGC7318B. 
The SF ridge (marked in Fig. \ref{fig:structures}) contains large amounts of ionised gas and an absence of an older stellar population; this is seen by the negligible stellar continuum emission, as is clear from the spectra (see for example the spectrum of region SF\_2 presented in Fig. \ref{fig:SF_2}). 
As \citet{duarte2019} show, there is a significant part of the SF ridge that contains dual velocity components, as we clarify and expand upon in our analysis and the tables presented in Appendix \ref{app:tables_bridge_ridge}. 
If we again turn our attention to region SF\_2, looking closer at the [OII]$\lambda3727$ line and the H$\alpha\_$[NII] sub-band and the fits of these, presented in Figs. \ref{fig:SF_2_Ha} and \ref{fig:SF_2_OII}, the gas congregations at two different velocities are clear. 

We find velocities spanning $5670-7100$ km/s in the SF ridge and components centring at multiple congregations, and as we move closer to NGC7318B the low velocity component similar to the line-of-sight velocity of NGC7318B becomes clearer, showing the mixed medium. 
Farther from NGC7318B, the ionised gas in the SF ridge shows dual velocities at $\sim6100$ km/s and $\sim6900$ km/s, combining gas of the higher velocity of NGC7319, and a gas mix of the lower velocity of NGC7320C, NGC7318A, NGC7318B, and potentially NGC7317. 
The CO gas indicates up to four velocity components; for example, in region SF\_ii these components centre around $5880\pm10$ km/s, $6554\pm24$ km/s, and $6785\pm17$ km/s, which can be related to NGC7318B, NGC7318A or NGC7317, and NGC7319, respectively. 

\begin{figure}[h]
        \centering
        \includegraphics[width=0.48\textwidth]{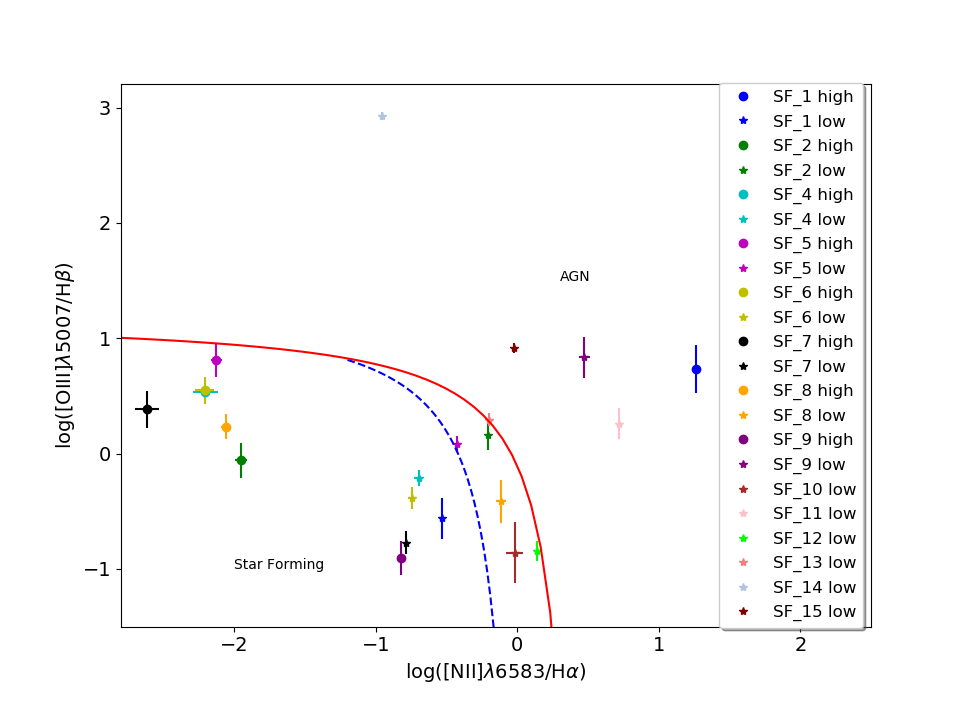}
        \caption{Optical diagnostic diagram, $\log($[NII]$\lambda 6585/$H$\alpha)$ vs. $\log($[OIII]$\lambda 5008/$H$\beta)$, for the SF ridge, region SF\_1-11, and the west ridge, region SF\_12-15. The low velocity component is marked `low' and the high `high'.}
        \label{fig:SF_BPTn}
    \end{figure}

\begin{figure}[h]
        \centering
        \includegraphics[width=0.48\textwidth]{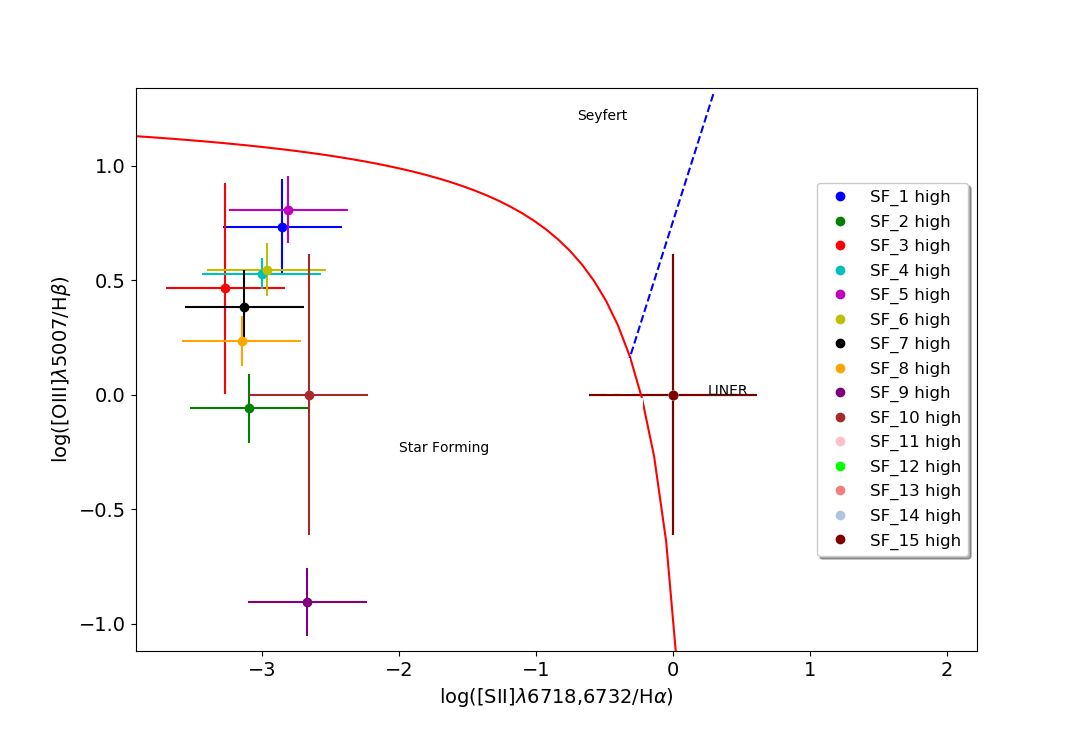}
        \caption{Optical diagnostic diagram, $\log($[SII]$\lambda6718,6732/$H$\alpha)$ vs. $\log($[OIII]$\lambda 5008/$H$\beta)$, for the SF ridge, region SF\_1-11, and the west ridge, region SF\_12-15. The low velocity component is marked `low' and the high `high'.}
        \label{fig:SF_BPTs}
    \end{figure}

\begin{figure}[h]
        \centering
        \includegraphics[width=0.48\textwidth]{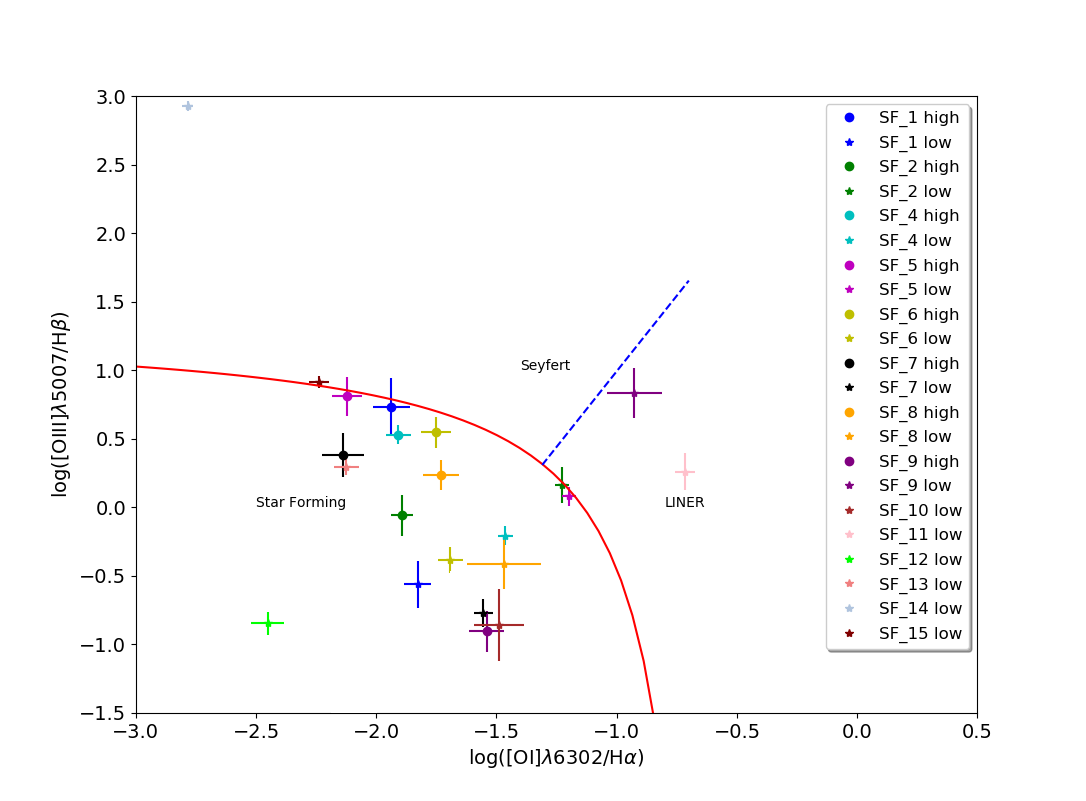}
        \caption{Optical diagnostic diagram, $\log($[OI]$\lambda6302/$H$\alpha)$ vs. $\log($[OIII]$\lambda5008/$H$\beta)$, for the SF ridge, region SF\_1-11, and the west ridge, region SF\_12-15. The low velocity component is marked `low' and the high `high'.}
        \label{fig:SF_BPTo}
    \end{figure}

Denoting the dual velocity components, when present, as `high' and `low' for each region we place them into the optical diagnostic diagrams (Eq.\ref{eq:bpt_SII}-\ref{eq:bpt_last}) in Fig.\ref{fig:SF_BPTn}-\ref{fig:SF_BPTo} (the regions of the west ridge, region SF\_12-15, which will be discussed in the subsequent section, section \ref{ch:west_ridge}, are also included in these plots). 
Primarily the regions gather in the part of the plots showing ionisation from star-formation, as expected from an area called the SF ridge. 
However, there are several regions that show LINER-like and even AGN line ratios. 
This high ionisation is caused by the shock that is created by the high relative velocity collision between the IGM and the intruder galaxy, NGC7318B, as LINER-like emission-line ratios can be incited by shocks. 
If the shock is travelling fast enough a photo-ionised precursor can incite the gas further into Seyfert-like emission line ratios \citep{allen2008,rod2014}. 
It is clear that the shock has a significant impact on the IGM gas of SQ. 

The SF ridge, region SF\_i-iv as marked in Fig.\ref{fig:radio_regions}, retains $20.7 \pm 0.5 \cdot 10^8 M_{\odot}$ (obtained by summing the spectral emission over the velocity range $5100-7500$ km/s), corresponding to $\sim20$\%, of the H$_2$ gas mass present in the analysed regions. 
Compared to the regions in or near NGC7319, the SF ridge shows distinctly more $^{12}\text{CO~}(2-1)$ emission and a higher complexity in the gas congregations present, considering the multiple velocities (see the spectra in Figs. \ref{fig:12CO10_SF_i}-\ref{fig:13CO10_SF_ii}). 
Furthermore, as seen in Table \ref{table:CO_SF}, the $^{12}\text{CO~}(2-1)$ indicates a preference to lower line-of-sight velocities than $^{12}\text{CO~}(1-0)$, allowing us to correlate the higher ionisation to the intruder galaxy and the lower ionisation to IGM gas deposited during past passages by NGC7317, NGC7318A, and NGC7320C.

\subsection{The NGC7318 pair}
As the spiral galaxy NGC7318B enters SQ from behind it interacts primarily with the IGM and NGC7318A. 
Due to the high line-of-sight velocity dispersion between NGC7318A and B, the pair is not expected to merge. 
NGC7318A and B are both quiescent galaxies, as clear from their optical spectra presented in Fig.\ref{fig:18a} and \ref{fig:18b}, and their low molecular gas content (listed in the tables in Appendix \ref{app:tables_7318}). 
The stellar continuum is strong with prominent absorption lines, which, fitted with the pPXF routine, provides maps of velocity dispersion and line-of-sight velocity as a function of position. 
These maps, at a S/N>2, are presented in Appendix \ref{app:7318_maps} and provide the line-of-sight velocity and velocity dispersion in the inner 1.5$''$ of NGC7318A as $6787\pm5$ km/s and $263.6\pm10.7$ km/s, and in the inner 1.5$''$ of NGC7318B as $5973\pm4$ km/s and $168.9\pm4.0$ km/s. 
Naturally, the velocities in the NGC7318B map increase as we move closer to the SF ridge. 

\begin{figure}[h]
    \centering
    \includegraphics[width=0.4\textwidth]{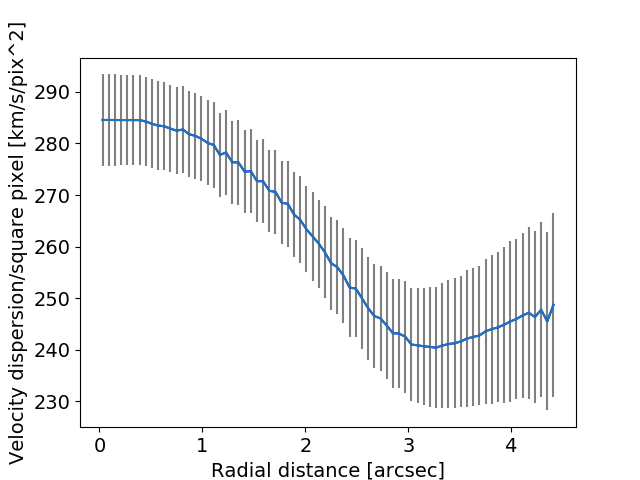}
    \caption{Velocity dispersion of NGC7318A as a function of radius.}
    \label{fig:7318a_sig}
\end{figure}

Studying the velocity dispersion as a function of radius in NGC7318A, Fig. \ref{fig:7318a_sig}, reveals a typical curve common for elliptical galaxies. 
From the line-of-sight velocity map in Appendix \ref{app:7318_maps}, a position angle of 62\degree is obtained. 

\begin{figure*}[ht!]
        \subfloat[\label{fig:7318_rotfield_vlos}]{%
        \includegraphics[width=0.3\textwidth]{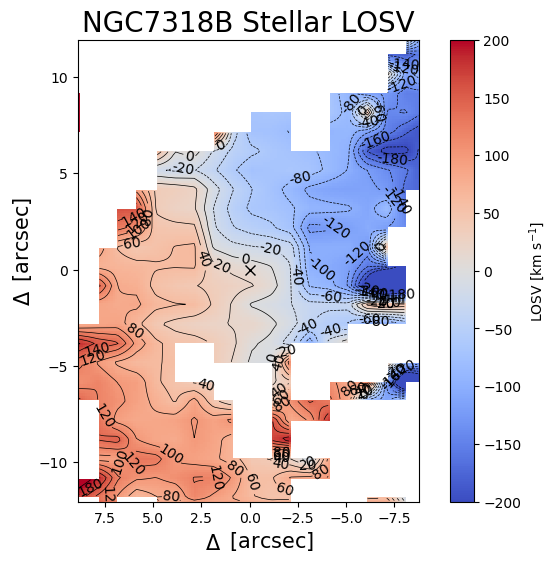}}
        \hspace{\fill}
        \subfloat[\label{fig:7318_rotfield_model} ]{%
            \includegraphics[width=0.3\textwidth]{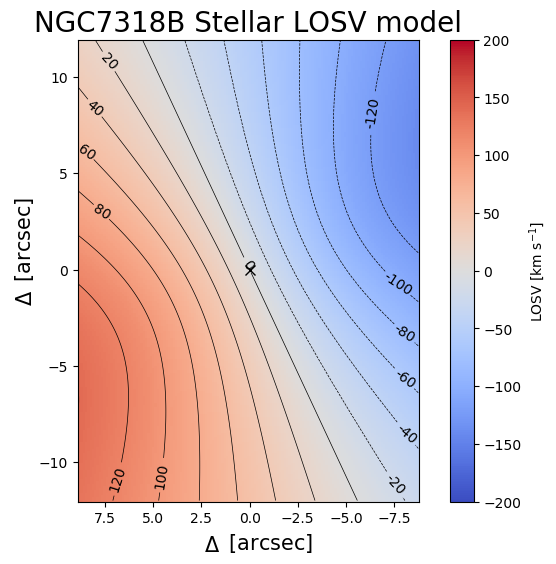}}
        \hspace{\fill}
        \subfloat[\label{fig:7318_rotfield_res}]{%
            \includegraphics[width=0.3\textwidth]{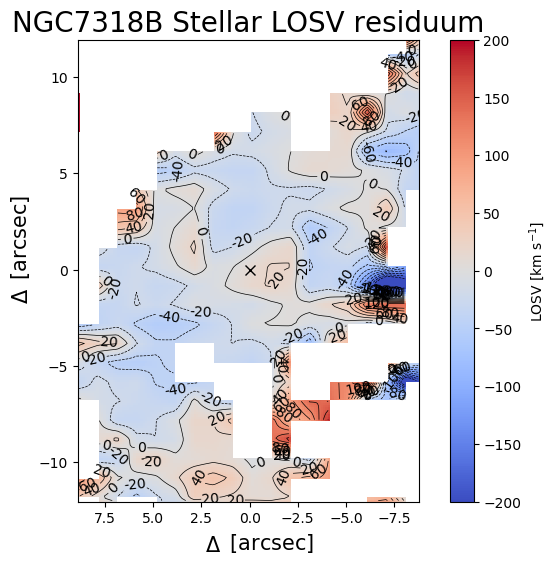}}
        \caption{\label{fig:7318_rotfield} Fit of the stellar line-of-sight velocity field in NGC7318B. (a) Observed stellar line-of-sight velocity at a S/N$>3$. (b) Model used to fit the rotation field. (c) Residual. }
    \end{figure*}

For the spiral galaxy, NGC7318B, we model the rotation field using the model presented by \citet{bertola1991}, for a rotating disk with circular orbits in the galaxy plane, 
        \begin{equation}\label{eq:rot_field}
            V_r = V_s +
            \frac{ARcos(\psi-\psi_0)sin(i)cos^p(i)}{\left(R^2\left[sin^2(\psi-\psi_0)+cos^2(\psi-\psi_0)\right]+c_0^2cos^2(i)\right)^{p/2}}
        ,\end{equation}
        where $V_r$ is the rotation velocity as a function of the polar coordinates R and $\psi$, $V_s$ is the systemic velocity, $\psi_0$ is the position angle, $A$ is the rotation field amplitude, $i$ is the disk inclination, $p$ is a measure of the slope of the rotation field, and $c_0$ is a concentration parameter. 
To ensure a proper fit, we increased the S/N restraint from 2 to 3 on the data used. 
The fit yields the results presented in Fig. \ref{fig:7318_rotfield} with a systemic line-of-sight velocity of $5973\pm4$ km/s, an inclination of $102\pm1.9\degree$, a rotation field amplitude of 290 km/s, and a position angle of $25\degree$. 

Although the gas emission lines are very faint, they can be fitted with Gaussian functions and provide a suggestion for the process of the interaction with NGC7318B. 
The ionised gas in NGC7318A and B presents line-of-sight velocities of $6003\pm28$ kms/s and $6142\pm26$ km/s, respectively, giving an indication that the ionised gas of NGC7318B is either mixed with a higher velocity IGM or decoupled from the stellar disk due to the interaction. 
The CO gas shows congregations at $5815\pm16$ km/s and $5800\pm25$ km/s at the location of NGC7317A and B, respectively, as well as fainter emission at higher velocities. 
In region 18\_i-iii we find a combination of gas at velocities matching both NGC7318A and B. 
The CO emission in the NW tail (marked in Fig.\ref{fig:structures}), region 18\_iii, indicates that there is a mix of high temperature molecular gas from NGC7318B coexisting with lower temperature molecular gas from NGC7318A. 
In addition, there is a lack of gas in the area between the two galaxies. 

Considering the line-of-sight velocities in the bridge, the SF ridge, and the west ridge, we can extrapolate a picture of a galaxy, NGC7318B, entering the group from behind, slightly from the SW. 
NGC7318B collides with the IGM at high speed, creating the shock-induced SF ridge. As NGC7318B passes through the group, its ISM is pulled out and back towards NGC7319 and NGC7318A, forming arms and tails that connect the structures and contributing to the bridge and the gas deposit at the lower velocity in front of NGC7318A. 
It is also possible that the lower velocity gas in front of NGC7318A originates from a past interaction with NGC7320C, although the aforementioned scenario, with NGC7318B in front, fits the data better.

\subsection{The west ridge}
\label{ch:west_ridge}
The west ridge, located to the west of the NGC7318 pair (as marked in Fig.\ref{fig:structures}), has ionised gas at velocities of $\sim5770$ km/s, $\sim6180$ km/s, and $\sim7000$ km/s (as shown in the tables in Appendix \ref{app:tables_bridge_ridge}) and no detectable CO emission, indicating that it is a mix of NGC7318A and NGC7318B gas (and potentially NGC7317, as the line-of-sight velocities are too similar to present an unambiguous statement). 
The line-of-sight velocities we obtain for H$\alpha$ emission in the west ridge coincide with those presented by \citet{duarte2019}. 
The west ridge may continue outside of the area mapped with the LBT (see \citealt{duarte2019}, whose H$\alpha$ maps show a number of SF regions scattered in the surrounding area). 
But whether regions SF\_14-15 and 18\_ii are a continuation of the NW and SW tails and influenced by the shock is yet to be determined. 
In the simulations of SQ by \citet{renaud2010} and \citet{hwang2012}, the west ridge has not been fully reproduced (nor has the SW tail). Although, \citet{renaud2010} indicate that the west ridge may have been created in the interaction between NGC7318A and B. Also, incorporating the work by \citet{struck2012} on tidal tail formation in disk galaxy collisions into the simulations of SQ may take us a step closer to a simulation that better matches the observations (this is, however, beyond the scope of this paper). 

The optical diagnostic diagrams show that a shock is involved in the ionisation of the west ridge as illustrated in Fig.\ref{fig:SF_BPTn}-\ref{fig:SF_BPTo}, where region SF\_14 reaches Seyfert-like emission line ratios, while region SF\_15 LINER-like. 
This high ionisation is likely due to a shock travelling through the medium caused by the emergence of NGC7318B, a shock that is formed in the collision of NGC7318A and NGC7318B gas or in the compressed leading edges of the tidal arms from NGC7318B \citep{struck2012}, supporting the \citet{renaud2010} indication that the west ridge is due to the interaction between NGC7318A and B. 
The west ridge is very similar in composition to the SF ridge, but with a higher oxygen content and higher ionisation. 
The west ridge also appears in X-ray \citep{trinchieri2003,trinchieri2005} and HI observations \citep{williams2002}; therefore, it is clear that the west ridge contains hot IGM, ionised gas and a significant amount of neutral hydrogen but negligible molecular gas.

\subsection{NGC7317, the past intruder?}
\label{ch:7317}
Whether NGC7317 has passed through the group in the past or not is still unclear, although the diffuse extended stellar halo observed by \citet{duc2018} suggests that an interaction has occurred. 
Our CO data indicate an extension in the $^{12}\text{CO~}(2-1)$ emission towards NGC7317, although the emission in the galaxy itself is negligible. 
Our CO emission hints at a spatial extension similar to that of the H$_2$ emission, as presented by \citet{cluver2010}, where the data indicate the areas in which the gas congregates: NGC7319, SQ-A, the SF ridge, and south of the NGC7318 pair. 
Our map covers a larger area and reveals the peak of the southern congregation. 
\citet{natale2010} also detected far-infrared dust emission in this region. 
The correlation between the CO and H$_2$ maps corroborates the southern molecular gas congregation, although it is, interestingly, primarily related to the $^{12}\text{CO~}(2-1)$ line, as seen in the spectra in Fig. \ref{fig:CO_Sblob}. These figures display the average spectra of the area marked with the magenta ellipse in Fig. \ref{fig:radio_regions}, covering 1024 $arcsec^2$. 

This southern CO gas may have remained in its high energy state due to its diffuse nature, disallowing interactions and de-excitation through collisions. 
Further observations are required to fully understand these processes and the effect of the harassment of NGC7317. 

\begin{figure}[h]
        \centering
        \includegraphics[width=0.48\textwidth]{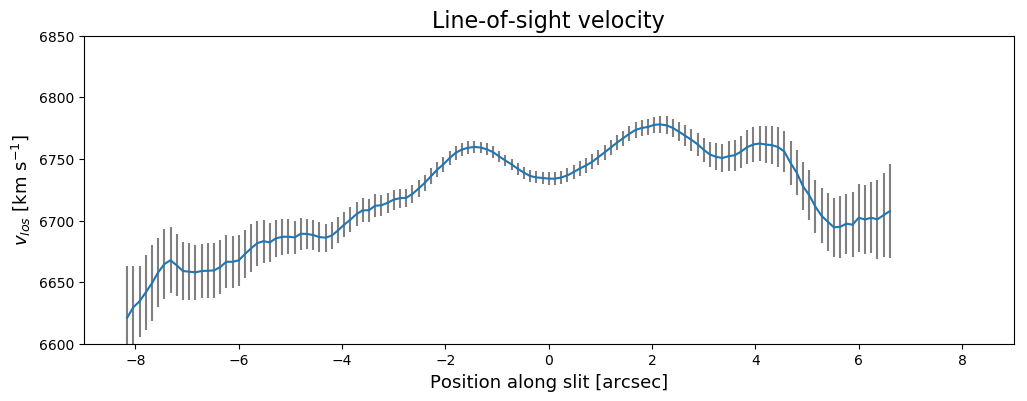}
        \caption{Line-of-sight velocity of the stellar component of NGC7317 as a function of position along the slit, offset from the 0 position, which is the centre of the galaxy, set to the peak stellar continuum emission.}
        \label{fig:7317_vlos}
    \end{figure}

\begin{figure}[h]
        \centering
        \includegraphics[width=0.48\textwidth]{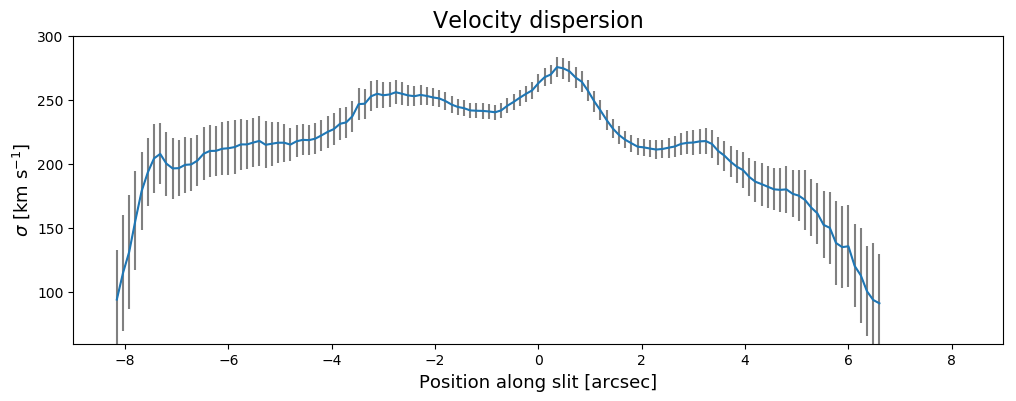}
        \caption{Velocity dispersion of the stellar component of NGC7317 as a function of position along the slit, offset from the 0 position, which is the centre of the galaxy.}
        \label{fig:7317_sig}
    \end{figure}

The optical spectrum of NGC7317 (Fig. \ref{fig:7317_spec}) reveals a quiescent galaxy with a line-of-sight velocity and velocity dispersion in the central 1.5$''$ of $6734\pm5$ km/s and $262.7\pm6.9$ km/s, respectively. 
Having observed NGC7317 at one slit position allows plotting the line-of-sight velocity and velocity dispersion as a function of the offset from the galaxy centre along the slit. 
These velocity-position plots are presented in Figs. \ref{fig:7317_vlos} and \ref{fig:7317_sig}, and they reveal an usual elliptical galaxy, mildly rotating and potentially still in morphological transition. 
These peculiarities may be due to the effects of galaxy harassment from when the galaxy passed through SQ in the past. 
NGC7317 may be an optimal candidate for studying the effect of galaxy harassment. 

\subsection{Optical depth and excitation temperature}
The optical depth and excitation temperature of the CO gas can be estimated using the ratios of the CO lines \citep{eckart1990,nishimura2015,zschaechner2018}. 
We note that the ratios derived in this section rely on the spectra extracted from the 22$''$ regions (as marked in Fig. \ref{fig:radio_regions}), which were extracted from the $^{12}\text{CO~}(1-0)$ and $^{12}\text{CO~}(2-1)$ maps that had been deconvolved and smoothed to a common resolution of a 50$''$ HPBW. 

The excitation temperature, $T_{\rm ex}$, can be estimated from the main beam brightness temperature, $T_{\rm mb}$, as 
\begin{equation} \label{eq:temp}
        \frac{T^{21}_{\rm mb}}{T^{10}_{\rm mb}} = \frac{1-e^{-\tau_{21}}}{1-e^{-\tau_{21}}} \frac{((h\nu_{21}/k)[exp(h\nu_{21}/kT_{\rm ex})-1]^{-1}-T^{21}_{\rm bg})}{((h\nu_{10}/k)[exp(h\nu_{10}/kT_{\rm ex})-1]^{-1}-T^{10}_{\rm bg})},
\end{equation}
assuming local thermal equilibrium, where 
$\nu$ is the transitional frequency and $\tau$ the optical depth of the respective lines and
$T_{\rm bg}$ is the contribution from the cosmic microwave background at the frequency in question. \\
Equation \ref{eq:temp} can, in the optically thick case, $\tau \gg 1$, be written as
\begin{equation} 
    \frac{T^{12\rm{CO~}(2- 1)}_{mb}}{T^{12\rm{CO~}(1- 0)}_{mb}} = \frac{11.03[\exp(11.03/T_{\rm ex})-1]^{-1}-0.20}{5.52[\exp(5.52/T_{\rm ex})-1]^{-1}-0.85},
\end{equation}
whereas in the optically thin case it can be estimated as
\begin{equation}
    \frac{T^{12\rm{CO~}(2-1)}_{mb}}{T^{12\rm{CO~}(1- 0)}_{\rm mb}} \approx 4 ~\exp \left( -\frac{11.03}{T_{\rm ex}} \right).
\end{equation}
These two cases show us that a line ratio of $(2-1)/(1-0)$ near 1 would indicate dense, warm, optically thick gas, while a line ratio near 4 would relate to dense, warm, optically thin gas, and a ratio value below 1 may reveal sub-thermally excited molecular gas at low temperature (T$\lesssim 10$K) \citep{eckart1990}. 

Studying the line ratios in our data reveal that the regions in or near NGC7319 and the bridge favours sub-thermally excited low temperature gas, while the SF ridge, SQ-A and region 18\_i retains warm, dense, optically thick gas. 
In addition the low velocity component in region SF\_iii and region 18\_i show a proclivity for warm, dense, optically thin gas as the ratio value approaches 4. 
The values of the main beam temperatures and the ratios for all regions with emission in both $^{12}\text{CO~}(1-0)$ and $^{12}\text{CO~}(2-1)$ within a velocity range of 3 channels (i.e. with a max separation of 120 km/s) are listed in Table \ref{table:CO_ratios}. 

\begin{table}[t]
\centering
\caption{\small{Line ratios of $^{12}\text{CO~}(2- 1)$ and $^{12}\text{CO~}(1- 0)$} main beam temperatures.}
    \label{table:CO_ratios}
\begin{tabular}{|p{0.82cm}p{1.25cm}|p{1.65cm}|p{1.65cm}|p{1.84cm}|} \hline \hline
        &          Vlos               &              &              &           \\
                &          average$^{(a)}$               &     $^{12}\text{CO~}(1- 0)$        &      $^{12}\text{CO~}(2- 1)$       &      $^{12}\text{CO~}(2- 1)$      \\
Reg.  & {(K$\,$km$\,$s$^{-1}$)} & {$T_{\mathrm{mb}}$ {(}mK{)}} & {$T_{\mathrm{mb}}$ {(}mK{)}} & $/^{12}\text{CO~}(1- 0)$ \\ \hline
19\_i   & 6645                    & 9.27          $\pm$ 0.84     & 6.45          $\pm$ 2.55     & 0.70          $\pm$ 0.28        \\ 
19\_iii & 6655                    & 9.12          $\pm$ 1.18     & 6.62          $\pm$ 3.18     & 0.73          $\pm$ 0.36        \\ \hline
b\_i    & 6640                    & 8.11          $\pm$ 1.28     & 4.06          $\pm$ 1.70     & 0.50          $\pm$ 0.22        \\
b\_ii   & 6420                    & 6.76          $\pm$ 1.36     & 6.25          $\pm$ 2.35     & 0.93          $\pm$ 0.39        \\ \hline
SQ-A    & 5845                    & 3.94          $\pm$ 1.69     & 6.42          $\pm$ 2.33     & 1.63          $\pm$ 0.92        \\ \hline
SF\_i   & 5845                    & 5.42          $\pm$ 1.45     & 6.43          $\pm$ 1.66     & 1.19          $\pm$ 0.44        \\
SF\_ii  & 5825                    & 6.16          $\pm$ 1.90     & 7.70          $\pm$ 1.95     & 1.25          $\pm$ 0.50        \\
        & 6575                    & 4.30          $\pm$ 0.75     & 5.09          $\pm$ 1.59     & 1.18          $\pm$ 0.42        \\
SF\_iii & 5805                    & 2.76          $\pm$ 0.88     & 8.88          $\pm$ 1.67     & 3.22          $\pm$ 1.19        \\
        & 6610                    & 4.44          $\pm$ 1.50     & 5.25          $\pm$ 1.63     & 1.18          $\pm$ 0.54        \\
SF\_iv  & 6465                    & 8.18          $\pm$ 1.87     & 5.49          $\pm$ 2.06     & 0.67          $\pm$ 0.29        \\ \hline
18\_i   & 5860                    & 3.90          $\pm$ 1.46     & 11.29$\pm$1.96     & 2.90          $\pm$ 1.20        \\
        & 6440                    & 4.36          $\pm$ 1.14     & 5.39          $\pm$ 1.93     & 1.24          $\pm$ 0.55       \\ \hline
\end{tabular}
\tablefoot{The main beam temperatures were obtained from the Gaussian fits.
\tablefoottext{a}{The velocities stated are the average of the two CO lines, separated by a maximum of three velocity channels, 120 km/s, the average separation is $\sim50$ km/s.}}
\end{table}

\section{Summary and conclusions}
\label{sum}
We have observed the compact galaxy group SQ in both optical and CO. 
We used the LBT in Tucson, Arizona, USA, to map the ionised gas and stellar kinematics in an important part of the group using multiple long slits, focusing on the nuclei of the galaxies and the large-scale dynamics of the group. 
The CO data were obtained with the IRAM 30m telescope in Sierra Nevada, Spain, using OTF mapping of an area of 5.67 arcmin$^2$ that covered the group. 
In this paper we have presented an extensive analysis, including gas and stellar distribution and kinematics, gas excitation, and molecular gas mass. 

The data reveal large amounts of gas in the IGM, wide-spread star formation, and a new puzzle in the form of the distribution of the CO gas. 
The group exhibits a total H$_2$ gas mass of $21 \pm 2 \cdot 10^{9} M_{\odot}$, obtained by summing the spectral emission over the velocity range $5100-7500$ km/s, in the observed area as marked in Fig. \ref{fig:iram_boxes}. 
Looking at the regions chosen for closer analysis, we find a total H$_2$ mass of $9.8 \pm 2 \cdot 10^{9} M_{\odot}$ from the Gaussian fits and $10 \pm 0.1 \cdot 10^{9} M_{\odot}$, determined by summing the spectra over the velocity range of $5100-7500$ km/s. 
Further, we find an H$_2$ gas mass of $52.9\pm1.1 \cdot 10^8 M_{\odot}$ in or near NGC7319, $8.8\pm0.3 \cdot 10^8 M_{\odot}$ in the bridge, $20.7\pm0.5 \cdot 10^8 M_{\odot}$ in the SF ridge, and $4.6\pm0.3 \cdot 10^8 M_{\odot}$ in SQ-A (when summing over the velocity range $5100-7500$ km/s). 
The ionised gas in the regions chosen for closer analysis amounts to a total of $20.1\pm0.2\cdot10^{10} M_{\odot}$ and favours the SF ridge, NGC7319, the bridge, and the west ridge, while the molecular gas primarily favours NGC7319 but also shows up significantly in the bridge and the SF ridge. 
Of the regions chosen for closer analysis in SQ, as marked in Fig. \ref{fig:radio_regions}, we find that $56$\% of the H$_2$ mass is in or near NGC7319 (from the masses estimated from the Gaussian fits). 
On the other hand, the $^{12}\text{CO~}(2-1)$ emission favours the areas in or near NGC7317, the NGC7318 pair, SQ-A, and the SF ridge. 

Our data detail an impressive complexity in the SF clouds throughout SQ, indicating up to four gas congregations of different velocities in multiple locations. 
The complexity of the clumpy IGM in the SF ridge, west ridge, and bridge provides an insight into the history of the group, where the mixed IGM indicates that at least three and perhaps four kinematical structures have been involved in the creation of the SF ridge, that is, NGC7319, NGC7320C, and potentially NGC7317, as well as the high relative velocity intruder NGC7318B. 
Further, for the SF ridge and the west ridge, we present the effect of the shock on the medium via the high line ratios. 

For NGC7319 we detail a fascinating interplay between AGN feeding and feedback in a galaxy with a decoupled gas and stellar disk. 
We observe a stellar disk and an ionised gas disk, which are approximately perpendicular to each other, as well as an ionised gas and stellar velocity field centre offset by $\sim3.2$ arcsec. 
As we map the gas and stellar emission in the galaxy, we trace a large-scale nuclear wind to the SW of the nucleus at a blueshifted velocity of $476\pm14$ km/s. 
All of the ionised gas observed in NGC7319 may be in the nuclear wind, although a significant contribution from a stellar population cannot be excluded. 
Furthermore, the molecular gas deposit present in the central 11$''$ of this galaxy may be important in the feeding of the AGN and/or the outflow. 

In addition, our high resolution data of the nuclear region of NGC7319 allow us, for the first time, to reveal the Seyfert 1 nature of this galaxy. 
We observe an FWHM of the NLR and BLR as $374\pm10$ km/s and $1265\pm43$ km/s, respectively. 

We confirm the existence of the bridge in both optical and millimetre emission and detail its kinematics. 
Our data of the bridge indicate a connection between NGC7319 and the intruder galaxy NGC7318B via (i) the line-of-sight velocities of two CO gas congregations at $\sim5800$ km/s and $\sim6700$ km/s in region b\_i (see Table \ref{table:CO_SF}) and (ii) the ionised gas with line-of-sight velocities ranging from $5650-7000$ km/s across region b\_1-4 (see Table \ref{table:bridge}). Therefore, it may be assumed that the bridge was likely created during the passage of NGC7318B through the group. 

While the $^{12}\text{CO~}(1-0)$ emission favours the area in or near NGC7319, the $^{12}\text{CO~}(2-1)$ emission shows a peak to the south of the NGC7318 pair and an extension towards NGC7317, corroborating the claim of \citet{duc2018} that NGC7317 passed through the centre of SQ in the past and in the process left a trail of a faint stellar halo and, as our data show, a tail of warm diffuse molecular gas. 
NGC7317 does not show the typical rotation and velocity dispersion curve of an elliptical galaxy, and its connection to the rest of the group via the $^{12}\text{CO~}(2-1)$ emission indicates that NGC7317 may be a prime candidate for the study of the effect of galaxy harassment. 

The west ridge is an interesting structure that may be connected to the NW and SW tails or the shocked ridge. 
We note a lack of molecular gas content in the area and a presence of a high oxygen content and high ionisation. 
The west ridge may provide further revelations regarding the ongoing interaction between NGC7318B and the rest of the group. 

It is clear that we are looking at a very complex and highly interactive structure, a galaxy group that shows that both external and internal processes have impressive effects on a galaxy's morphology and activity. 
Stephan's Quintet shows an extraordinary complex structure, and the higher the resolution and sensitivity that is used to observe it, the more details and fascinating aspects emerge. 
This group shows us the importance of detailed studies for our understanding of the evolution of galaxies.

\section*{Acknowledgements}
  The authors would like to thank the referee for their thorough and constructive suggestions and feedback. Their input has greatly improved the clarity of the paper. 
  This work was supported in part by SFB 956—Conditions and Impact of Star Formation. 
  We thank the Collaborative Research Centre 956, sub-project A02, funded by the Deutsche Forschungsgemeinschaft (DFG) – project ID 184018867. 
  Madeleine Yttergren was a member of the International Max Planck Research School for Astronomy and Astrophysics at the Universities of Bonn and Cologne, and have in part received financial support for this research from IMPRS.

\bibliographystyle{aa} 
\bibliography{bibliography} 

\begin{thebibliography}{75}
\expandafter\ifx\csname natexlab\endcsname\relax\def\natexlab#1{#1}\fi

\bibitem[{{Allen} {et~al.}(2008){Allen}, {Groves}, {Dopita}, {Sutherland}, \&
  {Kewley}}]{allen2008}
{Allen}, M.~G., {Groves}, B.~A., {Dopita}, M.~A., {Sutherland}, R.~S., \&
  {Kewley}, L.~J. 2008, \apjs, 178, 20

\bibitem[{{Allen} \& {Hartsuiker}(1972)}]{allen1972}
{Allen}, R.~J. \& {Hartsuiker}, J.~W. 1972, \nat, 239, 324

\bibitem[{{Allen} \& {Sullivan}(1980)}]{allen1980}
{Allen}, R.~J. \& {Sullivan}, W.~T., I. 1980, \aap, 84, 181

\bibitem[{{Aoki} {et~al.}(1996){Aoki}, {Ohtani}, {Yoshida}, \&
  {Kosugi}}]{aoki1996}
{Aoki}, K., {Ohtani}, H., {Yoshida}, M., \& {Kosugi}, G. 1996, \aj, 111, 140

\bibitem[{{Appleton} {et~al.}(2013){Appleton}, {Guillard}, {Boulanger},
  {Cluver}, {Ogle}, {Falgarone}, {Pineau des For{\^e}ts}, {O'Sullivan}, {Duc},
  {Gallagher}, {Gao}, {Jarrett}, {Konstantopoulos}, {Lisenfeld}, {Lord}, {Lu},
  {Peterson}, {Struck}, {Sturm}, {Tuffs}, {Valchanov}, {van der Werf}, \&
  {Xu}}]{appleton2013}
{Appleton}, P.~N., {Guillard}, P., {Boulanger}, F., {et~al.} 2013, \apj, 777,
  66

\bibitem[{{Appleton} {et~al.}(2017){Appleton}, {Guillard}, {Togi}, {Alatalo},
  {Boulanger}, {Cluver}, {Pineau des For{\^e}ts}, {Lisenfeld}, {Ogle}, \&
  {Xu}}]{appleton2017}
{Appleton}, P.~N., {Guillard}, P., {Togi}, A., {et~al.} 2017, \apj, 836, 76

\bibitem[{{Appleton} {et~al.}(2006){Appleton}, {Xu}, {Reach}, {Dopita}, {Gao},
  {Lu}, {Popescu}, {Sulentic}, {Tuffs}, \& {Yun}}]{appleton2006}
{Appleton}, P.~N., {Xu}, K.~C., {Reach}, W., {et~al.} 2006, \apjl, 639, L51

\bibitem[{{Baek} {et~al.}(2019){Baek}, {Chung}, {Schawinski}, {Oh}, {Wong},
  {Koss}, {Ricci}, {Trakhtenbrot}, {Smith}, \& {Ueda}}]{baek2019}
{Baek}, J., {Chung}, A., {Schawinski}, K., {et~al.} 2019, \mnras, 488, 4317

\bibitem[{{Baldwin} {et~al.}(1981){Baldwin}, {Phillips}, \&
  {Terlevich}}]{baldwin1981}
{Baldwin}, J.~A., {Phillips}, M.~M., \& {Terlevich}, R. 1981, \pasp, 93, 5

\bibitem[{{Bertola} {et~al.}(1991){Bertola}, {Bettoni}, {Danziger}, {Sadler},
  {Sparke}, \& {de Zeeuw}}]{bertola1991}
{Bertola}, F., {Bettoni}, D., {Danziger}, J., {et~al.} 1991, \apj, 373, 369

\bibitem[{{Boschetti} {et~al.}(2003){Boschetti}, {Rafanelli}, {Ciroi}, {Di
  Mille}, {. Afanasiev}, \& {Dodonov}}]{boschetti2003}
{Boschetti}, C.~S., {Rafanelli}, P., {Ciroi}, S., {et~al.} 2003, Memorie della
  Societa Astronomica Italiana Supplementi, 3, 226

\bibitem[{{Braine} {et~al.}(2001){Braine}, {Duc}, {Lisenfeld}, {Charmand aris},
  {Vallejo}, {Leon}, \& {Brinks}}]{braine2001}
{Braine}, J., {Duc}, P.~A., {Lisenfeld}, U., {et~al.} 2001, \aap, 378, 51

\bibitem[{{Burbidge} \& {Burbidge}(1961)}]{burbidge1961}
{Burbidge}, E.~M. \& {Burbidge}, G.~R. 1961, \apj, 134, 244

\bibitem[{{Cappellari}(2017)}]{cappellari2017}
{Cappellari}, M. 2017, \mnras, 466, 798

\bibitem[{{Cappellari} \& {Emsellem}(2004)}]{cappellari2004}
{Cappellari}, M. \& {Emsellem}, E. 2004, \pasp, 116, 138

\bibitem[{{Carter} {et~al.}(2012){Carter}, {Lazareff}, {Maier}, {Chenu},
  {Fontana}, {Bortolotti}, {Boucher}, {Navarrini}, {Blanchet}, {Greve}, {John},
  {Kramer}, {Morel}, {Navarro}, {Pe{\~n}alver}, {Schuster}, \&
  {Thum}}]{Carter2012}
{Carter}, M., {Lazareff}, B., {Maier}, D., {et~al.} 2012, \aap, 538, A89

\bibitem[{{Cluver} {et~al.}(2010){Cluver}, {Appleton}, {Boulanger}, {Guillard},
  {Ogle}, {Duc}, {Lu}, {Rasmussen}, {Reach}, {Smith}, {Tuffs}, {Xu}, \&
  {Yun}}]{cluver2010}
{Cluver}, M.~E., {Appleton}, P.~N., {Boulanger}, F., {et~al.} 2010, \apj, 710,
  248

\bibitem[{{de Mello} {et~al.}(2012){de Mello}, {Urrutia-Viscarra}, {Mendes de
  Oliveira}, {Torres-Flores}, {Carrasco}, \& {Cypriano}}]{demello2012}
{de Mello}, D.~F., {Urrutia-Viscarra}, F., {Mendes de Oliveira}, C., {et~al.}
  2012, \mnras, 426, 2441

\bibitem[{{Di Mille} {et~al.}(2008){Di Mille}, {Ciroi}, {Rafanelli}, {Moiseev},
  {Smirnova}, {Afanasiev}, \& {Dodonov}}]{dimille2008}
{Di Mille}, F., {Ciroi}, S., {Rafanelli}, P., {et~al.} 2008, Astronomical
  Society of the Pacific Conference Series, Vol. 396, {3D Spectroscopy of the
  Nuclear Environment of a Selected Sample of Nearby Active Galactic Nuclei:
  NGC 7319}, ed. J.~G. {Funes} \& E.~M. {Corsini}, 61

\bibitem[{{Duarte Puertas} {et~al.}(2019){Duarte Puertas},
  {Iglesias-P{\'a}ramo}, {Vilchez}, {Drissen}, {Kehrig}, \&
  {Martin}}]{duarte2019}
{Duarte Puertas}, S., {Iglesias-P{\'a}ramo}, J., {Vilchez}, J.~M., {et~al.}
  2019, \aap, 629, A102

\bibitem[{{Duc} {et~al.}(2018){Duc}, {Cuillandre}, \& {Renaud}}]{duc2018}
{Duc}, P.-A., {Cuillandre}, J.-C., \& {Renaud}, F. 2018, \mnras, 475, L40

\bibitem[{{Eckart} {et~al.}(1990){Eckart}, {Downes}, {Genzel}, {Harris},
  {Jaffe}, \& {Wild}}]{eckart1990}
{Eckart}, A., {Downes}, D., {Genzel}, R., {et~al.} 1990, \apj, 348, 434

\bibitem[{{Falc{\'o}n-Barroso} {et~al.}(2011){Falc{\'o}n-Barroso},
  {S{\'a}nchez-Bl{\'a}zquez}, {Vazdekis}, {Ricciardelli}, {Cardiel}, {Cenarro},
  {Gorgas}, \& {Peletier}}]{falcon2011}
{Falc{\'o}n-Barroso}, J., {S{\'a}nchez-Bl{\'a}zquez}, P., {Vazdekis}, A.,
  {et~al.} 2011, \aap, 532, A95

\bibitem[{{Fedotov} {et~al.}(2011){Fedotov}, {Gallagher}, {Konstantopoulos},
  {Chandar}, {Bastian}, {Charlton}, {Whitmore}, \& {Trancho}}]{fedotov2011}
{Fedotov}, K., {Gallagher}, S.~C., {Konstantopoulos}, I.~S., {et~al.} 2011,
  \aj, 142, 42

\bibitem[{{Flower} {et~al.}(2003){Flower}, {Le Bourlot}, {Pineau des
  For{\^e}ts}, \& {Cabrit}}]{flower2003}
{Flower}, D.~R., {Le Bourlot}, J., {Pineau des For{\^e}ts}, G., \& {Cabrit}, S.
  2003, \mnras, 341, 70

\bibitem[{{Gallagher} {et~al.}(2001){Gallagher}, {Charlton}, {Hunsberger},
  {Zaritsky}, \& {Whitmore}}]{gallagher2001}
{Gallagher}, S.~C., {Charlton}, J.~C., {Hunsberger}, S.~D., {Zaritsky}, D., \&
  {Whitmore}, B.~C. 2001, \aj, 122, 163

\bibitem[{{Gao} \& {Xu}(2000)}]{gao2000}
{Gao}, Y. \& {Xu}, C. 2000, \apjl, 542, L83

\bibitem[{{Greisen} {et~al.}(2006){Greisen}, {Calabretta}, {Valdes}, \&
  {Allen}}]{greisen2006}
{Greisen}, E.~W., {Calabretta}, M.~R., {Valdes}, F.~G., \& {Allen}, S.~L. 2006,
  \aap, 446, 747

\bibitem[{{Guillard} {et~al.}(2010){Guillard}, {Boulanger}, {Cluver},
  {Appleton}, {Pineau Des For{\^e}ts}, \& {Ogle}}]{guillard2010}
{Guillard}, P., {Boulanger}, F., {Cluver}, M.~E., {et~al.} 2010, \aap, 518, A59

\bibitem[{{Guillard} {et~al.}(2009){Guillard}, {Boulanger}, {Pineau Des
  For{\^e}ts}, \& {Appleton}}]{guillard2009}
{Guillard}, P., {Boulanger}, F., {Pineau Des For{\^e}ts}, G., \& {Appleton},
  P.~N. 2009, \aap, 502, 515

\bibitem[{{Guillard} {et~al.}(2012){Guillard}, {Boulanger}, {Pineau des
  For{\^e}ts}, {Falgarone}, {Gusdorf}, {Cluver}, {Appleton}, {Lisenfeld},
  {Duc}, {Ogle}, \& {Xu}}]{guillard2012}
{Guillard}, P., {Boulanger}, F., {Pineau des For{\^e}ts}, G., {et~al.} 2012,
  \apj, 749, 158

\bibitem[{{Guillet} {et~al.}(2009){Guillet}, {Jones}, \& {Pineau Des
  For{\^e}ts}}]{guillet2009}
{Guillet}, V., {Jones}, A.~P., \& {Pineau Des For{\^e}ts}, G. 2009, \aap, 497,
  145

\bibitem[{{Guillet} {et~al.}(2011){Guillet}, {Pineau Des For{\^e}ts}, \&
  {Jones}}]{guillet2011}
{Guillet}, V., {Pineau Des For{\^e}ts}, G., \& {Jones}, A.~P. 2011, \aap, 527,
  A123

\bibitem[{{Hickson}(1997)}]{hickson1997}
{Hickson}, P. 1997, \araa, 35, 357

\bibitem[{{Hwang} {et~al.}(2012){Hwang}, {Struck}, {Renaud}, \&
  {Appleton}}]{hwang2012}
{Hwang}, J.-S., {Struck}, C., {Renaud}, F., \& {Appleton}, P.~N. 2012, \mnras,
  419, 1780

\bibitem[{{Iglesias-P{\'a}ramo} {et~al.}(2012){Iglesias-P{\'a}ramo},
  {L{\'o}pez-Mart{\'\i}n}, {V{\'\i}lchez}, {Petropoulou}, \&
  {Sulentic}}]{iglesiasparamo2012}
{Iglesias-P{\'a}ramo}, J., {L{\'o}pez-Mart{\'\i}n}, L., {V{\'\i}lchez}, J.~M.,
  {Petropoulou}, V., \& {Sulentic}, J.~W. 2012, \aap, 539, A127

\bibitem[{{Iglesias-P{\'a}ramo} \& {V{\'\i}lchez}(2001)}]{iglesiasparamo2001}
{Iglesias-P{\'a}ramo}, J. \& {V{\'\i}lchez}, J.~M. 2001, \apj, 550, 204

\bibitem[{{Kauffmann} {et~al.}(2003){Kauffmann}, {Heckman}, {Tremonti},
  {Brinchmann}, {Charlot}, {White}, {Ridgway}, {Brinkmann}, {Fukugita}, {Hall},
  {Ivezi{\'c}}, {Richards}, \& {Schneider}}]{kauffmann2003}
{Kauffmann}, G., {Heckman}, T.~M., {Tremonti}, C., {et~al.} 2003, \mnras, 346,
  1055

\bibitem[{{Kewley} {et~al.}(2001){Kewley}, {Dopita}, {Sutherland}, {Heisler},
  \& {Trevena}}]{kewley2001}
{Kewley}, L.~J., {Dopita}, M.~A., {Sutherland}, R.~S., {Heisler}, C.~A., \&
  {Trevena}, J. 2001, \apj, 556, 121

\bibitem[{{Klein} {et~al.}(2012){Klein}, {Hochg{\"u}rtel}, {Kr{\"a}mer},
  {Bell}, {Meyer}, \& {G{\"u}sten}}]{Klein2012}
{Klein}, B., {Hochg{\"u}rtel}, S., {Kr{\"a}mer}, I., {et~al.} 2012, \aap, 542,
  L3

\bibitem[{{Konstantopoulos} {et~al.}(2014){Konstantopoulos}, {Appleton},
  {Guillard}, {Trancho}, {Cluver}, {Bastian}, {Charlton}, {Fedotov},
  {Gallagher}, {Smith}, \& {Struck}}]{konstantopoulos2014}
{Konstantopoulos}, I.~S., {Appleton}, P.~N., {Guillard}, P., {et~al.} 2014,
  \apj, 784, 1

\bibitem[{{Lisenfeld} {et~al.}(2004){Lisenfeld}, {Braine}, {Duc}, {Brinks},
  {Charmandaris}, \& {Leon}}]{lisenfeld2004}
{Lisenfeld}, U., {Braine}, J., {Duc}, P.~A., {et~al.} 2004, \aap, 426, 471

\bibitem[{{Lisenfeld} {et~al.}(2002){Lisenfeld}, {Braine}, {Duc}, {Leon},
  {Charmandaris}, \& {Brinks}}]{lisenfeld2002}
{Lisenfeld}, U., {Braine}, J., {Duc}, P.~A., {et~al.} 2002, \aap, 394, 823

\bibitem[{{McConnell} {et~al.}(2011){McConnell}, {Ma}, {Gebhardt}, {Wright},
  {Murphy}, {Lauer}, {Graham}, \& {Richstone}}]{mcconnell2011}
{McConnell}, N.~J., {Ma}, C.-P., {Gebhardt}, K., {et~al.} 2011, \nat, 480, 215

\bibitem[{{Moles} {et~al.}(1998){Moles}, {Marquez}, \& {Sulentic}}]{moles1998}
{Moles}, M., {Marquez}, I., \& {Sulentic}, J.~W. 1998, \aap, 334, 473

\bibitem[{{Moles} {et~al.}(1997){Moles}, {Sulentic}, \&
  {M{\'a}rquez}}]{moles1997}
{Moles}, M., {Sulentic}, J.~W., \& {M{\'a}rquez}, I. 1997, \apjl, 485, L69

\bibitem[{{Natale} {et~al.}(2010){Natale}, {Tuffs}, {Xu}, {Popescu},
  {Fischera}, {Lisenfeld}, {Lu}, {Appleton}, {Dopita}, {Duc}, {Gao}, {Reach},
  {Sulentic}, \& {Yun}}]{natale2010}
{Natale}, G., {Tuffs}, R.~J., {Xu}, C.~K., {et~al.} 2010, \apj, 725, 955

\bibitem[{{Nikiel-Wroczy{\'n}ski} {et~al.}(2013){Nikiel-Wroczy{\'n}ski},
  {Soida}, {Urbanik}, {Beck}, \& {Bomans}}]{nikiel2013}
{Nikiel-Wroczy{\'n}ski}, B., {Soida}, M., {Urbanik}, M., {Beck}, R., \&
  {Bomans}, D.~J. 2013, \mnras, 435, 149

\bibitem[{{Nishimura} {et~al.}(2015){Nishimura}, {Tokuda}, {Kimura}, {Muraoka},
  {Maezawa}, {Ogawa}, {Dobashi}, {Shimoikura}, {Mizuno}, {Fukui}, \&
  {Onishi}}]{nishimura2015}
{Nishimura}, A., {Tokuda}, K., {Kimura}, K., {et~al.} 2015, \apjs, 216, 18

\bibitem[{{Osterbrock} \& {Ferland}(2006)}]{osterbrockbible}
{Osterbrock}, D.~E. \& {Ferland}, G.~J. 2006, {Astrophysics of gaseous nebulae
  and active galactic nuclei}

\bibitem[{{O'Sullivan} {et~al.}(2009){O'Sullivan}, {Giacintucci}, {Vrtilek},
  {Raychaudhury}, \& {David}}]{osullivan2009}
{O'Sullivan}, E., {Giacintucci}, S., {Vrtilek}, J.~M., {Raychaudhury}, S., \&
  {David}, L.~P. 2009, \apj, 701, 1560

\bibitem[{{Petitpas} \& {Taylor}(2005)}]{pepitas2005}
{Petitpas}, G.~R. \& {Taylor}, C.~L. 2005, \apj, 633, 138

\bibitem[{{Pety}(2005)}]{Pety2005}
{Pety}, J. 2005, in SF2A-2005: Semaine de l'Astrophysique Francaise, ed.
  F.~{Casoli}, T.~{Contini}, J.~M. {Hameury}, \& L.~{Pagani}, 721

\bibitem[{{Pietsch} {et~al.}(1997){Pietsch}, {Trinchieri}, {Arp}, \&
  {Sulentic}}]{pietsch1997}
{Pietsch}, W., {Trinchieri}, G., {Arp}, H., \& {Sulentic}, J.~W. 1997, \aap,
  322, 89

\bibitem[{{Renaud} {et~al.}(2010){Renaud}, {Appleton}, \& {Xu}}]{renaud2010}
{Renaud}, F., {Appleton}, P.~N., \& {Xu}, C.~K. 2010, \apj, 724, 80

\bibitem[{{Rodr{\'\i}guez-Baras} {et~al.}(2014){Rodr{\'\i}guez-Baras},
  {Rosales-Ortega}, {D{\'\i}az}, {S{\'a}nchez}, \& {Pasquali}}]{rod2014}
{Rodr{\'\i}guez-Baras}, M., {Rosales-Ortega}, F.~F., {D{\'\i}az}, A.~I.,
  {S{\'a}nchez}, S.~F., \& {Pasquali}, A. 2014, \mnras, 442, 495

\bibitem[{{Rozas} {et~al.}(2006){Rozas}, {Richer}, {L{\'o}pez}, {Rela{\~n}o},
  \& {Beckman}}]{rozas2006}
{Rozas}, M., {Richer}, M.~G., {L{\'o}pez}, J.~A., {Rela{\~n}o}, M., \&
  {Beckman}, J.~E. 2006, \aap, 455, 539

\bibitem[{{S{\'a}nchez-Bl{\'a}zquez} {et~al.}(2006){S{\'a}nchez-Bl{\'a}zquez},
  {Peletier}, {Jim{\'e}nez-Vicente}, {Cardiel}, {Cenarro},
  {Falc{\'o}n-Barroso}, {Gorgas}, {Selam}, \& {Vazdekis}}]{sanchez2006}
{S{\'a}nchez-Bl{\'a}zquez}, P., {Peletier}, R.~F., {Jim{\'e}nez-Vicente}, J.,
  {et~al.} 2006, \mnras, 371, 703

\bibitem[{{Shostak} {et~al.}(1984){Shostak}, {Sullivan}, \&
  {Allen}}]{shostak1984}
{Shostak}, G.~S., {Sullivan}, W.~T., I., \& {Allen}, R.~J. 1984, \aap, 139, 15

\bibitem[{{Smith} \& {Struck}(2001)}]{smith2001}
{Smith}, B.~J. \& {Struck}, C. 2001, \aj, 121, 710

\bibitem[{{Stephan}(1877)}]{stephan1877}
{Stephan}, M. 1877, \mnras, 37, 334

\bibitem[{Struck \& Smith(2012)}]{struck2012}
Struck, C. \& Smith, B.~J. 2012, Monthly Notices of the Royal Astronomical
  Society, 422, 2444

\bibitem[{{Sulentic} {et~al.}(1995){Sulentic}, {Pietsch}, \&
  {Arp}}]{sulentic1995}
{Sulentic}, J.~W., {Pietsch}, W., \& {Arp}, H. 1995, \aap, 298, 420

\bibitem[{{Sulentic} {et~al.}(2001){Sulentic}, {Rosado}, {Dultzin-Hacyan},
  {Verdes-Montenegro}, {Trinchieri}, {Xu}, \& {Pietsch}}]{sulentic2001}
{Sulentic}, J.~W., {Rosado}, M., {Dultzin-Hacyan}, D., {et~al.} 2001, \aj, 122,
  2993

\bibitem[{{Trancho} {et~al.}(2012){Trancho}, {Konstantopoulos}, {Bastian},
  {Fedotov}, {Gallagher}, {Mullan}, \& {Charlton}}]{trancho2012}
{Trancho}, G., {Konstantopoulos}, I.~S., {Bastian}, N., {et~al.} 2012, \apj,
  748, 102

\bibitem[{{Trinchieri} {et~al.}(2003){Trinchieri}, {Sulentic}, {Breitschwerdt},
  \& {Pietsch}}]{trinchieri2003}
{Trinchieri}, G., {Sulentic}, J., {Breitschwerdt}, D., \& {Pietsch}, W. 2003,
  \aap, 401, 173

\bibitem[{{Trinchieri} {et~al.}(2005){Trinchieri}, {Sulentic}, {Pietsch}, \&
  {Breitschwerdt}}]{trinchieri2005}
{Trinchieri}, G., {Sulentic}, J., {Pietsch}, W., \& {Breitschwerdt}, D. 2005,
  \aap, 444, 697

\bibitem[{{Vazdekis} {et~al.}(2010){Vazdekis}, {S{\'a}nchez-Bl{\'a}zquez},
  {Falc{\'o}n-Barroso}, {Cenarro}, {Beasley}, {Cardiel}, {Gorgas}, \&
  {Peletier}}]{vazdekis2010}
{Vazdekis}, A., {S{\'a}nchez-Bl{\'a}zquez}, P., {Falc{\'o}n-Barroso}, J.,
  {et~al.} 2010, \mnras, 404, 1639

\bibitem[{{Veilleux} \& {Osterbrock}(1987)}]{veilleux1987}
{Veilleux}, S. \& {Osterbrock}, D.~E. 1987, \apjs, 63, 295

\bibitem[{{Verdes-Montenegro} {et~al.}(2001){Verdes-Montenegro}, {Yun},
  {Williams}, {Huchtmeier}, {Del Olmo}, \& {Perea}}]{verdesmontenegro2001}
{Verdes-Montenegro}, L., {Yun}, M.~S., {Williams}, B.~A., {et~al.} 2001, \aap,
  377, 812

\bibitem[{{Williams} {et~al.}(1999){Williams}, {Yun}, \&
  {Verdes-Montenegro}}]{williams1999}
{Williams}, B.~A., {Yun}, M., \& {Verdes-Montenegro}, L. 1999, in American
  Astronomical Society Meeting Abstracts, Vol. 194, American Astronomical
  Society Meeting Abstracts \#194, 19.05

\bibitem[{{Williams} {et~al.}(2002){Williams}, {Yun}, \&
  {Verdes-Montenegro}}]{williams2002}
{Williams}, B.~A., {Yun}, M.~S., \& {Verdes-Montenegro}, L. 2002, \aj, 123,
  2417

\bibitem[{{Xu} {et~al.}(2005){Xu}, {Iglesias-P{\'a}ramo}, {Burgarella}, {Rich},
  {Neff}, {Lauger}, {Barlow}, {Bianchi}, {Byun}, {Forster}, {Friedman},
  {Heckman}, {Jelinsky}, {Lee}, {Madore}, {Malina}, {Martin}, {Milliard},
  {Morrissey}, {Schiminovich}, {Siegmund}, {Small}, {Szalay}, {Welsh}, \&
  {Wyder}}]{xu2005}
{Xu}, C.~K., {Iglesias-P{\'a}ramo}, J., {Burgarella}, D., {et~al.} 2005, \apjl,
  619, L95

\bibitem[{{Xu} {et~al.}(2003){Xu}, {Lu}, {Condon}, {Dopita}, \&
  {Tuffs}}]{xu2003}
{Xu}, C.~K., {Lu}, N., {Condon}, J.~J., {Dopita}, M., \& {Tuffs}, R.~J. 2003,
  \apj, 595, 665

\bibitem[{{Zschaechner} {et~al.}(2018){Zschaechner}, {Bolatto}, {Walter},
  {Leroy}, {Herrera}, {Krieger}, {Kruijssen}, {Meier}, {Mills}, {Ott},
  {Veilleux}, \& {Weiss}}]{zschaechner2018}
{Zschaechner}, L.~K., {Bolatto}, A.~D., {Walter}, F., {et~al.} 2018, \apj, 867,
  111

\end{thebibliography}

\begin{appendix} 

\onecolumn
\section{NGC7319: Additional gas emission maps}
\label{app:7319_maps}
\begin{figure*}[ht!]
   \subfloat[\label{fig:7319_OII_flux}]{%
      \includegraphics[width=0.3\textwidth]{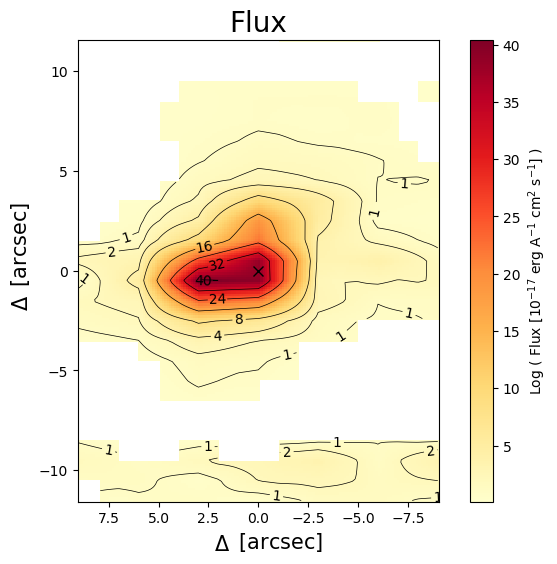}}
\hspace{\fill}
   \subfloat[\label{fig:7319_OII_vlos} ]{%
      \includegraphics[width=0.3\textwidth]{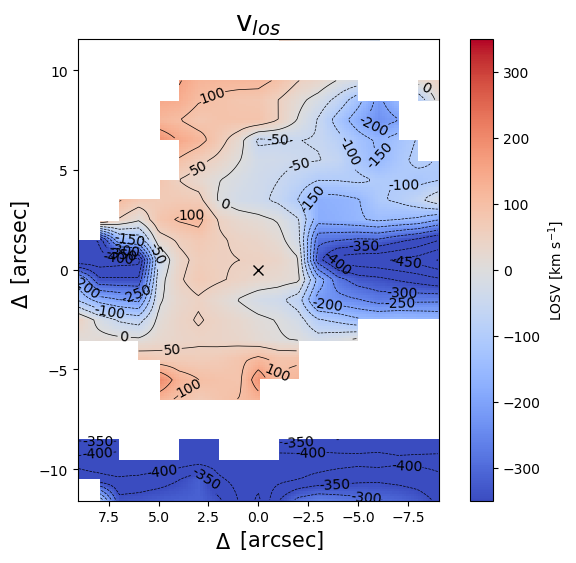}}
\hspace{\fill}
   \subfloat[\label{fig:7319_OII_sigma}]{%
      \includegraphics[width=0.3\textwidth]{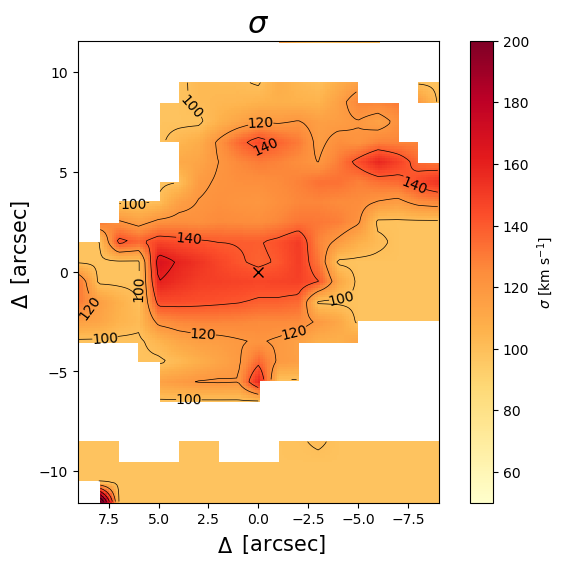}}
\caption{\label{fig:7319_OII}Maps of the [OII]$\lambda$3727 emission in NGC7319. (a) Flux. (b) Line-of-sight velocity. (c) Velocity dispersion.}
\end{figure*}

\begin{figure*}[ht!]
   \subfloat[\label{fig:7319_Hbeta_NLR_flux}]{%
      \includegraphics[width=0.3\textwidth]{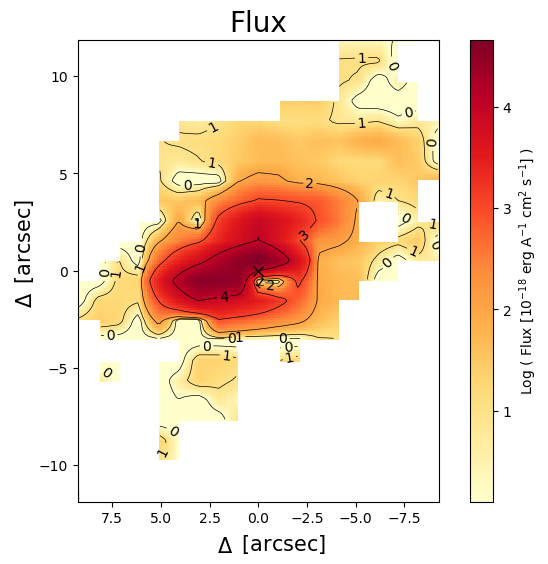}}
\hspace{\fill}
   \subfloat[\label{fig:7319_Hbeta_NLR_vlos} ]{%
      \includegraphics[width=0.3\textwidth]{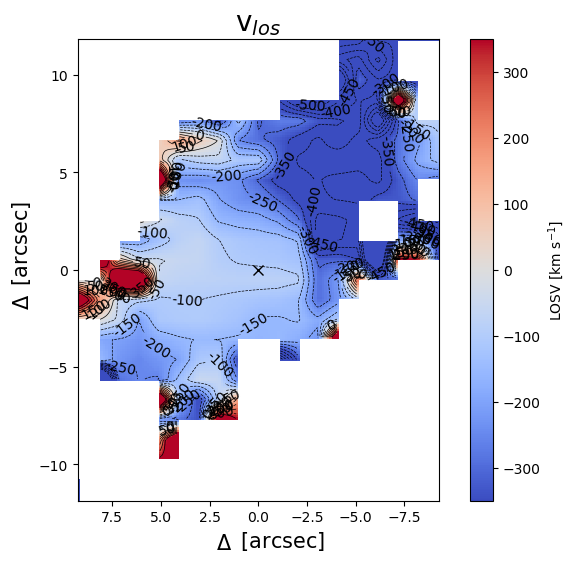}}
\hspace{\fill}
   \subfloat[\label{fig:7319_Hbeta_NLR_sigma}]{%
      \includegraphics[width=0.3\textwidth]{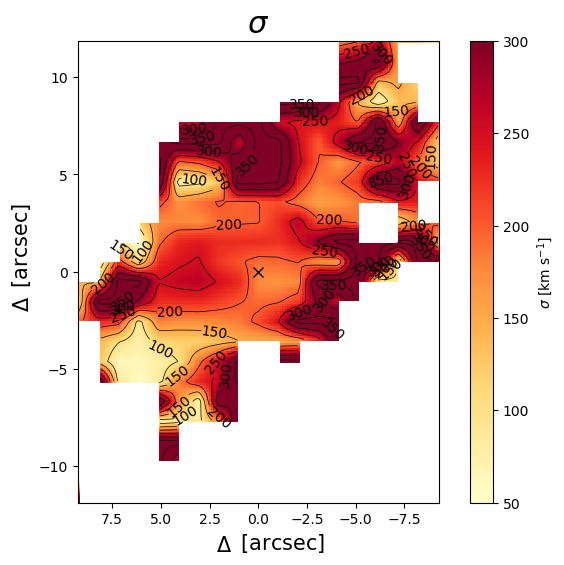}}
\caption{\label{fig:7319_Hbeta_NLR}Maps of the H$\beta$ NLR emission in NGC7319. (a) Flux. (b) Line-of-sight velocity. (c) Velocity dispersion.}
\end{figure*}
 
    \begin{figure*}[ht!]
        \subfloat[\label{fig:7319_OI_flux}]{%
            \includegraphics[width=0.3\textwidth]{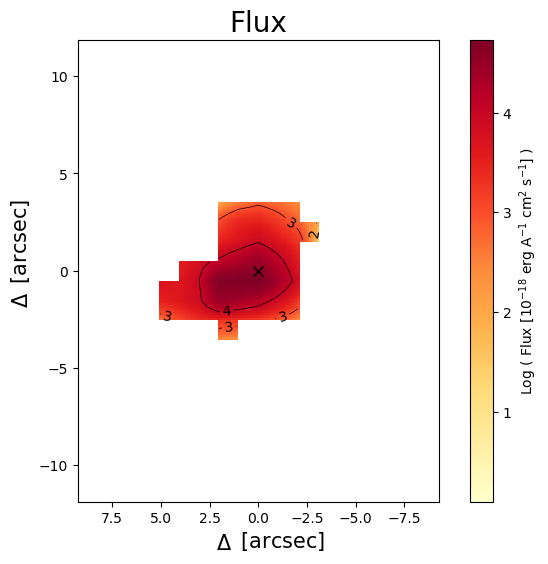}}
            \hspace{\fill}
        \subfloat[\label{fig:7319_OI_vlos} ]{%
            \includegraphics[width=0.3\textwidth]{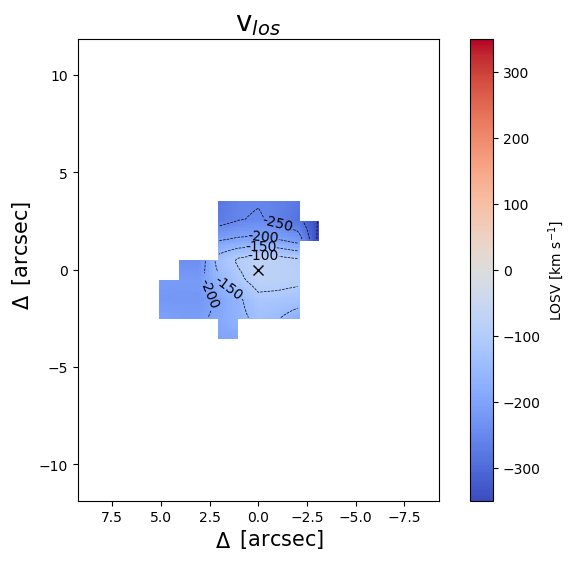}}
            \hspace{\fill}
        \subfloat[\label{fig:7319_OI_sigma}]{%
            \includegraphics[width=0.3\textwidth]{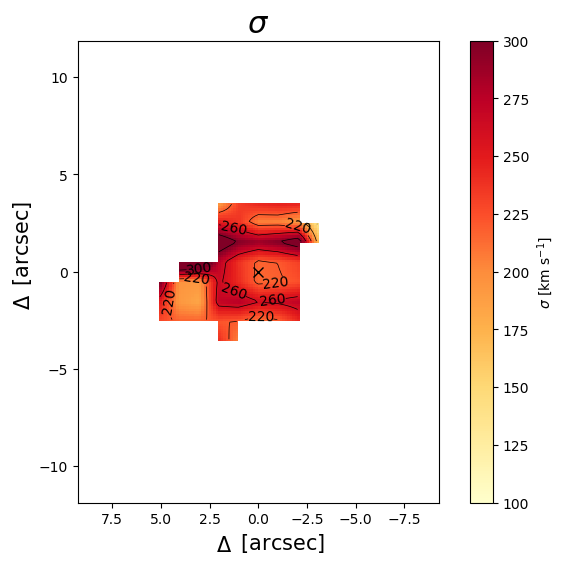}}
        \caption{\label{fig:7319_OI}Maps of the [OI]$\lambda$6302 emission in NGC7319. (a) Flux. (b) Line-of-sight velocity. (c) Velocity dispersion.}
    \end{figure*}

\begin{figure*}[ht!]
   \subfloat[\label{fig:7319_Halpha_NLR_flux}]{%
      \includegraphics[width=0.3\textwidth]{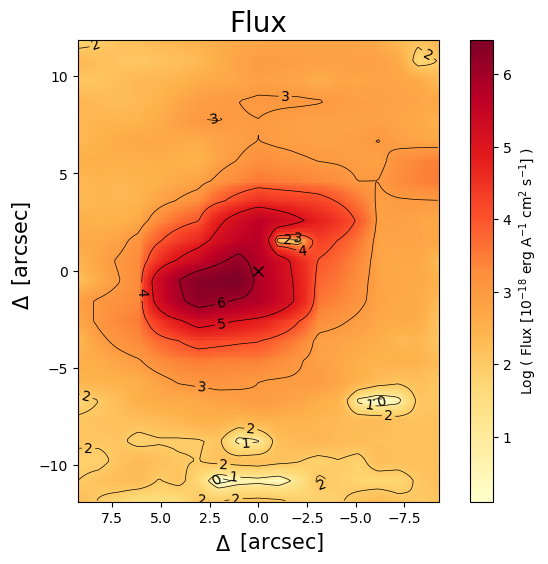}}
\hspace{\fill}
   \subfloat[\label{fig:7319_Halpha_NLR_vlos} ]{%
      \includegraphics[width=0.3\textwidth]{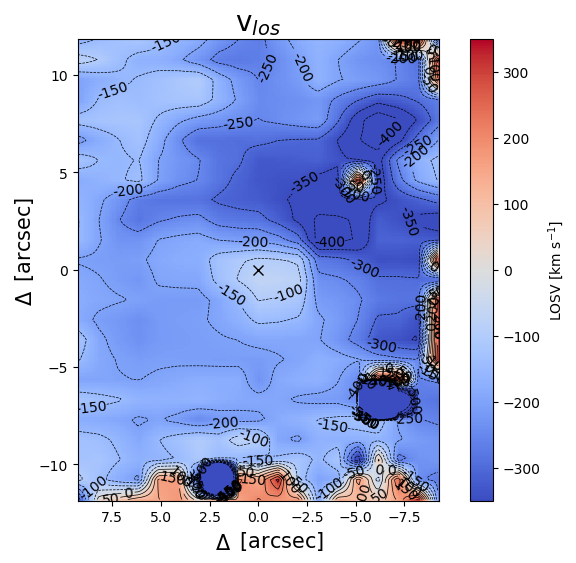}}
\hspace{\fill}
   \subfloat[\label{fig:7319_Halpha_NLR_sigma}]{%
      \includegraphics[width=0.3\textwidth]{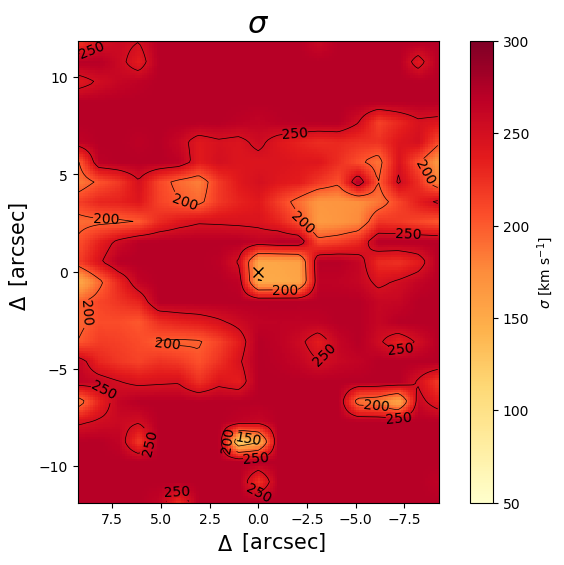}}
\caption{\label{fig:7319_Halpha_NLR}Maps of the  H$\alpha$ NLR emission in NGC7319. (a) Flux. (b) Line-of-sight velocity. (c) Velocity dispersion.}
\end{figure*}

\begin{figure*}[ht!]
   \subfloat[\label{fig:7319_NII_flux}]{%
      \includegraphics[width=0.3\textwidth]{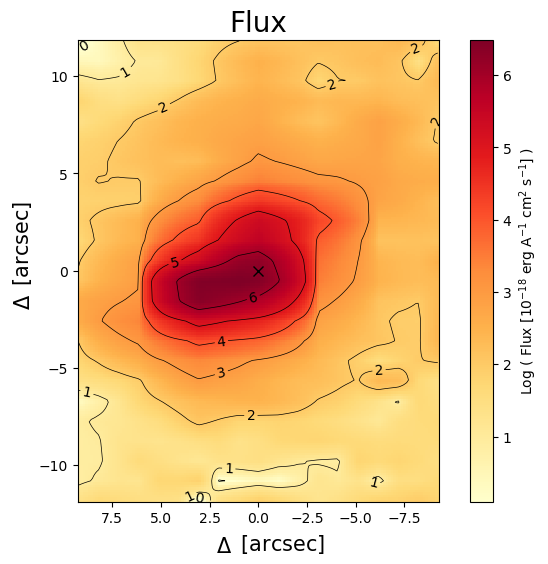}}
\hspace{\fill}
   \subfloat[\label{fig:7319_NII_vlos} ]{%
      \includegraphics[width=0.3\textwidth]{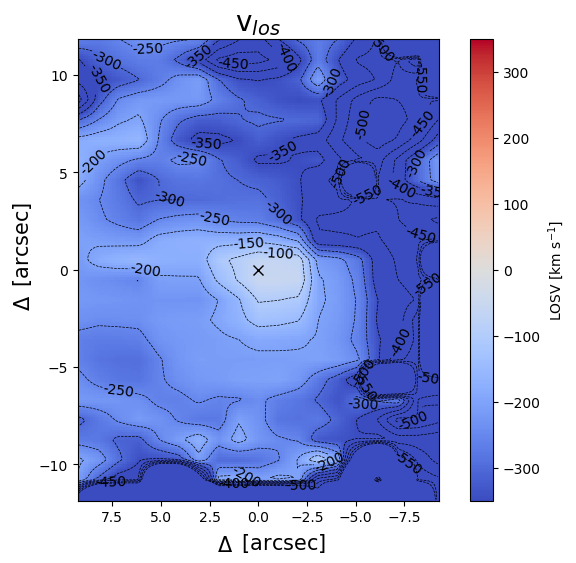}}
\hspace{\fill}
   \subfloat[\label{fig:7319_NII_sigma}]{%
      \includegraphics[width=0.3\textwidth]{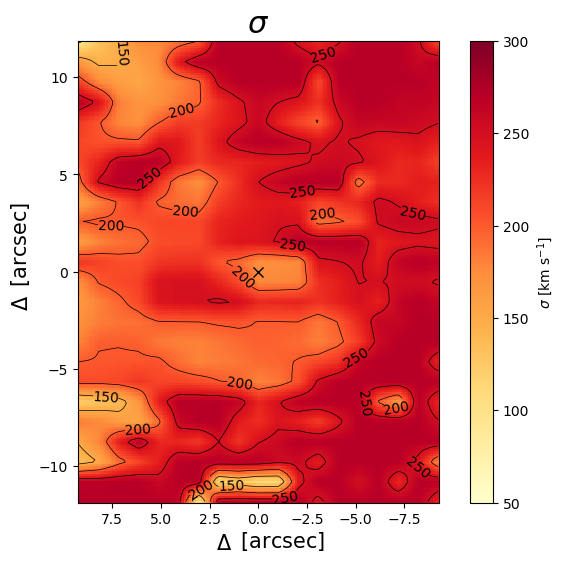}}
\caption{\label{fig:7319_NII}Maps of the  [NII]$\lambda$6585 emission in NGC7319. (a) Flux. (b) Line-of-sight velocity. (c) Velocity dispersion.}
\end{figure*}

\begin{figure*}[ht!]
   \subfloat[\label{fig:7319_SII_flux}]{%
      \includegraphics[width=0.3\textwidth]{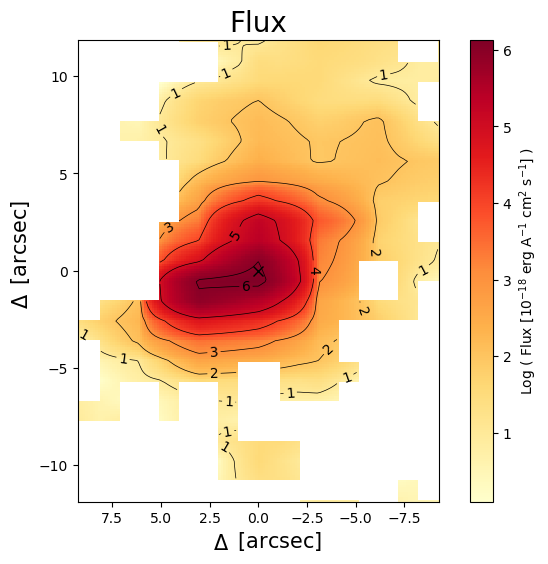}}
\hspace{\fill}
   \subfloat[\label{fig:7319_SII_vlos} ]{%
      \includegraphics[width=0.3\textwidth]{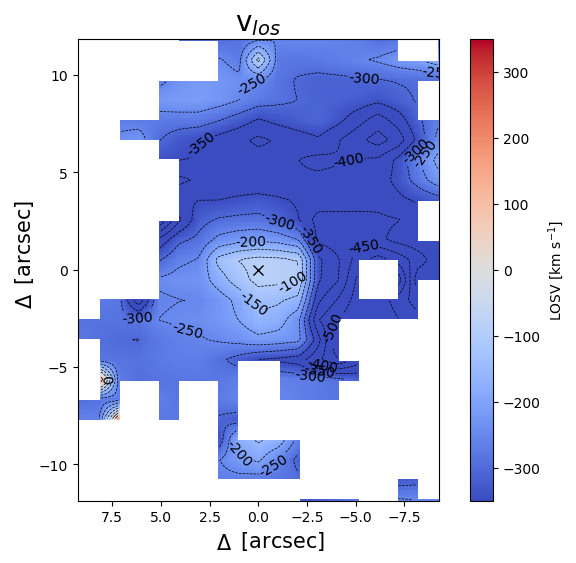}}
\hspace{\fill}
   \subfloat[\label{fig:7319_SII_sigma}]{%
      \includegraphics[width=0.3\textwidth]{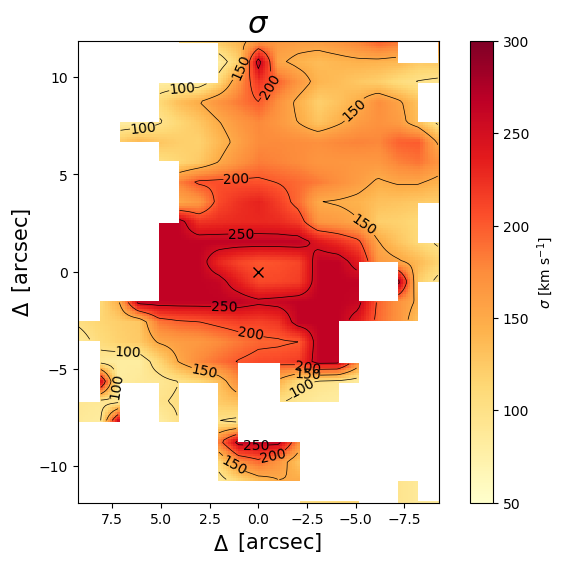}}
\caption{\label{fig:7319_SII}Maps of the  [SII]$\lambda$6718,6732 emission in NGC7319. (a) Flux. (b) Line-of-sight velocity. (c) Velocity dispersion.}
\end{figure*}

\clearpage

\section{Maps of the stellar kinematics in the NGC7318 pair}
\label{app:7318_maps}
\begin{figure} [h!]
  \begin{minipage}[c]{0.49\textwidth}
      \includegraphics[width=\textwidth]{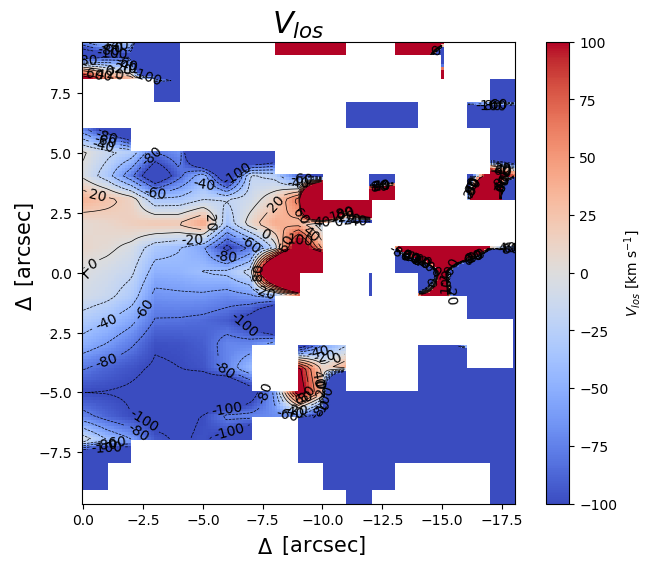}
      
      \includegraphics[width=\textwidth]{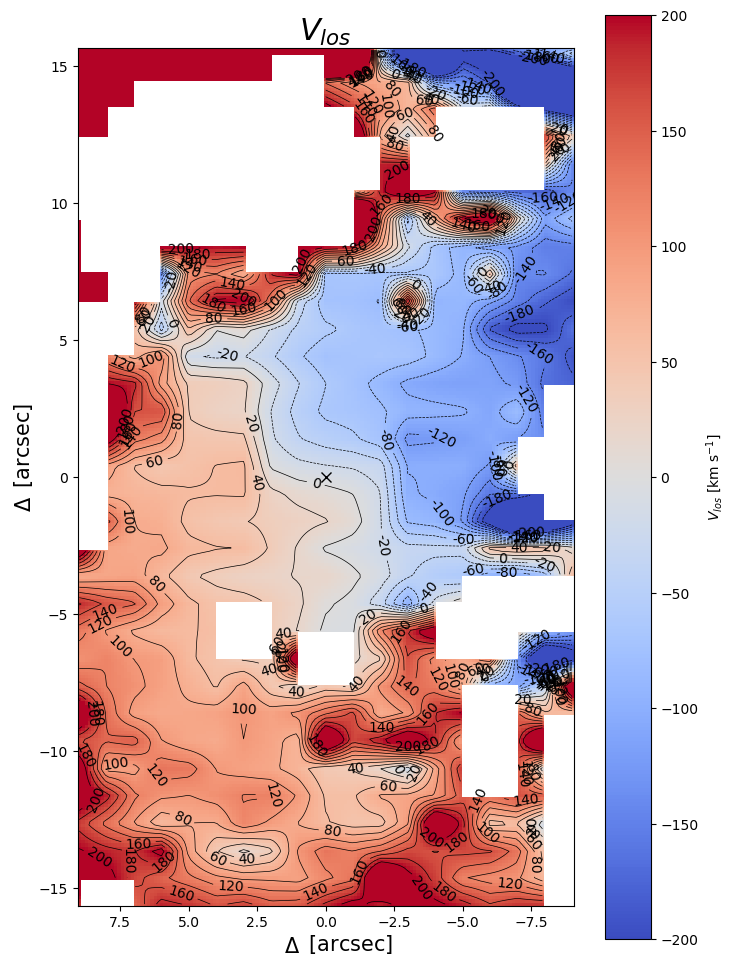}
  \end{minipage}\hfill 
    \begin{minipage}[c]{0.49\textwidth}
    \includegraphics[width=\textwidth]{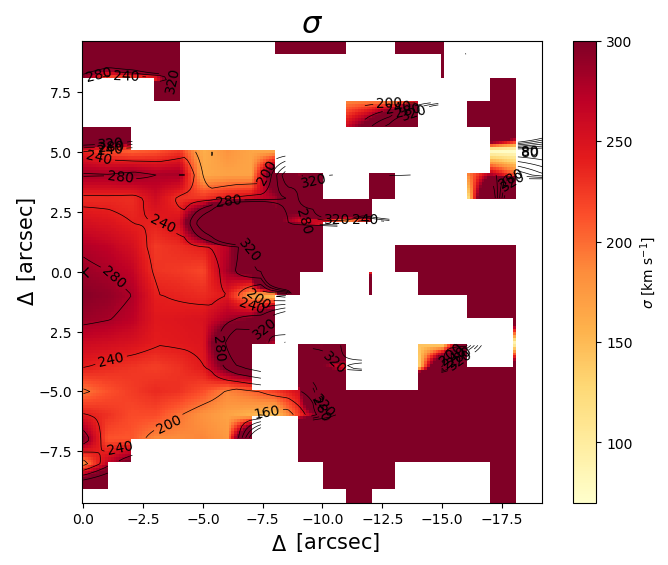}
    
    \includegraphics[width=\textwidth]{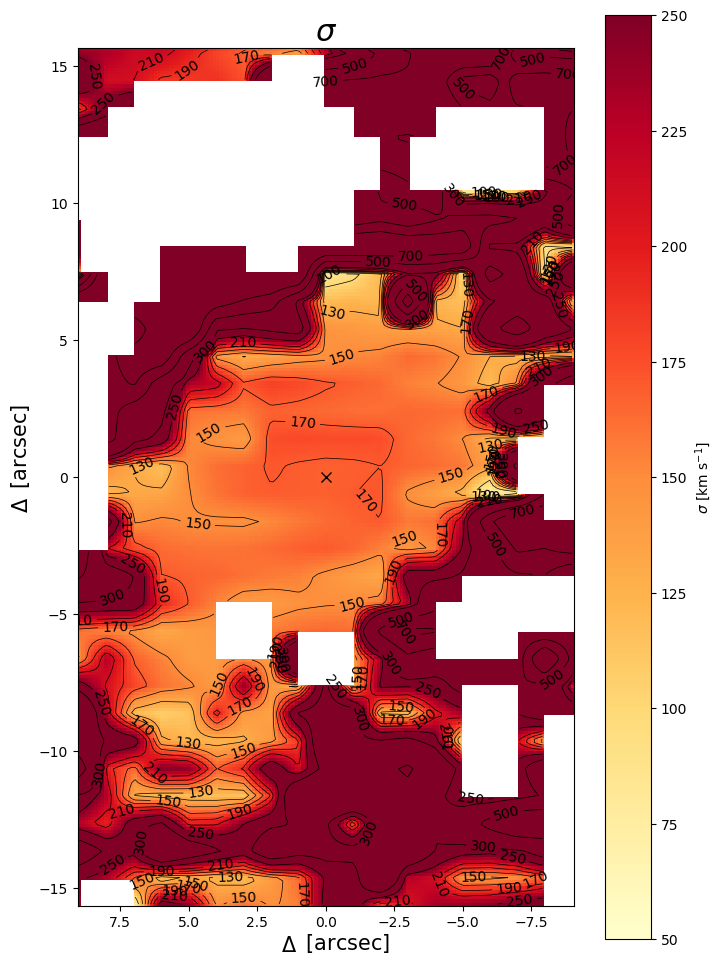}
  \end{minipage}
  \caption{\textit{Top left:} Map of the line-of-sight velocity in NGC7318A as a function of position. The 0 velocity is set to the average of the central $1.5''$, $6787\pm5$ km/s, while the cross at 0 RA, 0 Dec. is the centre of the galaxy as seen by the peak stellar continuum emission. textit{Bottom left:} Map of the line-of-sight velocity in NGC7318B as a function of position. Again, the 0 velocity is set to the average of the central $1.5''$, $5973\pm4$ km/s, and the cross in the centre of the galaxy marks the peak stellar continuum emission. \textit{Top right:} Map of the stellar velocity dispersion in NGC7318A as a function of position. As in the left image, the cross in the centre of the galaxy pinpoints the peak stellar continuum emission. \textit{Bottom right:} Map of the stellar velocity dispersion in NGC7318B as a function of position. The cross in the centre of the galaxy again marks the peak stellar continuum emission.}\label{fig:7318_stellar}
\end{figure}

\clearpage

\section{The regions, RA and Dec.}
\label{app:regions_ra_dec}
\begin{table}[htp]
\caption{\label{table:regions_RA_Dec} Equatorial (J2000) RA and Dec. coordinates of the centres of the regions adopted for the closer analysis and their respective ionised gas mass estimate and H$_2$ gas mass estimate. }
\begin{minipage}[t]{.48\linewidth}%
\centering
\begin{tabular}[t]{l|ccc} \hline \hline
Optical         & RA            & Dec.               & $ M_{HII}$ $^{(b)}$\\ 
 regions$^{(a)}$        &  (hh:mm:ss)   & (dd:mm:ss)        & ($ 10^9M_{\odot}$) \\ \hline
17              & 22:35:51.87   & 33:56:41.80       & \\
18a             & 22:35:56.53   & 33:57:56.40       & \\
18b             & 22:35:58.29   & 33:57:57.85       & \\
19\_0           & 22:36:03.55   & 33:58:32.62       &  3.72$\pm$0.0\\
19\_1           & 22:36:03.45   & 33:58:28.49       &  4.51$\pm$0.05\\
19\_2           & 22:36:03.23   & 33:58:33.80       &  8.29$\pm$0.07\\
19\_3           & 22:36:03.65   & 33:58:36.61       &  10.31$\pm$0.08\\
19\_4           & 22:36:03.88   & 33:58:31.31       &  9.65$\pm$0.08\\
19\_5           & 22:36:03.01   & 33:58:32.39       &  8.38$\pm$0.07\\
19\_6           & 22:36:03.32   & 33:58:25.32       &  8.26$\pm$0.07\\
19\_7           & 22:36:02.96   & 33:58:22.97       &  11.32$\pm$0.08\\
b\_1            & 22:36:02.09   & 33:58:26.28       &  11.32$\pm$0.08\\
b\_2            & 22:36:01.85   & 33:58:19.04       &  11.32$\pm$0.08\\
b\_3            & 22:36:01.12   & 33:58:23.29       &  6.51$\pm$0.06\\
b\_4            & 22:36:00.63   & 33:58:20.00       &  10.52$\pm$0.08\\
SF\_1           & 22:35:59.96   & 33:58:12.18       &  11.32$\pm$0.08\\
SF\_10          & 22:35:59.31   & 33:57:55.41       &  3.95$\pm$0.05\\
SF\_11          & 22:35:58.70   & 33:57:52.53       &  6.15$\pm$0.06\\
SF\_12          & 22:35:56.82   & 33:57:38.96       & 
 2.35$\pm$0.04\\
SF\_13          & 22:35:56.33   & 33:57:44.21       & 
 2.71$\pm0.04$\\
SF\_14          & 22:35:55.52   & 33:57:43.92       & 
 2.02$\pm$0.03\\
SF\_15          & 22:35:55.57   & 33:57:36.32       & 
 2.19$\pm$0.03\\
SF\_2           & 22:35:59.97   & 33:58:05.47       &  11.32$\pm$0.08\\
SF\_3           & 22:35:59.45   & 33:58:15.67       &  11.32$\pm$0.08\\
SF\_4           & 22:35:59.60   & 33:58:09.83       &  4.75$\pm$0.05\\
SF\_5           & 22:35:59.95   & 33:57:59.64       &  11.32$\pm$0.08\\
SF\_6           & 22:35:59.65   & 33:58:06.71       &  5.12$\pm$0.06\\
SF\_7           & 22:35:59.39   & 33:58:08.42       &  9.51$\pm$0.07\\
SF\_8           & 22:35:59.48   & 33:58:02.18       &  6.16$\pm$0.06\\
SF\_9           & 22:35:59.15   & 33:58:03.42       &  11.32$\pm$0.08\\ 
\hline 
\end{tabular}
\tablefoot{
\tablefoottext{a}{The optical regions have been extracted via 1.5$''$ circles from the interpolated map of the optical spectra. }
\tablefoottext{a}{The ionised gas mass has been estimated as described in Sect. \ref{obs}. }}
\end{minipage}%
\begin{minipage}[t]{.48\linewidth}%
\centering
\begin{tabular}[t]{l|ccc} \hline \hline
CO              & RA            & Dec.               & $ M_{H_2}$ $^{(b)}$            \\ 
 regions$^{(a)}$        & (hh:mm:ss)    & (dd:mm:ss)        & ($ 10^8M_{\odot}$)     \\  \hline
17              & 22:35:51.86   & 33:56:40.82       &                       \\
18\_i           & 22:35:55.92   & 33:57:39.08       & $ 1.65\pm0.18$         \\
18\_ii          & 22:35:57.74   & 33:57:37.89       & $ 1.17\pm0.15$         \\
18\_iii         & 22:35:56.40   & 33:58:18.50       & $ 0.85\pm0.17$         \\
18\_iv          & 22:35:54.86   & 33:57:20.57       &                       \\
18a             & 22:35:56.71   & 33:57:55.58       & $ 1.47\pm0.15$         \\
18b             & 22:35:58.32   & 33:57:55.58       & $ 3.10\pm0.19$         \\
19\_            & 22:36:02.59   & 33:58:46.56       & $ 9.62\pm0.42$         \\
19\_i           & 22:36:03.55   & 33:58:32.55       & $ 8.73\pm0.43$         \\
19\_ii          & 22:36:04.65   & 33:58:17.89       & $ 6.61\pm0.34$         \\
19\_iii         & 22:36:04.15   & 33:58:53.70       & $ 6.97\pm0.48$         \\
19\_iv          & 22:36:05.76   & 33:58:54.82       & $ 4.34\pm0.42$         \\
19\_v           & 22:36:05.21   & 33:58:38.54       & $ 6.28\pm0.40$         \\
19\_vi          & 22:36:06.60   & 33:58:26.69       & $ 1.93\pm0.30$         \\
19\_o           & 22:36:03.12   & 33:58:13.79       & $ 8.37\pm0.43$         \\
b\_i            & 22:36:01.58   & 33:58:02.08       & $ 8.84\pm0.32$         \\
b\_ii           & 22:36:01.56   & 33:58:23.30       & $ 6.57\pm0.26$         \\
SF\_i           & 22:35:59.54   & 33:58:34.60       & $ 6.32\pm0.28$         \\
SF\_ii          & 22:35:59.85   & 33:58:16.55       & $ 6.27\pm0.24$         \\
SF\_iii         & 22:35:59.85   & 33:58:04.00       & $ 5.45\pm0.22$         \\
SF\_iv          & 22:35:59.73   & 33:57:37.16       & $ 2.63\pm0.29$         \\
SQ-A            & 22:35:58.88   & 33:58:50.69       & $ 4.64\pm0.30$         \\  \hline
\end{tabular}
\tablefoot{
\tablefoottext{a}{The CO regions have been extracted with 11$''$ radius circles from the smoothed CO maps. }
\tablefoottext{b}{The H$_2$ gas mass obtained from integrating the spectra over the velocity range covering SQ ($5100-7500$ km/s). }}
\end{minipage}%
\end{table}
\clearpage

\section{Optical spectra for a selection of regions}
\label{app:spec}

\begin{figure}[ht!]
    \centering
    \includegraphics[width=0.78\textwidth]{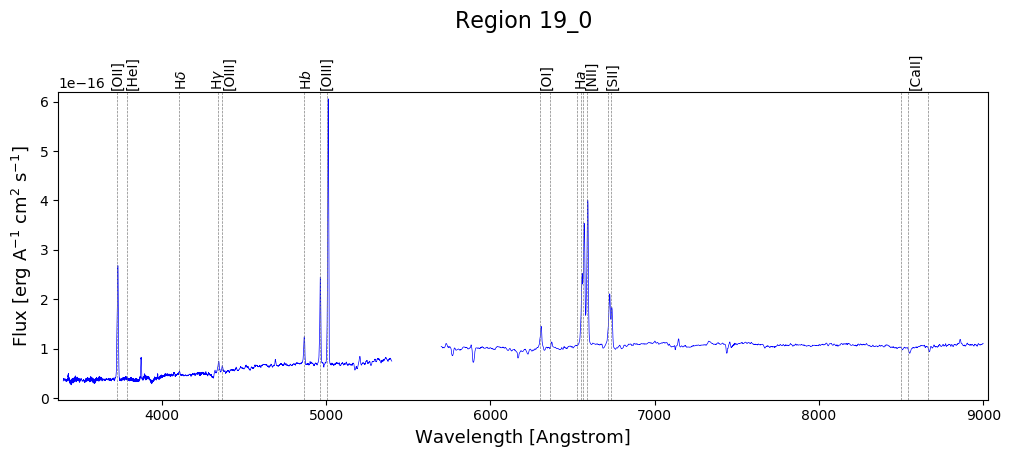}
    \caption{Spectrum of the central 1.5$''$ of NGC7319. The spectrum has been corrected to the average redshift of the group, 0.0215, and the vertical dotted lines mark the expected line centres of several interesting emission lines at this average redshift.}
    \label{fig:19_0}
\end{figure}

\begin{figure}[ht!]
    \centering
    \includegraphics[width=0.78\textwidth]{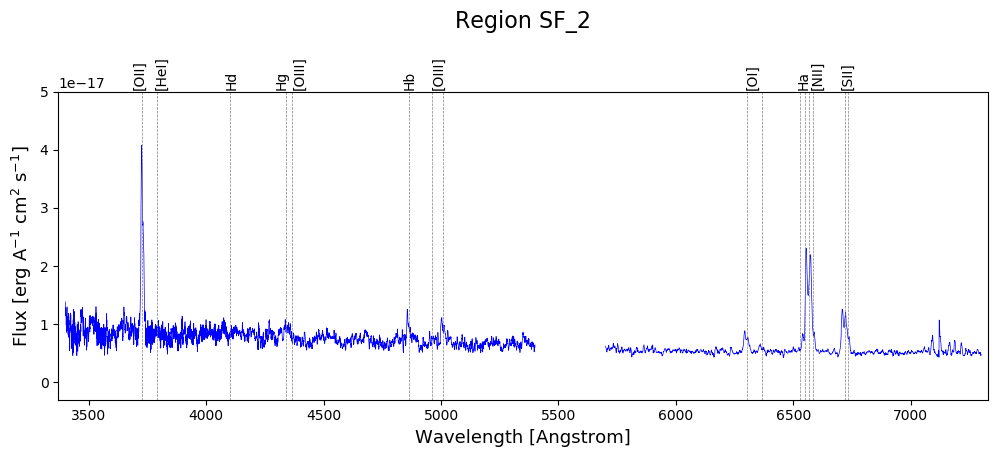}
    \caption{As in Fig. \ref{fig:19_0} but for region SF\_2.}
    \label{fig:SF_2}
\end{figure}

\begin{figure}[ht!]
    \centering
    \includegraphics[width=0.78\textwidth]{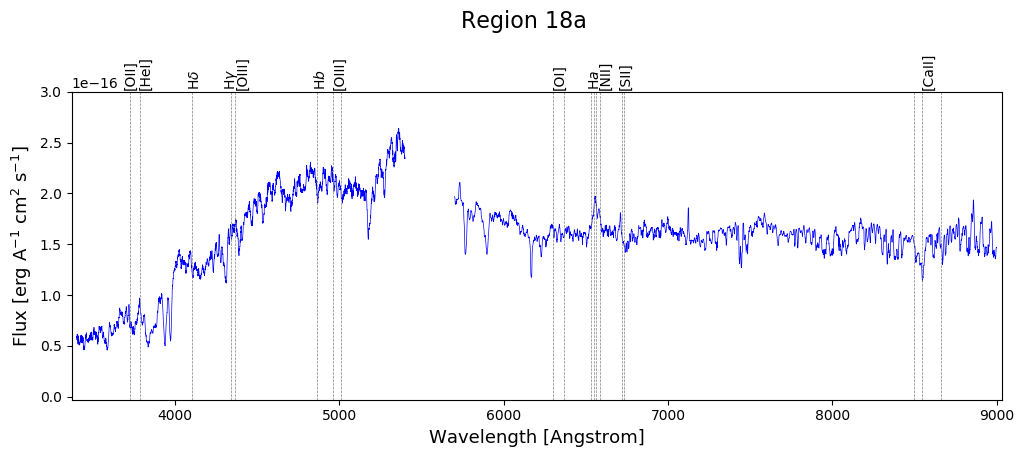}
    \caption{As in Fig. \ref{fig:19_0} but for the central $1.5''$ of NGC7318A, region 18a.}
    \label{fig:18a}
\end{figure}

\begin{figure}[ht!]
    \centering
    \includegraphics[width=0.78\textwidth]{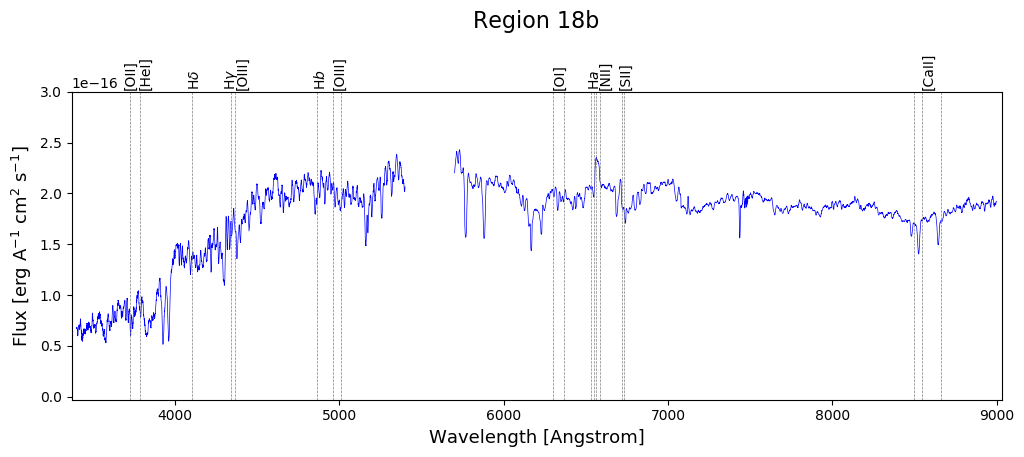}
    \caption{As in Fig. \ref{fig:19_0} but for the central $1.5''$ of NGC7318B, region 18b.}
    \label{fig:18b}
\end{figure}

\begin{figure}[h]
    \centering
    \includegraphics[width=0.78\textwidth]{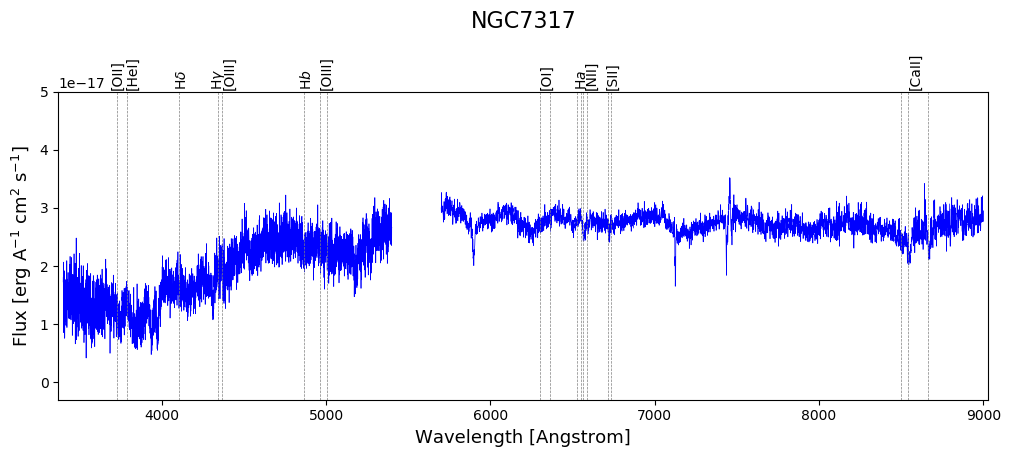}
    \caption{As in Fig. \ref{fig:19_0} but for NGC7317. }
    \label{fig:7317_spec}
\end{figure}

\clearpage

\section{$^{12}\text{CO~}(1-0)$, $^{12}\text{CO~}(2-1),$ and $^{13}\text{CO~}(1-0)$ emission and noise as a function of position}
\label{app:CO_maps}
\begin{figure}[h]
    \centering
    \begin{minipage}{0.48\textwidth}
    \centering
    \includegraphics[width=\textwidth]{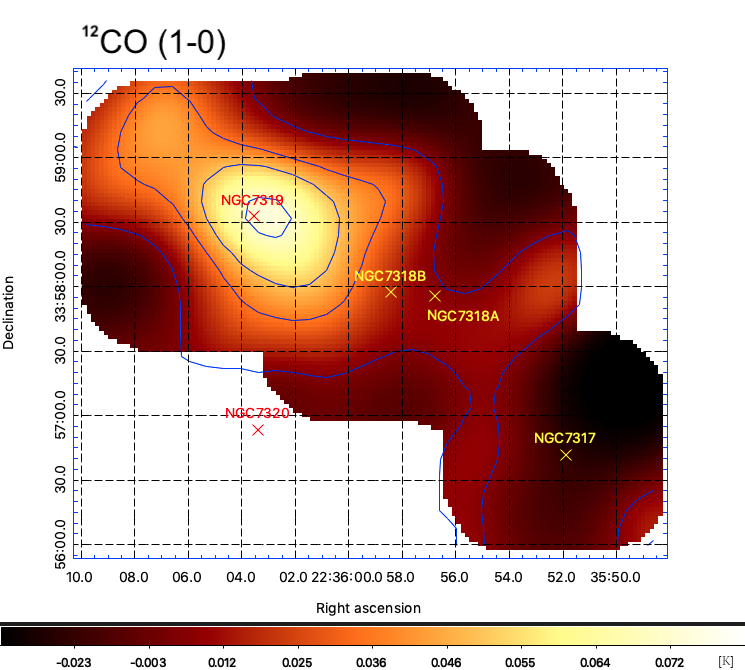}
    \end{minipage}
    \begin{minipage}{0.48\textwidth}
    \centering
    \includegraphics[width=\textwidth]{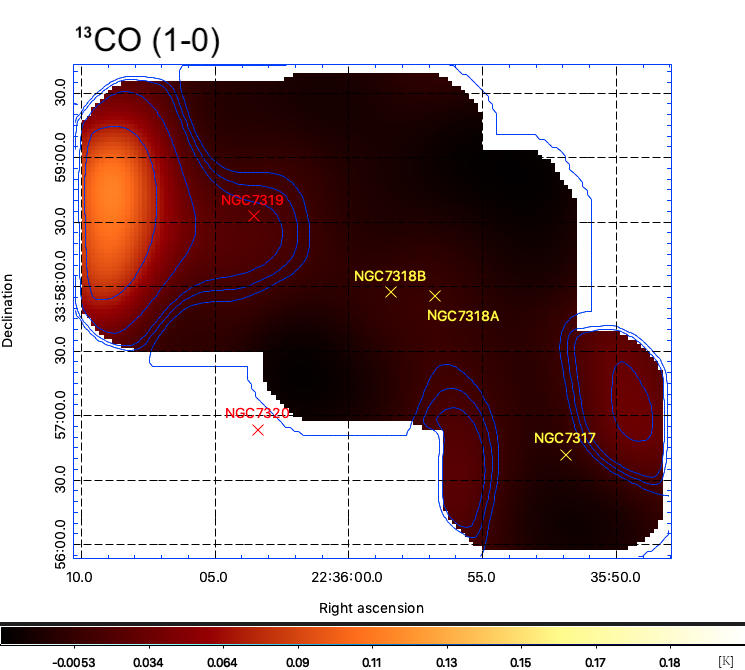}
    \end{minipage}\\
     \begin{minipage}{0.48\textwidth}
    \centering
    \includegraphics[width=\textwidth]{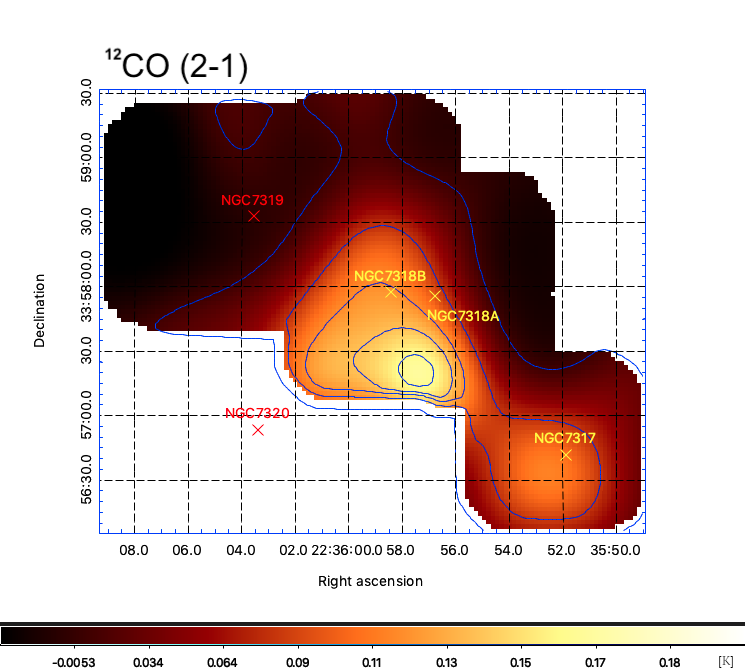}
    \end{minipage}
    \begin{minipage}{0.48\textwidth}
    \centering
    \caption{Maps showing the summed emission per spaxel of the CO channels over the velocity range $5100-7500$ km/s. Contours in blue show the flux variations, while the red and yellow crosses pinpoint the galaxies. \textit{Top Left:} $^{12}\text{CO\,}(1-0)$ contours at 4, 35, 50, and 70 mK;.  \textit{Bottom Left:} $^{12}\text{CO\,}(2-1)$, 4, 80, 120, 140, and 160 mK. \textit{Top Right: }$^{13}\text{CO\,}(1-0)$ contours at 0, 5, 10, 30, and 70 mK.}
    \label{fig:CO_maps}
    \end{minipage}   
\end{figure}

\begin{figure}[h]
    \centering
    \begin{minipage}{0.48\textwidth}
    \centering
    \includegraphics[width=\textwidth]{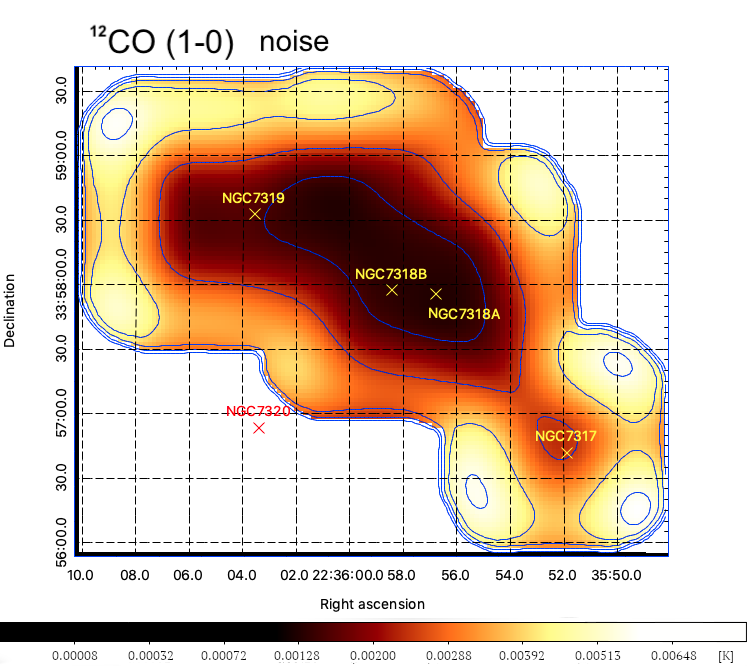}
    \end{minipage}
    \begin{minipage}{0.48\textwidth}
    \centering
    \includegraphics[width=\textwidth]{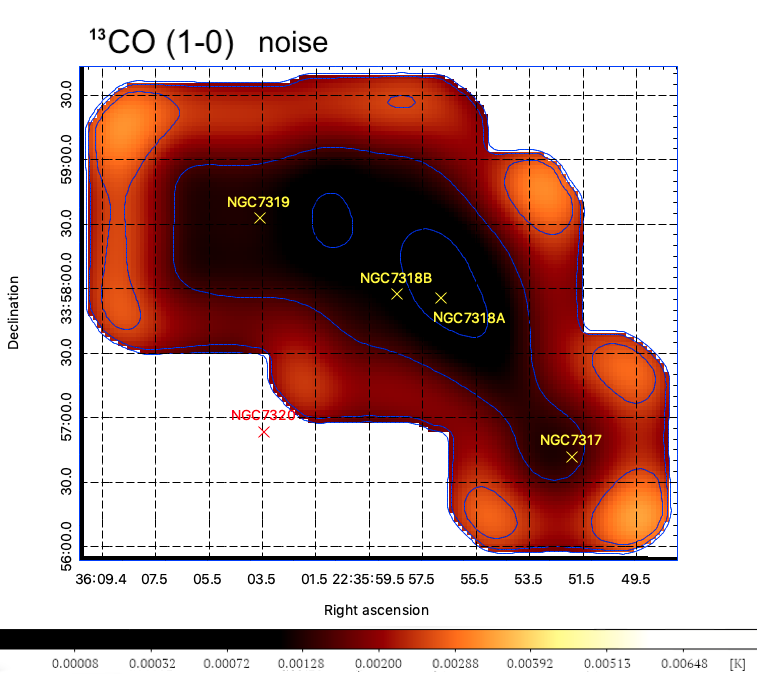}
    \end{minipage}\\
     \begin{minipage}{0.48\textwidth}
    \centering
    \includegraphics[width=\textwidth]{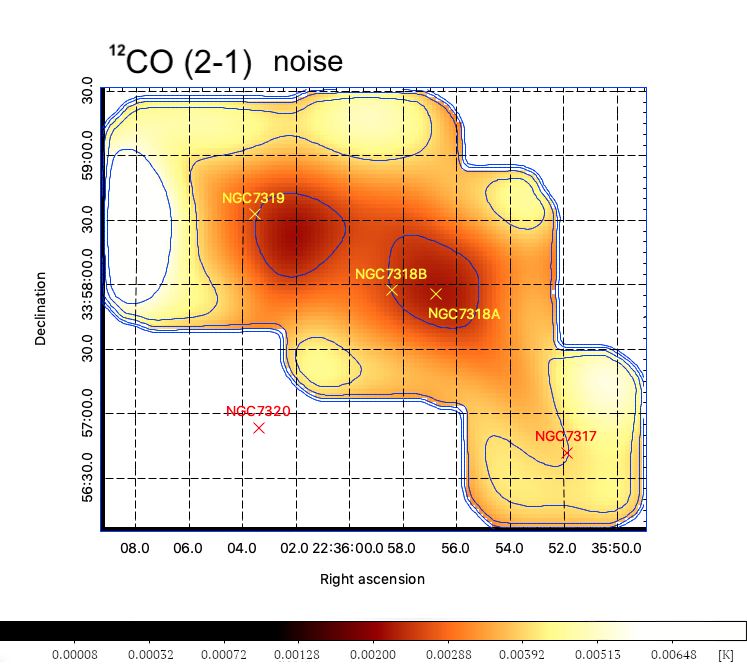}
    \end{minipage}
    \begin{minipage}{0.48\textwidth}
    \centering
    \caption{Maps showing the rms noise in the CO channels. The red and yellow crosses pinpoint the galaxies.  \textit{Top Left:} $^{12}\text{CO\,}(1-0)$; Contours at 1.5, 2.5, 4.0 and 5.5 mK are displayed in blue. \textit{Bottom Left:} $^{12}\text{CO\,}(2-1)$. Contours at 2.5, 4.0, and 5.5 mK are displayed in blue.\ \textit{Top Right: }$^{13}\text{CO\,}(1-0)$. Contours at 0.9, 1.5, and 2.5 mK are displayed in blue.}
    \label{fig:CO_noise}
    \end{minipage}   
\end{figure}
\clearpage

\section{$^{12}\text{CO~}(1-0)$, $^{12}\text{CO~}(2-1),$ and $^{13}\text{CO~}(1-0)$ spectra for a selection of regions}
\label{app:CO_spec}

\begin{figure*}[ht!]
   \subfloat[\label{fig:12CO10_19}]{%
      \includegraphics[width=0.3\textwidth]{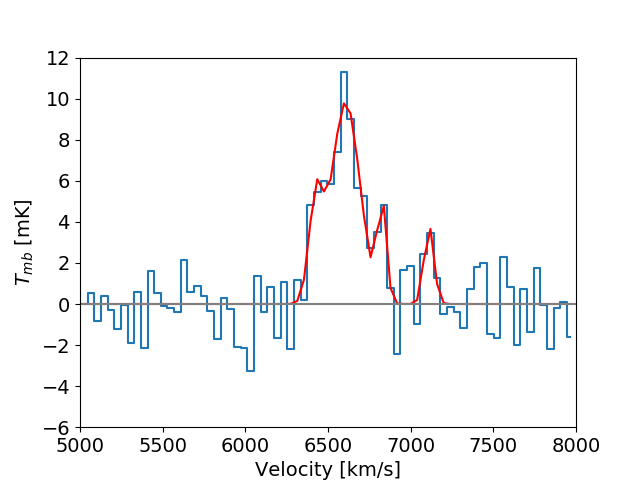}}
\hspace{\fill}
   \subfloat[\label{fig:12CO21_19} ]{%
      \includegraphics[width=0.3\textwidth]{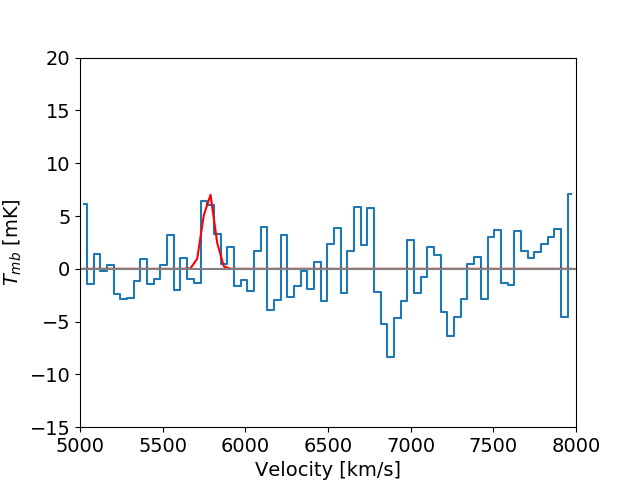}}
\hspace{\fill}
   \subfloat[\label{fig:13CO10_19}]{%
      \includegraphics[width=0.3\textwidth]{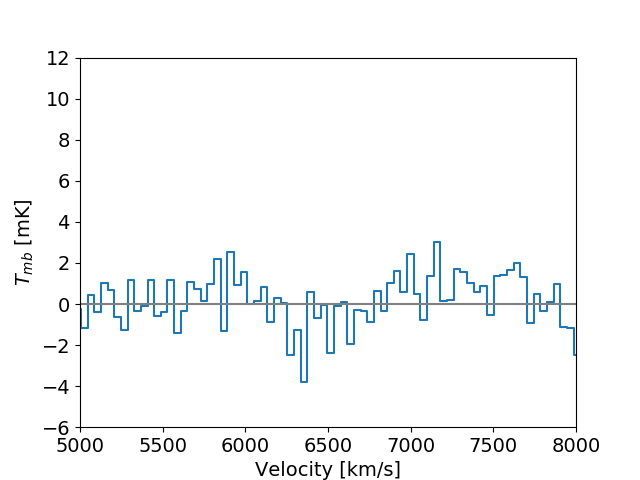}}
\caption{\label{fig:CO_19}Spectra of the CO emission extracted using an HPBW of a 22$''$ diameter from the smoothed maps (i.e. the central 11$''$) of NGC7319, region 19. The fitted lines are in red. (a) $^{12}\text{CO~}(1- 0)$ line. (b) $^{12}\text{CO~}(2- 1)$ line. (c)  $^{13}\text{CO~}(1- 0)$ line.}
\end{figure*}

\begin{figure*}[ht!]
   \subfloat[\label{fig:12CO10_19_o}]{%
      \includegraphics[width=0.3\textwidth]{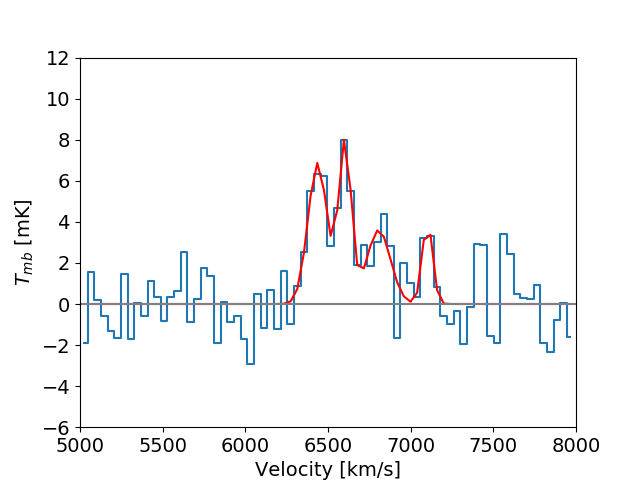}}
\hspace{\fill}
   \subfloat[\label{fig:12CO21_19_o} ]{%
      \includegraphics[width=0.3\textwidth]{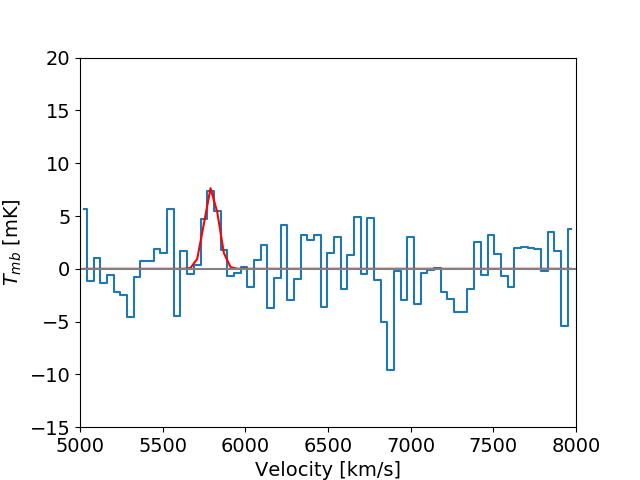}}
\hspace{\fill}
   \subfloat[\label{fig:13CO10_19_o}]{%
      \includegraphics[width=0.3\textwidth]{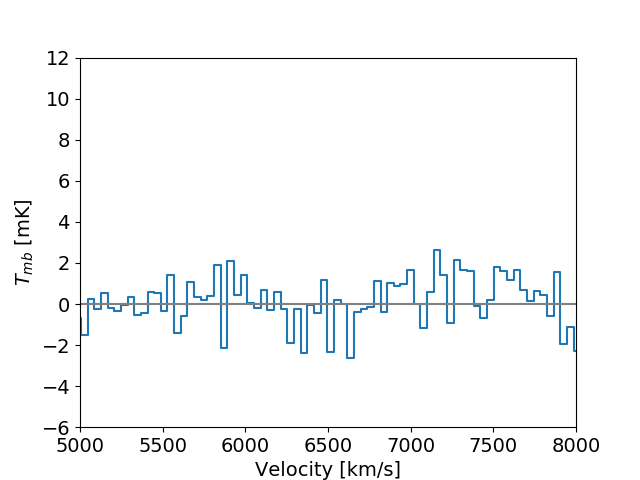}}
\caption{\label{fig:CO_19_o}As in Fig. \ref{fig:CO_19} but for region 19\_o. (a)  $^{12}\text{CO~}(1- 0)$ line. (b) $^{12}\text{CO~}(2- 1)$ line. (c)  $^{13}\text{CO~}(1- 0)$ line.}
\end{figure*}

\begin{figure*}[ht!]
   \subfloat[\label{fig:12CO10_19_i}]{%
      \includegraphics[width=0.3\textwidth]{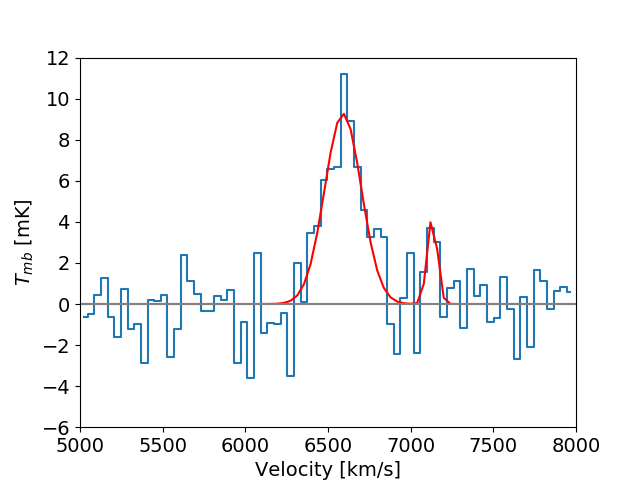}}
\hspace{\fill}
   \subfloat[\label{fig:12CO21_19_i} ]{%
      \includegraphics[width=0.3\textwidth]{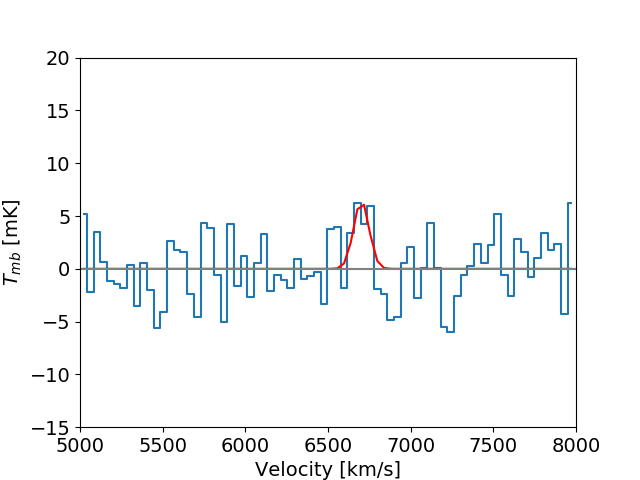}}
\hspace{\fill}
   \subfloat[\label{fig:13CO10_19_i}]{%
      \includegraphics[width=0.3\textwidth]{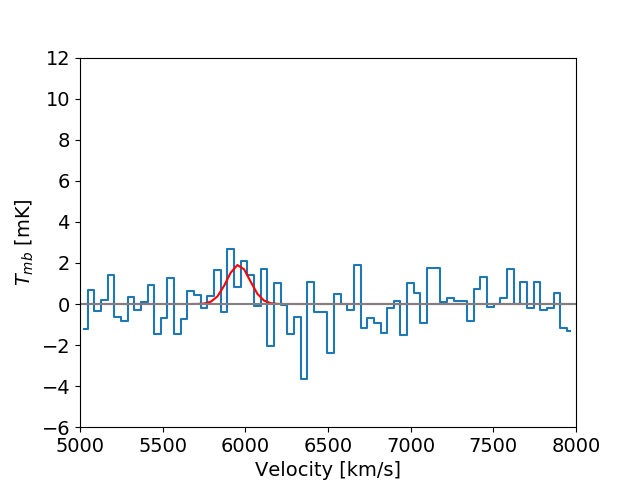}}
\caption{\label{fig:CO_19_i}As in Fig. \ref{fig:CO_19} but for region 19\_i. (a)  $^{12}\text{CO~}(1- 0)$ line. (b)  $^{12}\text{CO~}(2- 1)$ line. (c)  $^{13}\text{CO~}(1- 0)$ line.}
\end{figure*}

\begin{figure*}[ht!]
   \subfloat[\label{fig:12CO10_19_ii}]{%
      \includegraphics[width=0.3\textwidth]{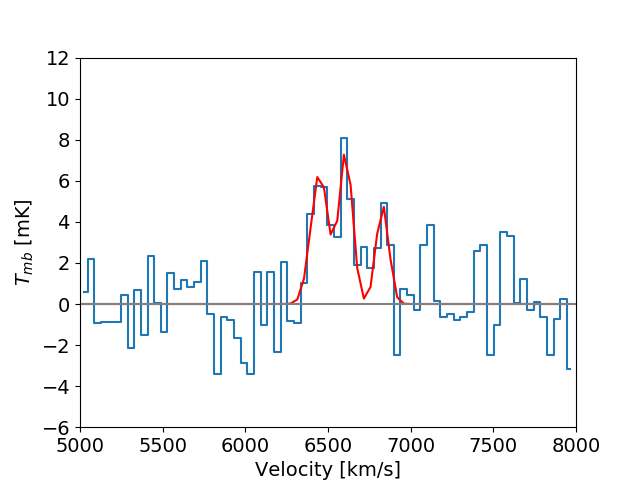}}
\hspace{\fill}
   \subfloat[\label{fig:12CO21_19_ii} ]{%
      \includegraphics[width=0.3\textwidth]{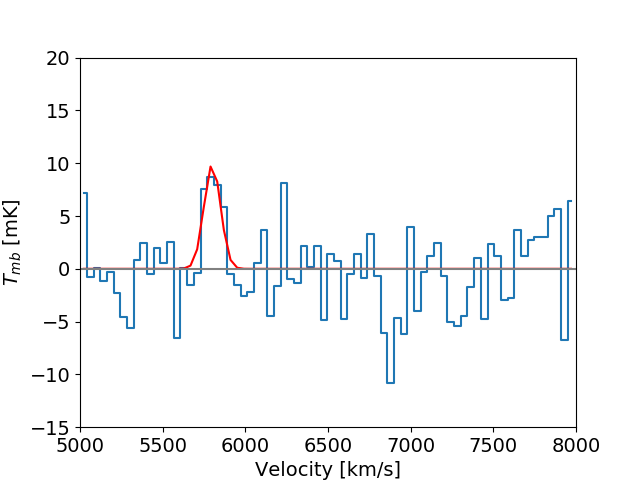}}
\hspace{\fill}
   \subfloat[\label{fig:13CO10_19_ii}]{%
      \includegraphics[width=0.3\textwidth]{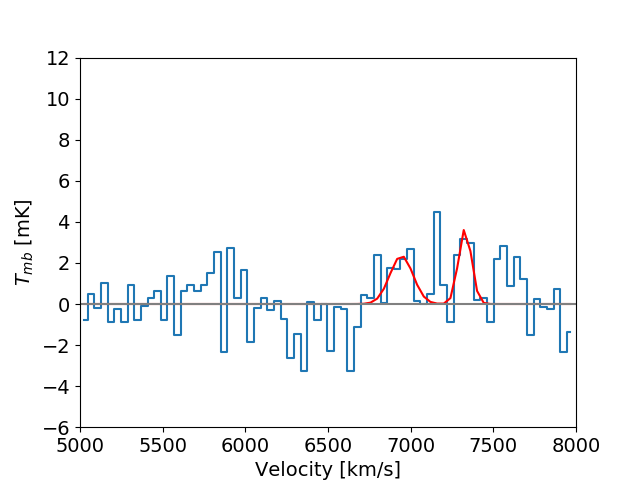}}
\caption{\label{fig:CO_19_ii}As in Fig. \ref{fig:CO_19} but for region 19\_ii. (a)  $^{12}\text{CO~}(1- 0)$ line. (b)  $^{12}\text{CO~}(2- 1)$ line. (c)  $^{13}\text{CO~}(1- 0)$ line.}
\end{figure*}

\begin{figure*}[ht!]
   \subfloat[\label{fig:12CO10_19_iii}]{%
      \includegraphics[width=0.3\textwidth]{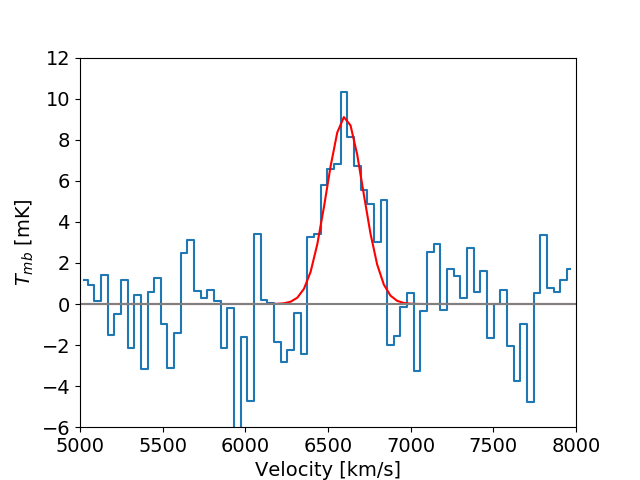}}
\hspace{\fill}
   \subfloat[\label{fig:12CO21_19_iii} ]{%
      \includegraphics[width=0.3\textwidth]{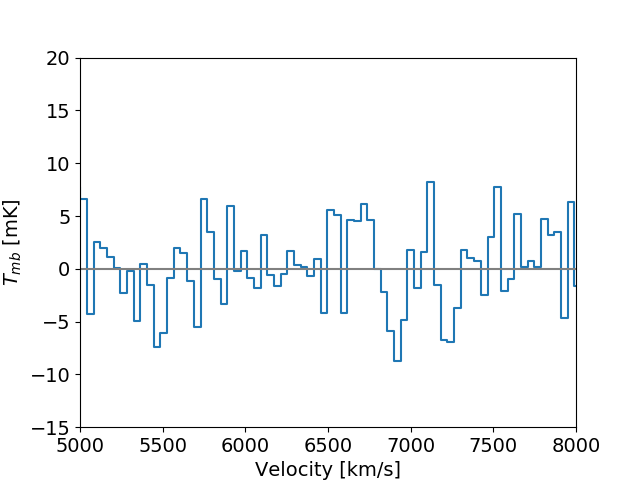}}
\hspace{\fill}
   \subfloat[\label{fig:13CO10_19_iii}]{%
      \includegraphics[width=0.3\textwidth]{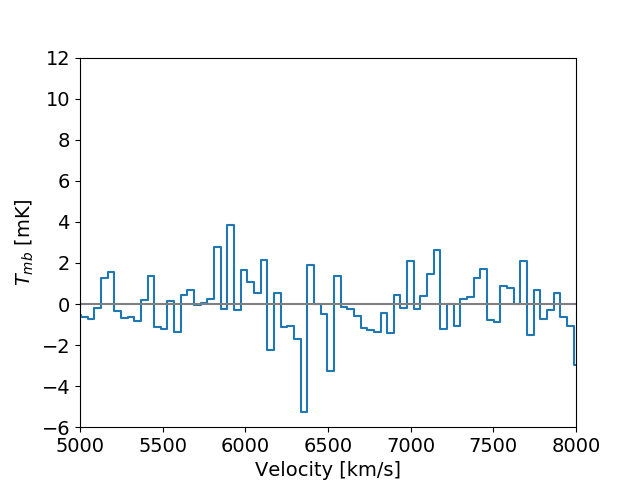}}
\caption{\label{fig:CO_19_iii}As in Fig. \ref{fig:CO_19} but for region 19\_iii. (a)  $^{12}\text{CO}(1\to 0)$ line. (b) $\text{CO}(2\to 1)$ line. (c)  $^{13}\text{CO}(1\to 0)$ line.}
\end{figure*}

\begin{figure*}[ht!]
   \subfloat[\label{fig:12CO10_19_iv}]{%
      \includegraphics[width=0.3\textwidth]{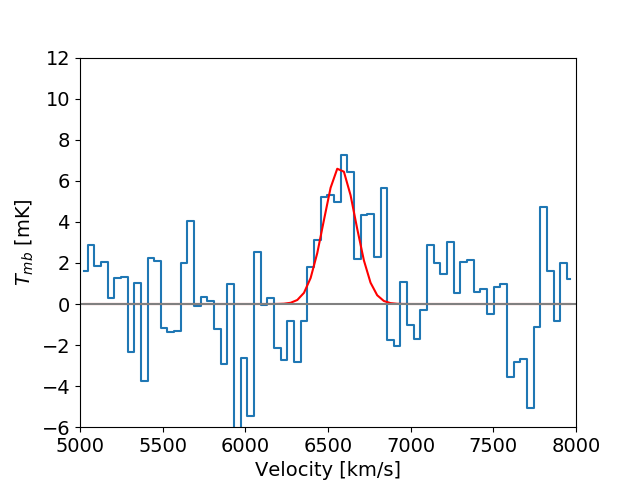}}
\hspace{\fill}
   \subfloat[\label{fig:12CO21_19_iv} ]{%
      \includegraphics[width=0.3\textwidth]{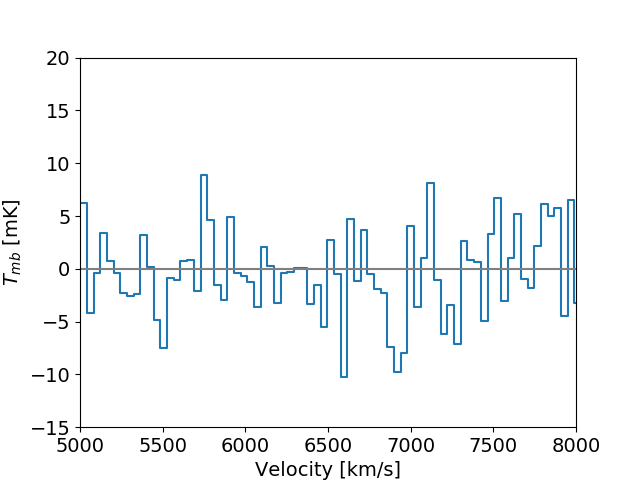}}
\hspace{\fill}
   \subfloat[\label{fig:13CO10_19_iv}]{%
      \includegraphics[width=0.3\textwidth]{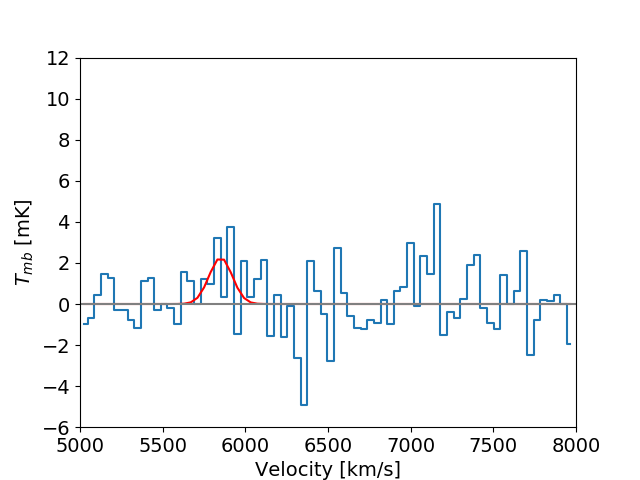}}
\caption{\label{fig:CO_19_iv}As in Fig. \ref{fig:CO_19} but for region 19\_iv. (a)  $^{12}\text{CO}(1\to 0)$ line. (b)  $^{12}\text{CO}(2\to 1)$ line. (c)  $^{13}\text{CO}(1\to 0)$ line.}
\end{figure*}

\begin{figure*}[ht!]
   \subfloat[\label{fig:12CO10_19_v}]{%
      \includegraphics[width=0.3\textwidth]{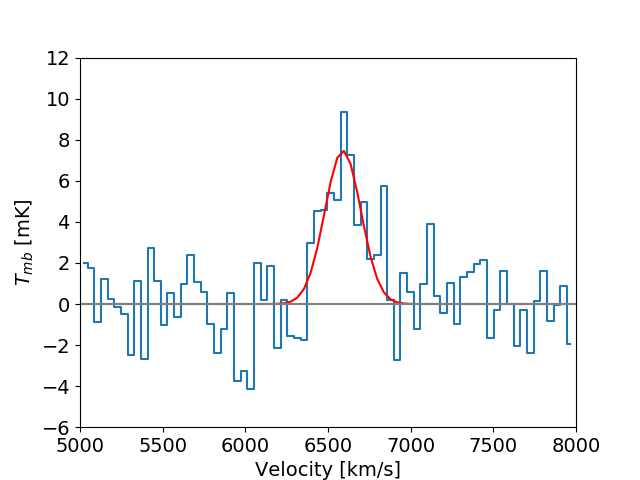}}
\hspace{\fill}
   \subfloat[\label{fig:12CO21_19_v} ]{%
      \includegraphics[width=0.3\textwidth]{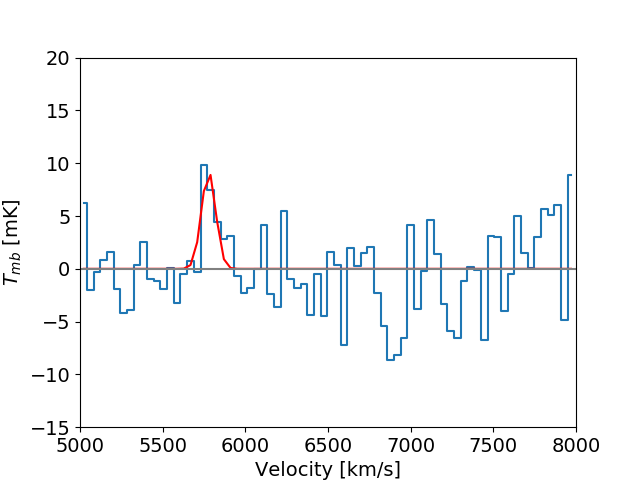}}
\hspace{\fill}
   \subfloat[\label{fig:13CO10_19_v}]{%
      \includegraphics[width=0.3\textwidth]{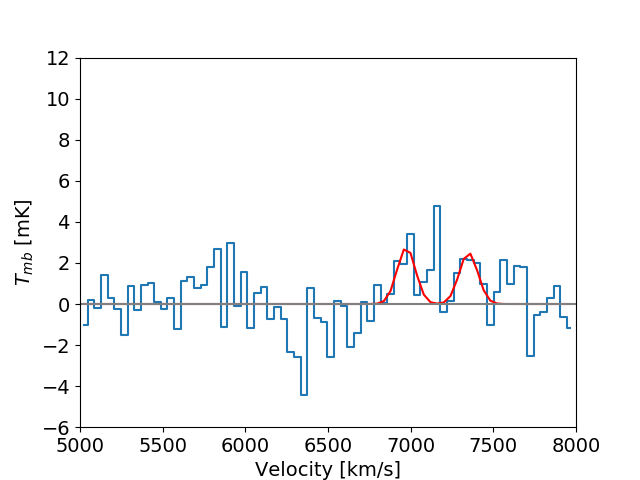}}
\caption{\label{fig:CO_19_v}As in Fig. \ref{fig:CO_19} but for region 19\_v. (a)  $^{12}\text{CO~}(1- 0)$ line. (b)  $^{12}\text{CO~}(2- 1)$ line. (c)  $^{13}\text{CO~}(1- 0)$ line.}
\end{figure*}

\begin{figure*}[ht!]
   \subfloat[\label{fig:12CO10_19_vi}]{%
      \includegraphics[width=0.3\textwidth]{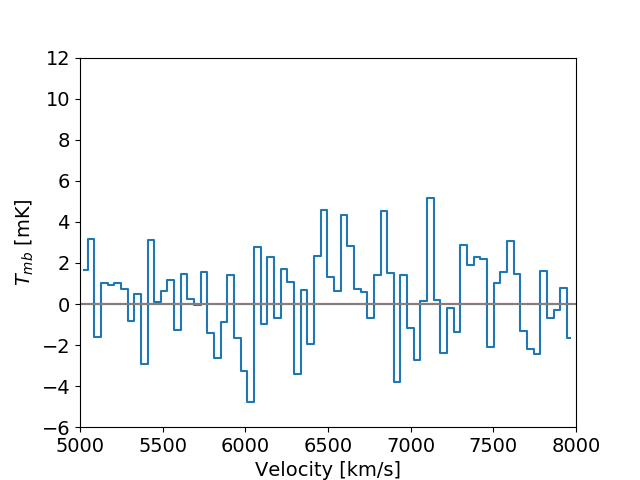}}
\hspace{\fill}
   \subfloat[\label{fig:12CO21_19_vi} ]{%
      \includegraphics[width=0.3\textwidth]{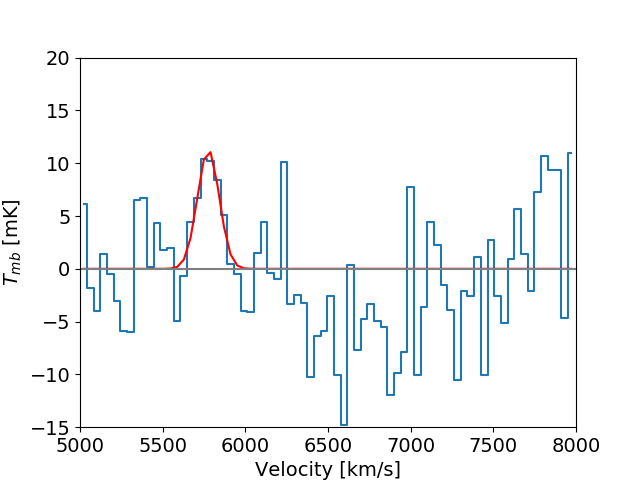}}
\hspace{\fill}
   \subfloat[\label{fig:13CO10_19_vi}]{%
      \includegraphics[width=0.3\textwidth]{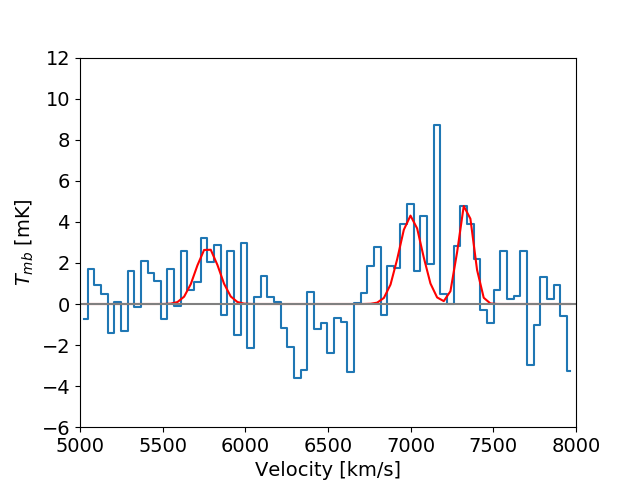}}
\caption{\label{fig:CO_19_vi}As in Fig. \ref{fig:CO_19} but for region 19\_vi. (a)  $^{12}\text{CO~}(1- 0)$ line. (b)  $^{12}\text{CO~}(2- 1)$ line. (c)  $^{13}\text{CO~}(1- 0)$ line.}
\end{figure*}

\begin{figure*}[ht!]
   \subfloat[\label{fig:12CO10_b_i}]{%
      \includegraphics[width=0.3\textwidth]{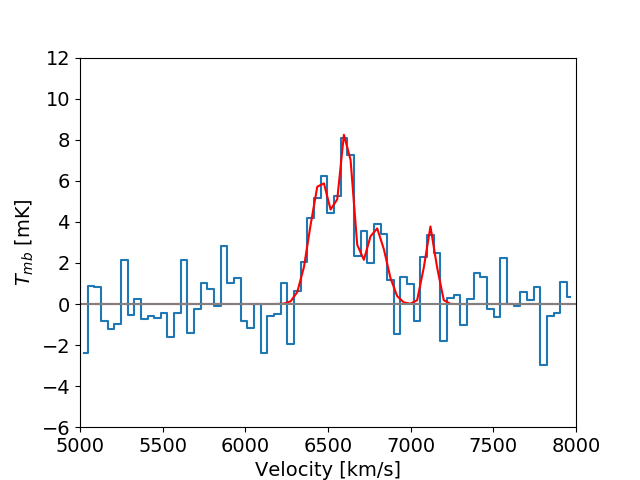}}
\hspace{\fill}
   \subfloat[\label{fig:12CO21_b_i} ]{%
      \includegraphics[width=0.3\textwidth]{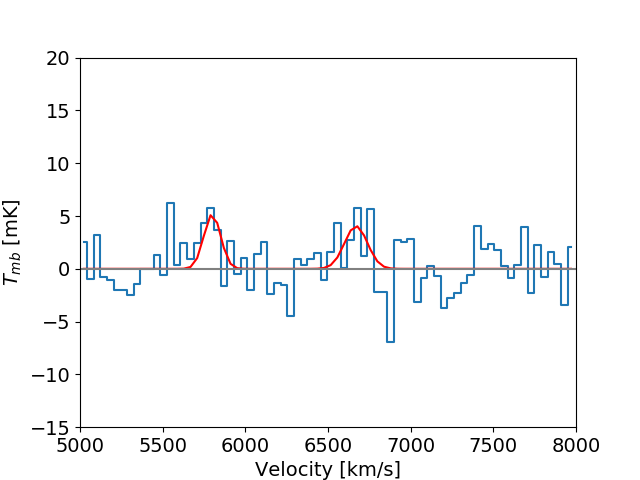}}
\hspace{\fill}
   \subfloat[\label{fig:13CO10_b_i}]{%
      \includegraphics[width=0.3\textwidth]{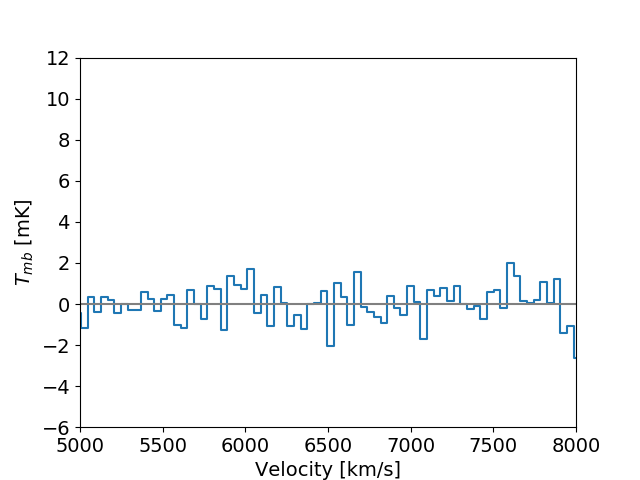}}
\caption{\label{fig:CO_b_i}As in Fig. \ref{fig:CO_19} but for region b\_i. (a)  $^{12}\text{CO~}(1- 0)$ line. (b)  $^{12}\text{CO~}(2- 1)$ line. (c)  $^{13}\text{CO~}(1- 0)$ line.}
\end{figure*}

\begin{figure*}[ht!]
   \subfloat[\label{fig:12CO10_b_ii}]{%
      \includegraphics[width=0.3\textwidth]{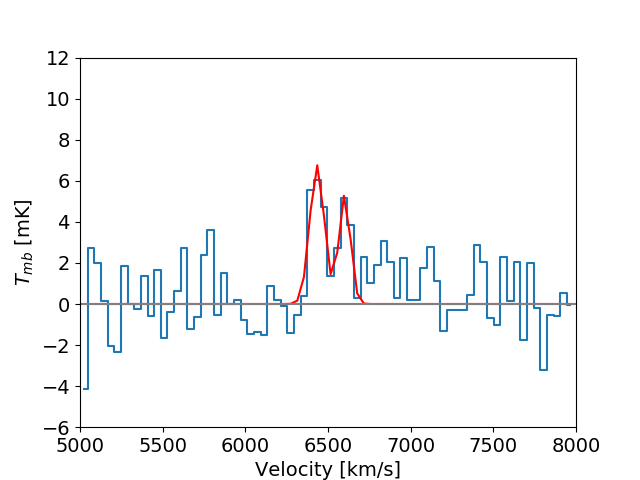}}
\hspace{\fill}
   \subfloat[\label{fig:12CO21_b_ii} ]{%
      \includegraphics[width=0.3\textwidth]{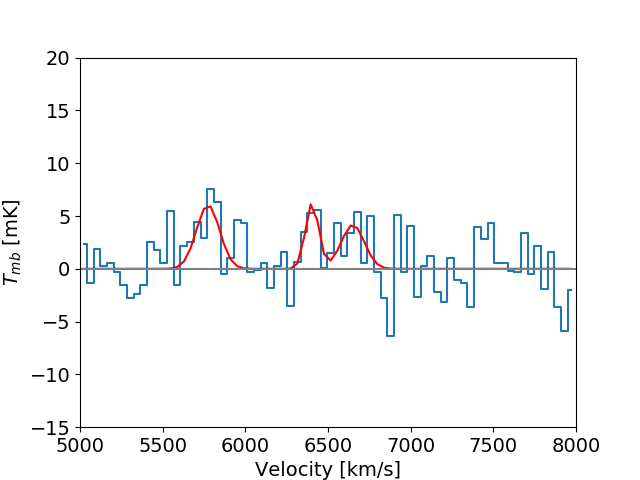}}
\hspace{\fill}
   \subfloat[\label{fig:13CO10_b_ii}]{%
      \includegraphics[width=0.3\textwidth]{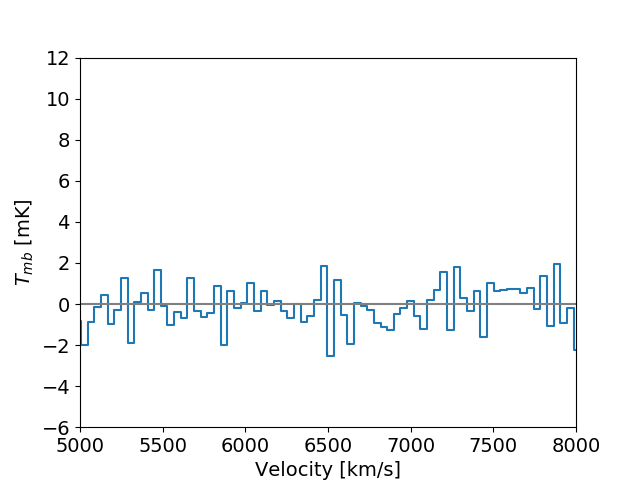}}
\caption{\label{fig:CO_b_ii}As in Fig. \ref{fig:CO_19} but for region b\_ii. (a)  $^{12}\text{CO~}(1- 0)$ line. (b)  $^{12}\text{CO~}(2- 1)$ line. (c)  $^{13}\text{CO~}(1- 0)$ line.}
\end{figure*}

\begin{figure*}[ht!]
   \subfloat[\label{fig:12CO10_sqa}]{%
      \includegraphics[width=0.3\textwidth]{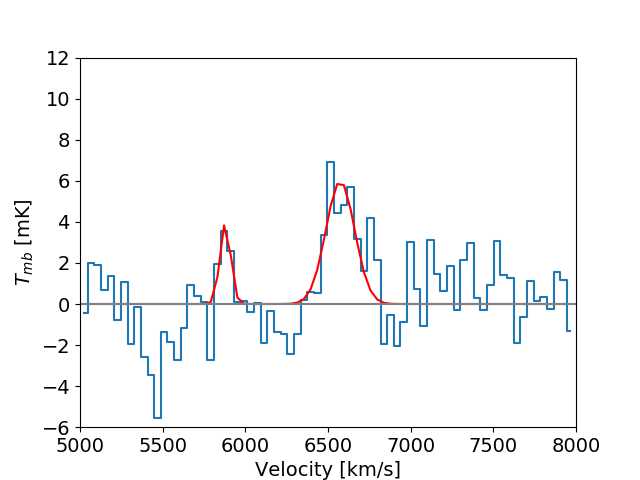}}
\hspace{\fill}
   \subfloat[\label{fig:12CO21_sqa} ]{%
      \includegraphics[width=0.3\textwidth]{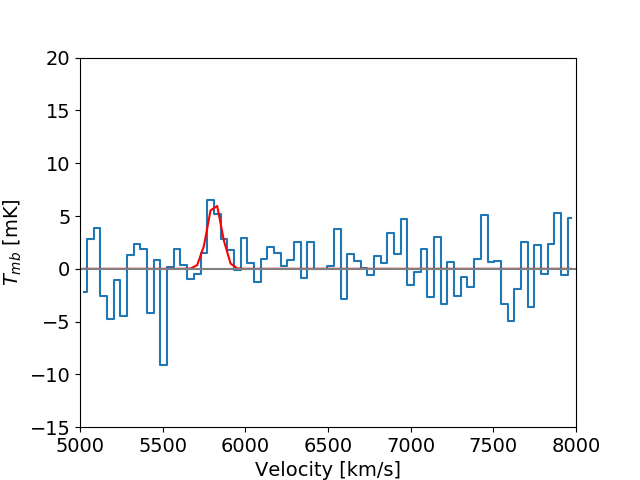}}
\hspace{\fill}
   \subfloat[\label{fig:13CO10_sqa}]{%
      \includegraphics[width=0.3\textwidth]{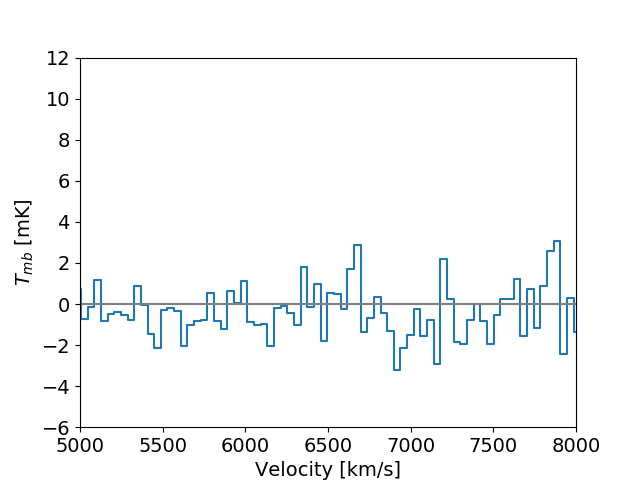}}
\caption{\label{fig:CO_sqa}As in Fig. \ref{fig:CO_19} but for region SQ-A. (a)   $^{12}\text{CO~}(1- 0)$ line. (b)  $^{12}\text{CO~}(2- 1)$ line. (c) $^{13}\text{CO~}(1- 0)$ line.}
\end{figure*}

\begin{figure*}[ht!]
   \subfloat[\label{fig:12CO10_SF_i}]{%
      \includegraphics[width=0.3\textwidth]{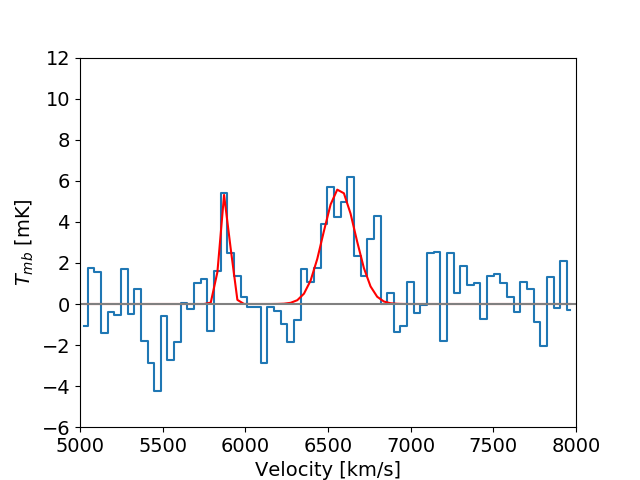}}
\hspace{\fill}
   \subfloat[\label{fig:12CO21_SF_i} ]{%
      \includegraphics[width=0.3\textwidth]{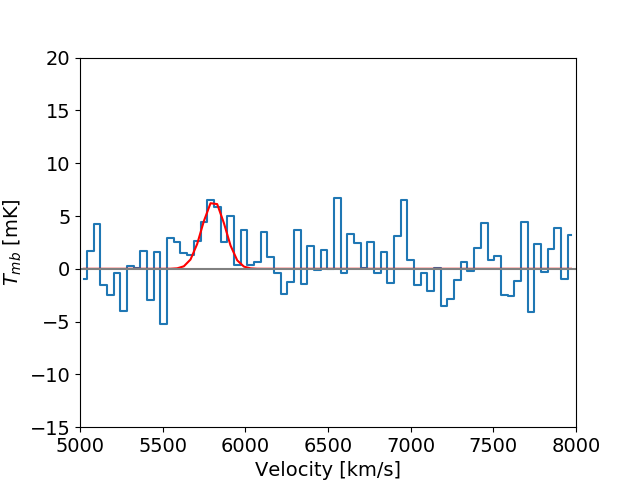}}
\hspace{\fill}
   \subfloat[\label{fig:13CO10_SF_i}]{%
      \includegraphics[width=0.3\textwidth]{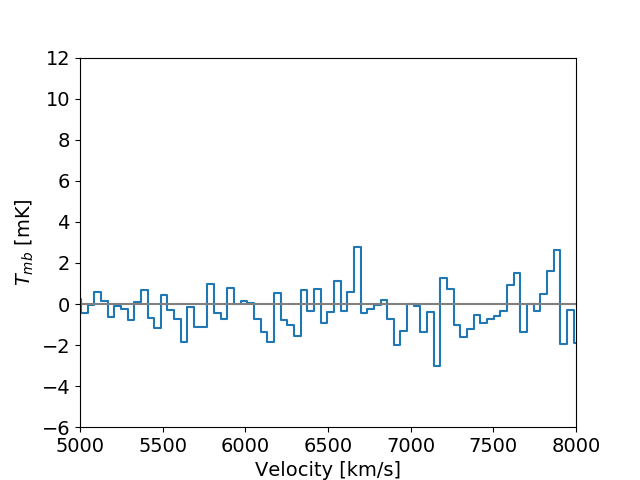}}
\caption{\label{fig:CO_SF_i}As in Fig. \ref{fig:CO_19} but for region SF\_i. (a)  $^{12}\text{CO~}(1- 0)$ line. (b)  $^{12}\text{CO~}(2- 1)$ line. (c)  $^{13}\text{CO~}(1- 0)$ line.}
\end{figure*}

\begin{figure*}[ht!]
   \subfloat[\label{fig:12CO10_SF_ii}]{%
      \includegraphics[width=0.3\textwidth]{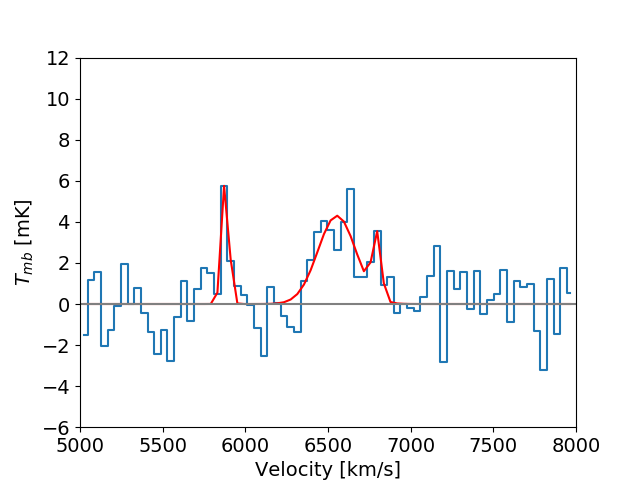}}
\hspace{\fill}
   \subfloat[\label{fig:12CO21_SF_ii} ]{%
      \includegraphics[width=0.3\textwidth]{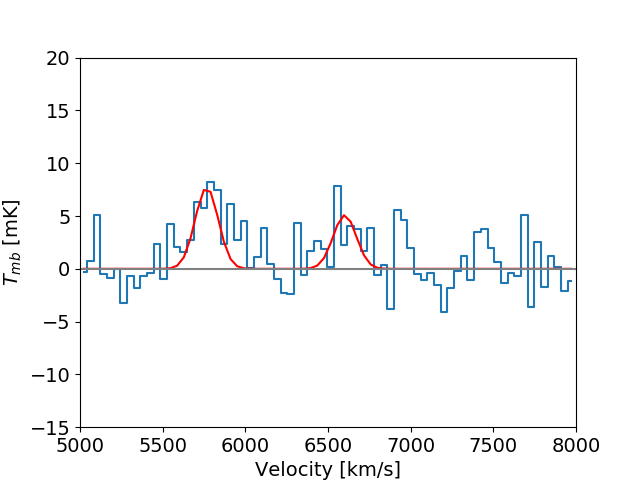}}
\hspace{\fill}
   \subfloat[\label{fig:13CO10_SF_ii}]{%
      \includegraphics[width=0.3\textwidth]{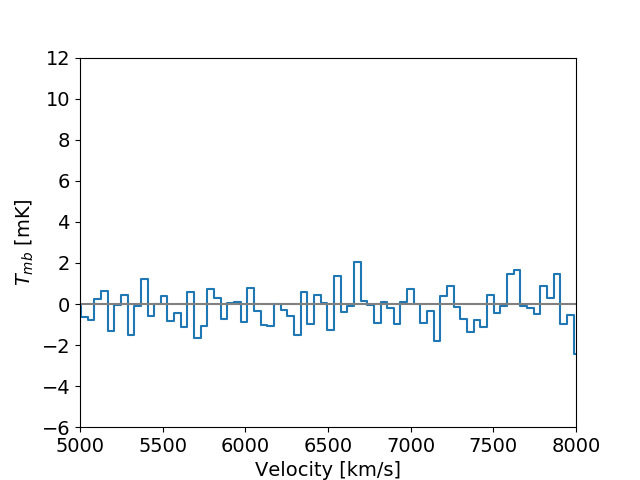}}
\caption{\label{fig:CO_SF_ii}As in Fig. \ref{fig:CO_19} but for region SF\_ii. (a)  $^{12}\text{CO~}(1- 0)$ line. (b)  $^{12}\text{CO~}(2- 1)$ line. (c)  $^{13}\text{CO~}(1- 0)$ line.}
\end{figure*}

\begin{figure*}[ht!]
   \subfloat[\label{fig:12CO10_SF_iii}]{%
      \includegraphics[width=0.3\textwidth]{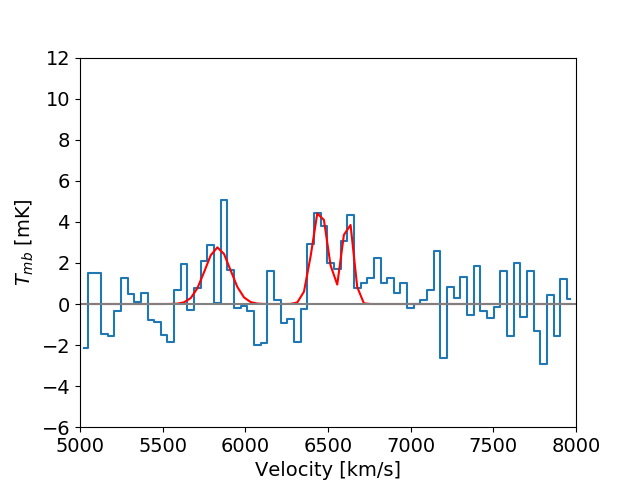}}
\hspace{\fill}
   \subfloat[\label{fig:12CO21_SF_iii} ]{%
      \includegraphics[width=0.3\textwidth]{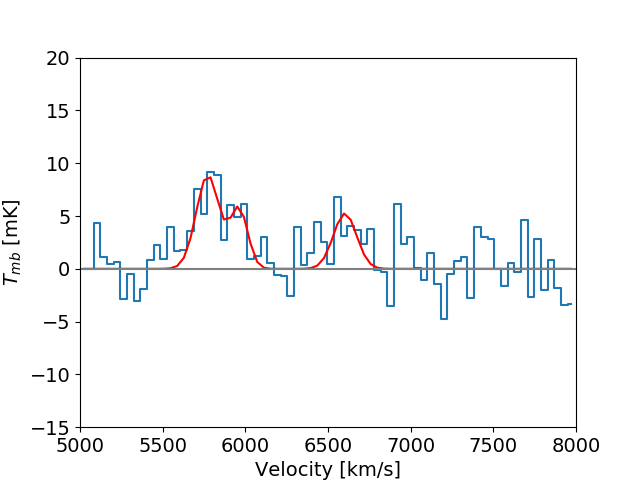}}
\hspace{\fill}
   \subfloat[\label{fig:13CO10_SF_iii}]{%
      \includegraphics[width=0.3\textwidth]{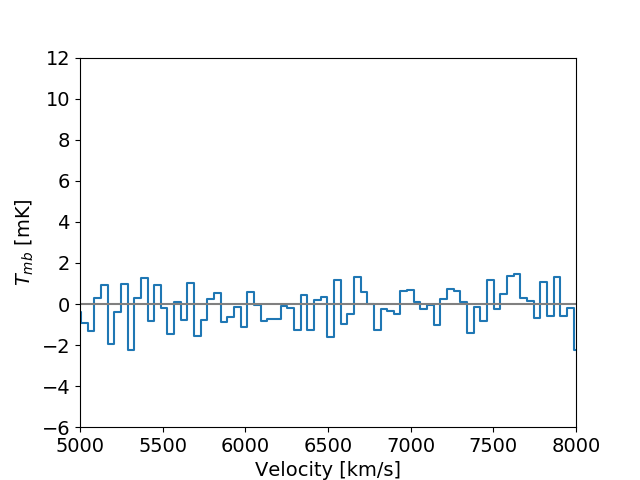}}
\caption{\label{fig:CO_SF_iii}As in Fig. \ref{fig:CO_19} but for region SF\_iii. (a)  $^{12}\text{CO}(1\to 0)$ line. (b)  $^{12}\text{CO}(2\to 1)$ line. (c)  $^{13}\text{CO}(1\to 0)$ line.}
\end{figure*}

\begin{figure*}[ht!]
   \subfloat[\label{fig:12CO10_SF_iv}]{%
      \includegraphics[width=0.3\textwidth]{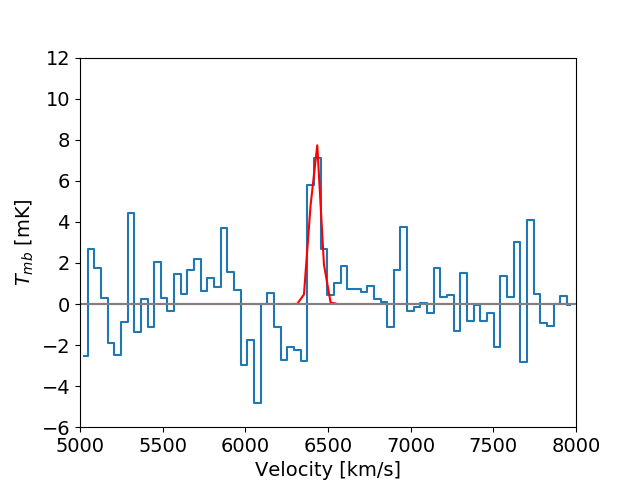}}
\hspace{\fill}
   \subfloat[\label{fig:12CO21_SF_iv} ]{%
      \includegraphics[width=0.3\textwidth]{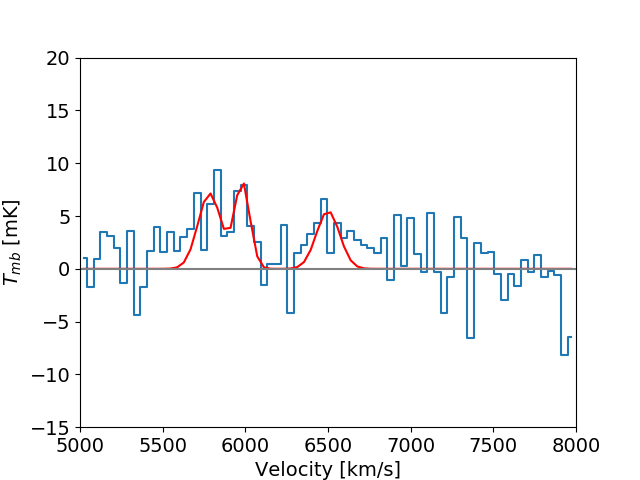}}
\hspace{\fill}
   \subfloat[\label{fig:13CO10_SF_iv}]{%
      \includegraphics[width=0.3\textwidth]{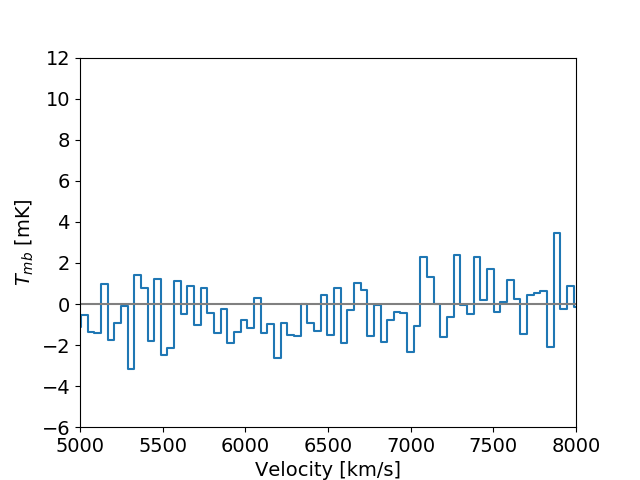}}
\caption{\label{fig:CO_SF_iv}As in Fig. \ref{fig:CO_19} but for region SF\_iv. (a)  $^{12}\text{CO}(1\to 0)$ line. (b)  $^{12}\text{CO}(2\to 1)$ line. (c)  $^{13}\text{CO}(1\to 0)$ line.}
\end{figure*}

\begin{figure*}[ht!]
   \subfloat[\label{fig:12CO10_18a}]{%
      \includegraphics[width=0.3\textwidth]{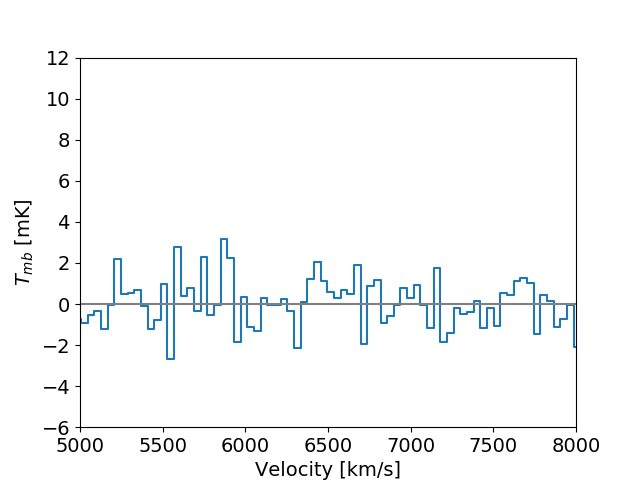}}
\hspace{\fill}
   \subfloat[\label{fig:12CO21_18a} ]{%
      \includegraphics[width=0.3\textwidth]{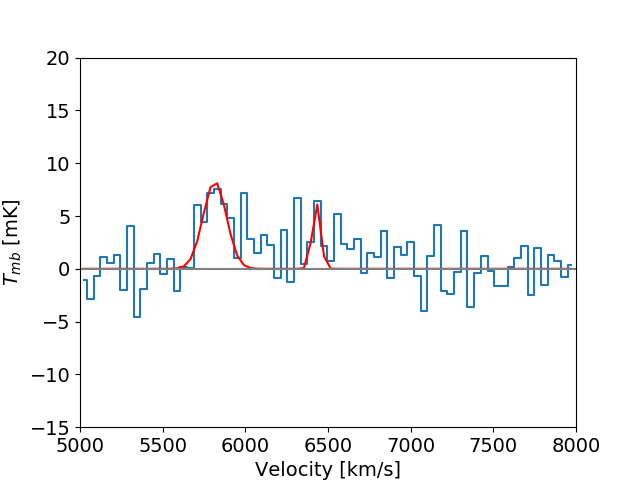}}
\hspace{\fill}
   \subfloat[\label{fig:13CO10_18a}]{%
      \includegraphics[width=0.3\textwidth]{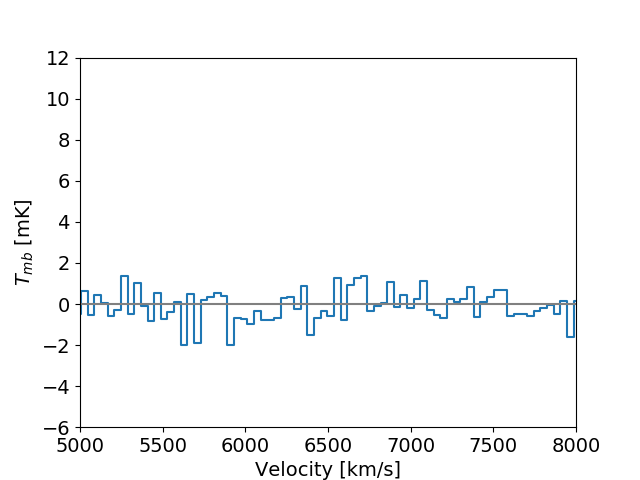}}
\caption{\label{fig:CO_18a}As in Fig. \ref{fig:CO_19} but for NGC7318A, region 18a. (a)   $^{12}\text{CO~}(1- 0)$ line. (b)   $^{12}\text{CO~}(2- 1)$ line. (c)   $^{13}\text{CO~}(1- 0)$ line.}
\end{figure*}

\begin{figure*}[ht!]
   \subfloat[\label{fig:12CO10_18b}]{%
      \includegraphics[width=0.3\textwidth]{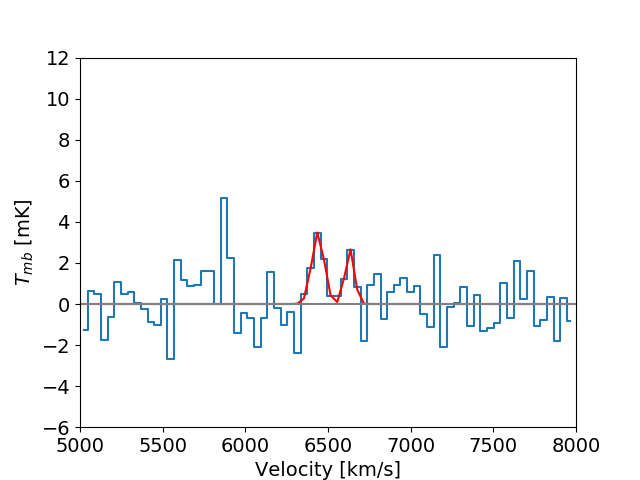}}
\hspace{\fill}
   \subfloat[\label{fig:12CO21_18b} ]{%
      \includegraphics[width=0.3\textwidth]{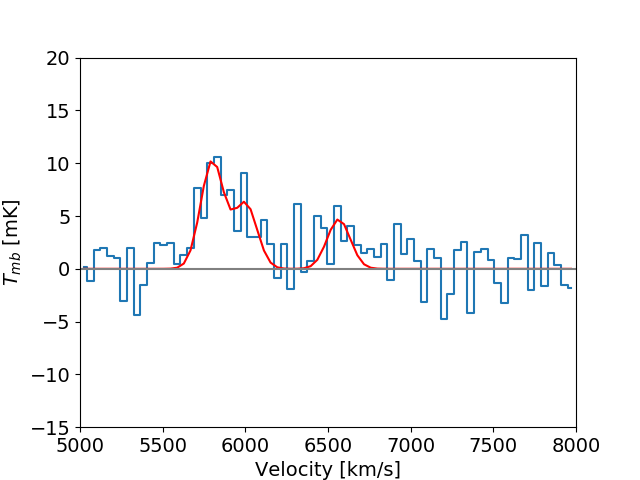}}
\hspace{\fill}
   \subfloat[\label{fig:13CO10_18b}]{%
      \includegraphics[width=0.3\textwidth]{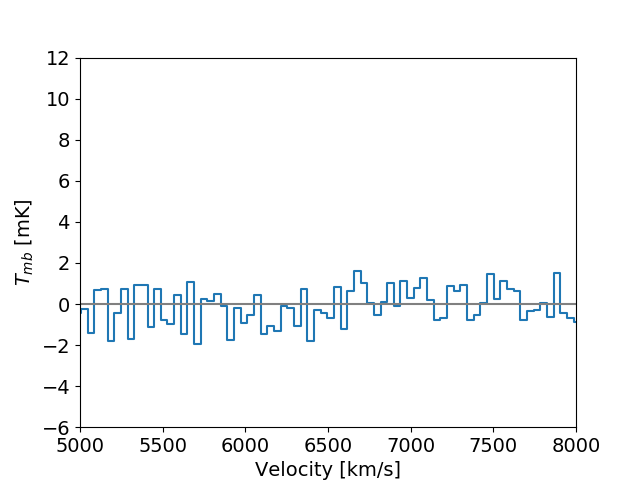}}
\caption{\label{fig:CO_18_b}As in Fig. \ref{fig:CO_19} but for NGC7318B, region 18b. (a)  $^{12}\text{CO~}(1- 0)$ line. (b)  $^{12}\text{CO~}(2- 1)$ line. (c)  $^{13}\text{CO~}(1- 0)$ line.}
\end{figure*}

\begin{figure*}[ht!]
   \subfloat[\label{fig:12CO10_18_i}]{%
      \includegraphics[width=0.3\textwidth]{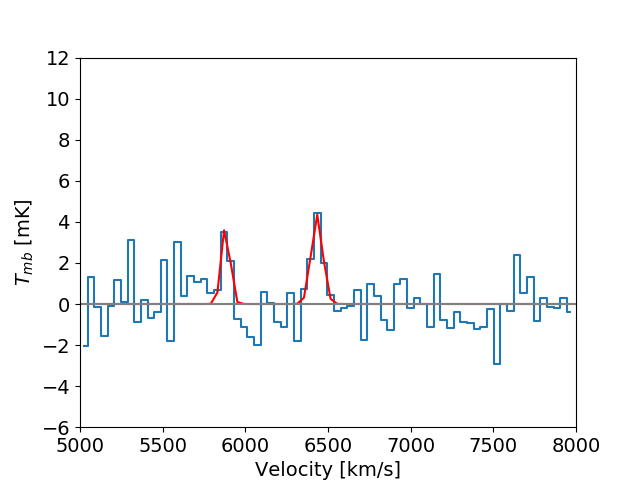}}
\hspace{\fill}
   \subfloat[\label{fig:12CO21_18_i} ]{%
      \includegraphics[width=0.3\textwidth]{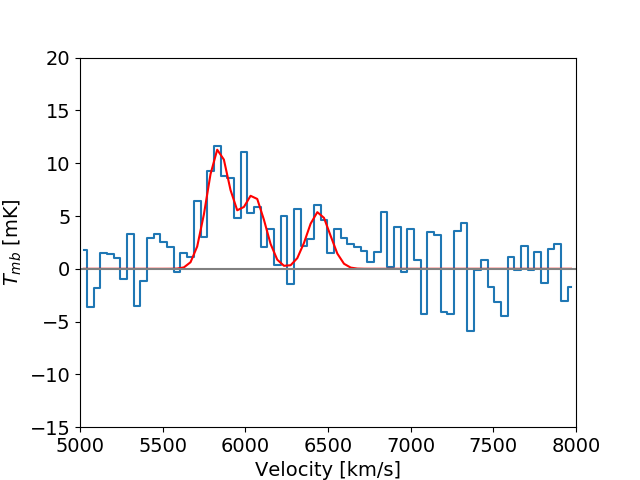}}
\hspace{\fill}
   \subfloat[\label{fig:13CO10_18_i}]{%
      \includegraphics[width=0.3\textwidth]{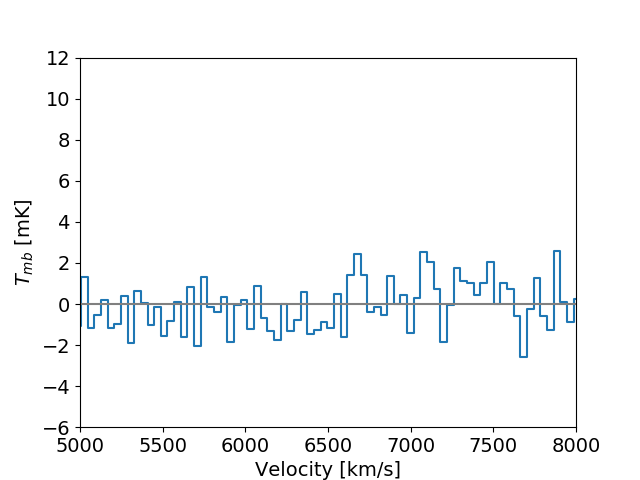}}
\caption{\label{fig:CO_18_i}As in Fig. \ref{fig:CO_19} but for region 18\_i. (a)  $^{12}\text{CO~}(1- 0)$ line. (b)  $^{12}\text{CO~}(2- 1)$ line. (c)  $^{13}\text{CO~}(1- 0)$ line.}
\end{figure*}

\begin{figure*}[ht!]
   \subfloat[\label{fig:12CO10_18_ii}]{%
      \includegraphics[width=0.3\textwidth]{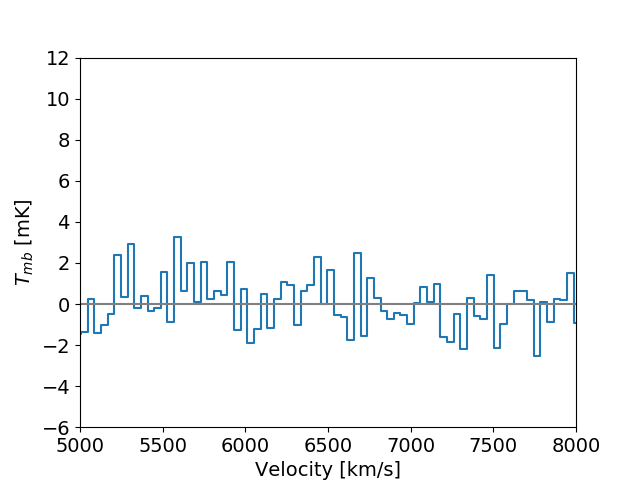}}
\hspace{\fill}
   \subfloat[\label{fig:12CO21_18_ii} ]{%
      \includegraphics[width=0.3\textwidth]{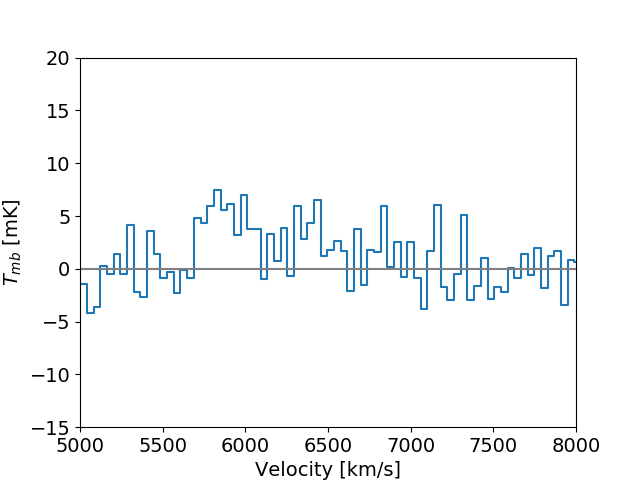}}
\hspace{\fill}
   \subfloat[\label{fig:13CO10_18_ii}]{%
      \includegraphics[width=0.3\textwidth]{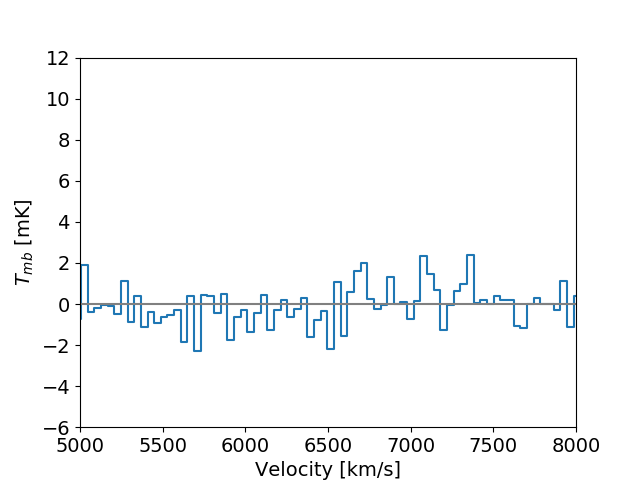}}
\caption{\label{fig:CO_18_ii}As in Fig. \ref{fig:CO_19} but for region 18\_ii. (a)  $^{12}\text{CO~}(1- 0)$ line; (b)  $^{12}\text{CO~}(2- 1)$ line; (c)  $^{13}\text{CO~}(1- 0)$ line.}
\end{figure*}

\begin{figure*}[ht!]
   \subfloat[\label{fig:12CO10_18_iii}]{%
      \includegraphics[width=0.3\textwidth]{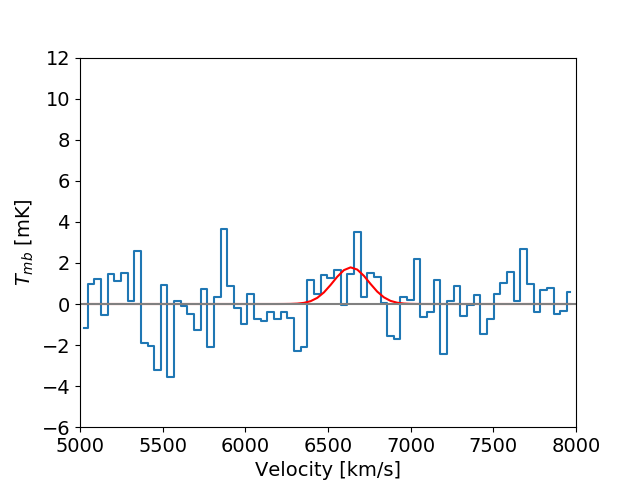}}
\hspace{\fill}
   \subfloat[\label{fig:12CO21_18_iii} ]{%
      \includegraphics[width=0.3\textwidth]{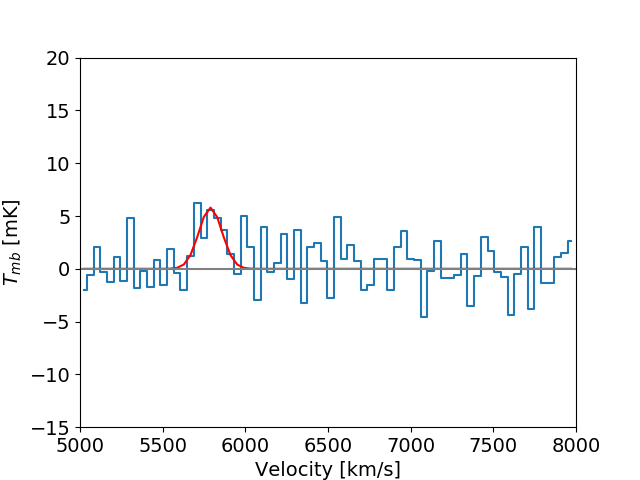}}
\hspace{\fill}
   \subfloat[\label{fig:13CO10_18_iii}]{%
      \includegraphics[width=0.3\textwidth]{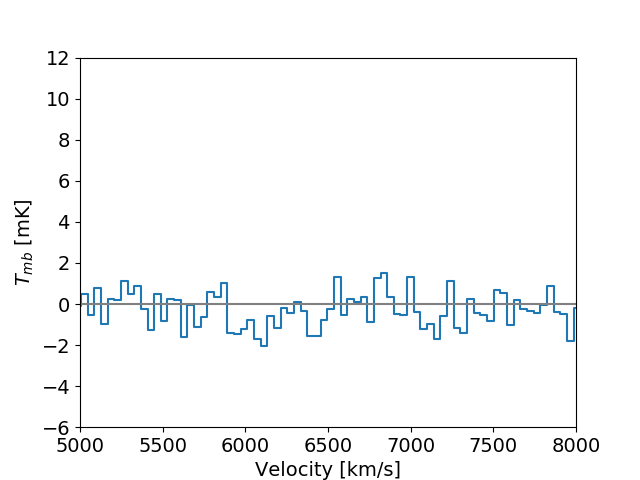}}
\caption{\label{fig:CO_18_iii}As in Fig. \ref{fig:CO_19} but for region 18\_iii. (a) $^{12}\text{CO~}(1- 0)$ line. (b) $^{12}\text{CO~}(2- 1)$ line. (c)  $^{13}\text{CO~}(1- 0)$ line.}
\end{figure*}

\begin{figure*}[ht!]
   \subfloat[\label{fig:12CO10_18_iv}]{%
      \includegraphics[width=0.3\textwidth]{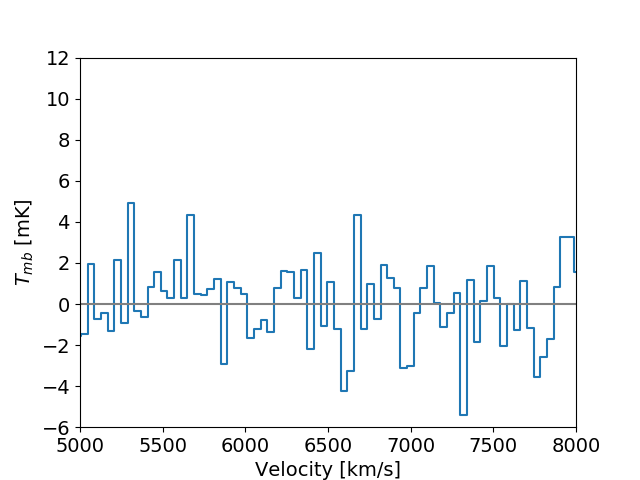}}
\hspace{\fill}
   \subfloat[\label{fig:12CO21_18_iv} ]{%
      \includegraphics[width=0.3\textwidth]{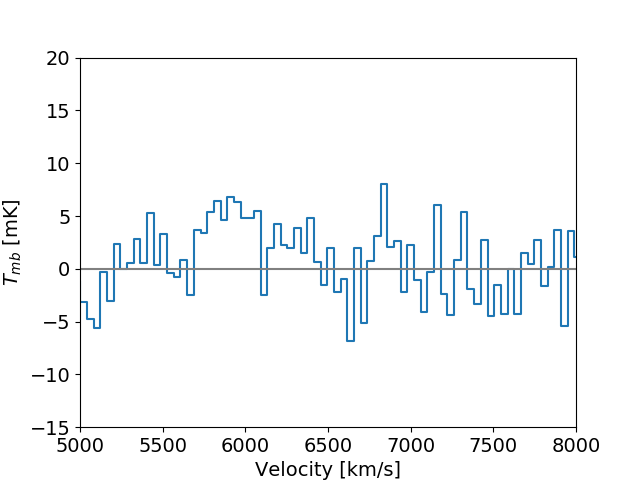}}
\hspace{\fill}
   \subfloat[\label{fig:13CO10_18_iv}]{%
      \includegraphics[width=0.3\textwidth]{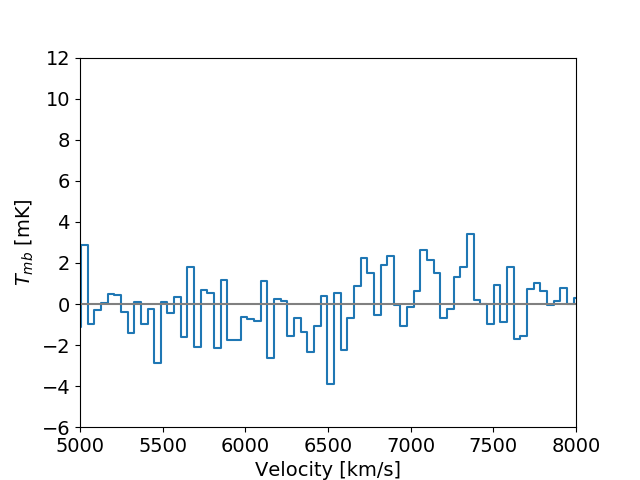}}
\caption{\label{fig:CO_18_iv}As in Fig. \ref{fig:CO_19} but for region 18\_iv. (a) $^{12}\text{CO~}(1- 0)$ line. (b) $^{12}\text{CO~}(2- 1)$ line. (c) $^{13}\text{CO~}(1- 0)$ line.}
\end{figure*}

\begin{figure*}[ht!]
   \subfloat[\label{fig:12CO10_17}]{%
      \includegraphics[width=0.3\textwidth]{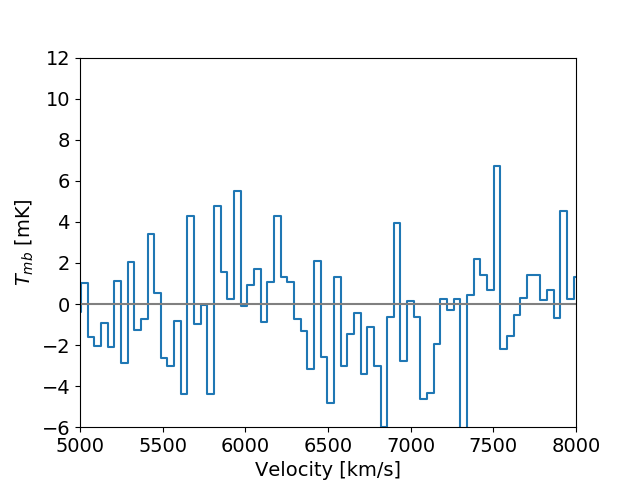}}
\hspace{\fill}
   \subfloat[\label{fig:12CO21_17} ]{%
      \includegraphics[width=0.3\textwidth]{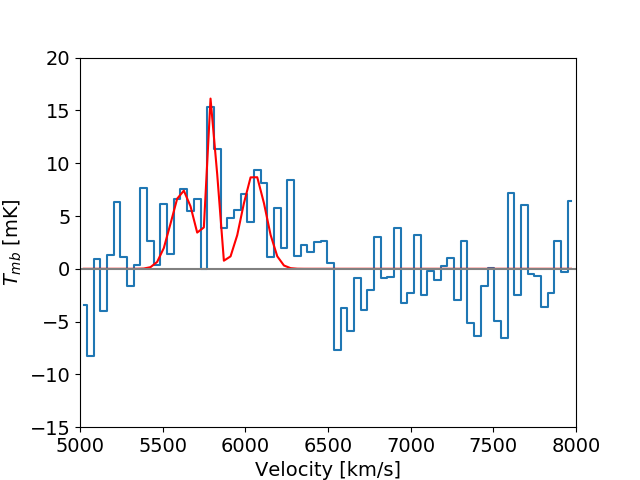}}
\hspace{\fill}
   \subfloat[\label{fig:13CO10_17}]{%
      \includegraphics[width=0.3\textwidth]{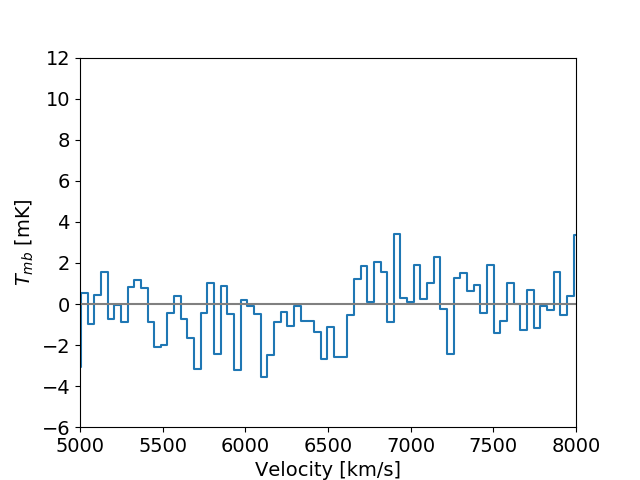}}
\caption{\label{fig:CO_17}As in Fig. \ref{fig:CO_19} but for NGC7317, region 17. (a)  $^{12}\text{CO~}(1- 0)$ line. (b)  $^{12}\text{CO~}(2- 1)$ line. (c)  $^{13}\text{CO~}(1- 0)$ line.}
\end{figure*}

\begin{figure*}[ht!]
   \subfloat[\label{fig:12CO10_Sblob}]{%
      \includegraphics[width=0.3\textwidth]{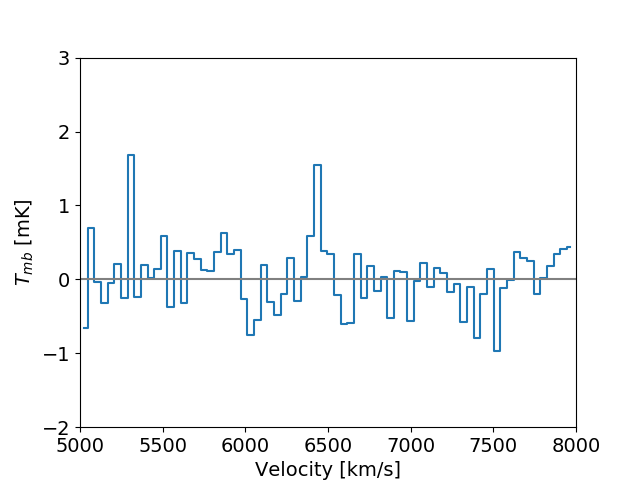}}
\hspace{\fill}
   \subfloat[\label{fig:12CO21_Sblob} ]{%
      \includegraphics[width=0.3\textwidth]{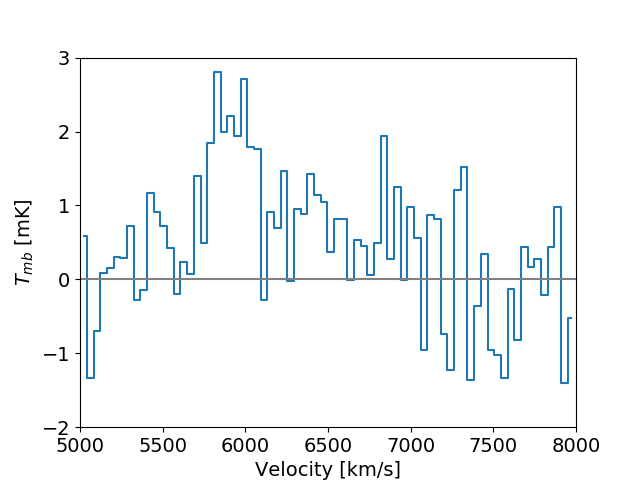}}
\hspace{\fill}
   \subfloat[\label{fig:13CO10_Sblob}]{%
      \includegraphics[width=0.3\textwidth]{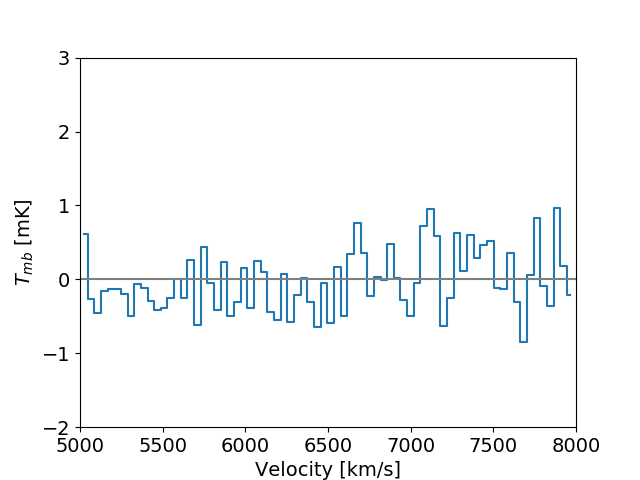}}
\caption{\label{fig:CO_Sblob}Spectra of the average CO emission in the southern congregation per $arcsec^2$ as marked by the magenta ellipse in Fig. \ref{fig:radio_regions}. (a)  $^{12}\text{CO~}(1- 0)$ line. (b)  $^{12}\text{CO~}(2- 1)$ line. (c)  $^{13}\text{CO~}(1- 0)$ line.}
\end{figure*}

\clearpage

\section{Tables}
\label{app:tables}
\subsection{Regions in or near NGC7319}
\label{app:tables_7319}
\begin{table}[h!]
\centering
\caption{\small{Ionised gas and stellar flux, line-of-sight velocity, and velocity dispersion in region 19\_0-3.}}
    \label{table:7319}
\begin{tabular}{|lll|c|c|c|c|}
\hline \hline
                & Vel &  &  &   &   &   \\ 
                & comp &  & {Region 19\_0} & {19\_1} & {19\_2} & {19\_3} \\  \hline
H$\beta$\_NLR         & high              & Flux      & 5.28E-16       $\pm$ 2.54E-17       & 8.98E-17     $\pm$ 7.99E-18    & 1.13E-16     $\pm$ 5.76E-18    & 2.19E-16     $\pm$ 8.56E-18    \\
                &                   & Vlos      & 6709         $\pm$ 9            & 6394       $\pm$ 15        & 6642       $\pm$ 9         & 6704       $\pm$ 7         \\
                &                   & Vel.Disp  & 211.1          $\pm$ 8.7            & 192.4        $\pm$ 14.7        & 195.8        $\pm$ 8.6         & 197.1        $\pm$ 6.6         \\
                &     low              &   &                  -              &                -           &                -           &                -           \\ \hline
H$\alpha$\_BLR         & high              & Flux      & 2.86E-15       $\pm$ 1.24E-16       &                -           &                -           &                -           \\
                &                   & Vlos      & 6714         $\pm$ 2            &                -           &                -           &                -           \\
\hline
\end{tabular}
\tablefoot{
Flux in ($erg\,cm^{2}\,s^{-1}$), line-of-sight velocity, and velocity dispersion in (km$\,$s$^{-1}$). The regions are marked in Fig.\ref{fig:regions}.
}
\end{table}

\begin{table}[h!]
\centering
\caption{\small{As in Table\, \ref{table:7319} but for region 19\_4-7.}}
\label{table:7319_2}
\begin{tabular}{|lll|c|c|c|c|}
\hline \hline
                & Vel&  &  &   &   &   \\ 
                & comp&  & {Region 19\_4}& {19\_5}& {19\_6}& {19\_7}\\  \hline
H $\beta$\_NLR        & high & Flux     & 2.22E-17  $\pm$ 5.42E-18 & 1.75E-17  $\pm$ 4.50E-18 & 1.11E-17  $\pm$ 3.17E-18 &          -        \\
                &      & Vlos     & 7502    $\pm$ 59     & 7502    $\pm$ 62     & 7502    $\pm$ 43     &          -        \\
                &      & Vel.Disp & 305.1     $\pm$ 64.6     & 305.1     $\pm$ 68.1     & 178.6     $\pm$ 44.1     &          -        \\
                & low  & Flux     & 2.64E-17  $\pm$ 4.29E-18 & 2.62E-17  $\pm$ 3.47E-18 & 3.99E-17  $\pm$ 3.97E-18 & 1.55E-17  $\pm$ 3.02E-18 \\
                &      & Vlos     & 6596    $\pm$ 27     & 6622    $\pm$ 21     & 6450    $\pm$ 24     & 6444    $\pm$ 35     \\
                &      & Vel.Disp & 196.7     $\pm$ 27.2     & 187.4     $\pm$ 21.0     & 290.4     $\pm$ 24.8     & 207.0     $\pm$ 34.7     \\
\hline  
\end{tabular}
\end{table}

\begin{table}[h!]
\caption{\label{table:CO_19} \small{ CO line-of-sight velocity, velocity dispersion, integrated flux, and H$_2$ molecular gas mass for regions in or near NGC7319.}}
$^{12}\text{CO~}(1- 0)$ \\
\begin{tabular}{|cc|c|c|c|c|c|} 
\hline\hline
Region  &  & {Vlos        } & {Vel.Disp     } & {Flux          } & {M(H$_2$)      } \\ 
  &  & {(km$\,$s$^{-1}$)} & {(km$\,$s$^{-1}$)} & {(K km$\,$s$^{-1}$)} & {      ($10^8M_{\odot}$)} \\\hline 
19\_    &           & 6427                $\pm$             16         & 41.4             $\pm$                16.1            & 0.50              $\pm$                 0.19             & 1.62            $\pm$                0.62            \\
        &           & 6607                $\pm$             13         & 85.6             $\pm$                15.7            & 2.11              $\pm$                 0.30             & 6.81            $\pm$                0.95            \\
        &           & 6826                $\pm$             10         & 26.4             $\pm$                11.6            & 0.33              $\pm$                 0.14             & 1.06            $\pm$                0.46            \\
        &           & 7111                $\pm$             14         & 29.5             $\pm$                12.8            & 0.28              $\pm$                 0.13             & 0.91            $\pm$                0.41            \\ \hline 
19\_o   &           & 6437                $\pm$ 13         & 58.0             $\pm$ 13.9            & 1.00              $\pm$ 0.21             & 3.22            $\pm$ 0.69            \\
        &           & 6603                $\pm$ 9         & 38.1             $\pm$ 9.9             & 0.76              $\pm$ 0.17             & 2.43            $\pm$ 0.55            \\
\hline                         
\end{tabular}
\vspace{0.5cm}
\\
\begin{minipage}{.48\linewidth}
$^{12}\text{CO~}(2- 1)$\\
\begin{tabular}{|cc|c|c|c|c|}
\hline\hline
Region  &  & {Vlos        } & {Vel.Disp     } & {Flux          }  \\ 
  &  & {(km$\,$s$^{-1}$)} & {(km$\,$s$^{-1}$)} & {(K km$\,$s$^{-1}$)}  \\\hline 
19\_    &           & 5779                $\pm$ 17         & 34.6             $\pm$ 16.6            &              $<0.64$              \\  \hline
19\_o   &           & 5795                $\pm$ 16         & 41.7             $\pm$ 16.2            & 0.80              $\pm$ 0.32             \\
        &           &         -              &      -        &    -  \\
        &           &          -             &       -       &    -     \\  \hline
19\_i   &           &          -             &       -       &    -  \\
        &           & 6702                $\pm$ 21         & 46.5             $\pm$ 22.0            & 0.75              $\pm$ 0.36             \\
\hline     
\end{tabular}
\tablefoot{
The regions are marked in Fig. \ref{fig:radio_regions}.
}
\end{minipage}
\begin{minipage}{.48\linewidth}
$^{13}\text{CO~}(1- 0)$\\
\begin{tabular}{|cc|c|c|c|c|}
\hline\hline
Region  &  & {Vlos        } & {Vel.Disp     } & {Flux          }  \\ 
  &  & {(km$\,$s$^{-1}$)} & {(km$\,$s$^{-1}$)} & {(K km$\,$s$^{-1}$)}  \\\hline 

19\_    &  &      -        &     -       &      -      \\ \hline
19\_o   &  &      -        &      -      &      -      \\ \hline
19\_i   &  & 5958 $\pm$ 32 & 70.0 $\pm$ 31.9 &  $<0.32$  \\
        &  &      -        &     -       &      -      \\ \hline
19\_ii  &  &      -        &     -       &       -     \\
        &  & 6944 $\pm$ 34 & 70.0 $\pm$ 34.1 &  $<0.42$  \\
        &  & 7329 $\pm$ 16 & 39.0 $\pm$ 16.3 & 0.36 $\pm$ 0.15 \\ \hline
19\_iii &  &     -         &      -      &      -      \\ \hline
19\_iv  &  & 5851 $\pm$ 37 & 70.0 $\pm$ 36.5 & $<0.42$ \\ \hline
19\_v   &  &      -        &      -      &       -     \\
        &  & 6974 $\pm$ 24 & 55.7 $\pm$ 24.3 & 0.38 $\pm$ 0.17 \\
        &  & 7350 $\pm$ 26 & 56.5 $\pm$ 26.8 & $<0.36$ \\ \hline
19\_vi  &  & 5771 $\pm$ 33 & 70.0 $\pm$ 33.2 & 0.48 $\pm$ 0.24 \\
        &  & 7000 $\pm$ 23 & 70.0 $\pm$ 27.5 & 0.76 $\pm$ 0.27 \\
        &  & 7334 $\pm$ 15 & 45.7 $\pm$ 14.9 & 0.57 $\pm$ 0.19 \\  \hline
\end{tabular}
\end{minipage}
\end{table}
\clearpage

\subsection{Regions in the bridge and the star-forming ridge}
\label{app:tables_bridge_ridge}
\begin{table}[h]
\centering
\caption{\small{Ionised gas and stellar flux, line-of-sight velocity, and velocity dispersion in region b\_1-4.}}
    \label{table:bridge}
\begin{tabular}{|lll|c|c|c|c|}
\hline \hline
                & Vel &  &  &   &   &   \\ 
                & comp &          & {Region b\_1} & {b\_2} & {b\_3} & {b\_4} \\ \hline
H$\beta$\_NLR         & high     & Flux     &             -          & 5.95E-18  $\pm$ 1.71E-18 & 3.18E-17  $\pm$ 3.03E-18 & 4.77E-17  $\pm$ 3.49E-18 \\
                &          & Vlos     &             -          & 6434    $\pm$ 20     & 6398    $\pm$ 20     & 6467    $\pm$ 23     \\
                &          & Vel.Disp &             -          & 78.1      $\pm$ 19.8     & 245.6     $\pm$ 20.0     & 367.3     $\pm$ 22.9     \\
                & low      & Flux     &             -          &           -        &           -        &           -        \\
                &          & Vlos     &             -          &           -        &           -        &           -        \\
                &          & Vel.Disp &             -          &           -        &           -        &           -        \\
\hline
\end{tabular}
\tablefoot{
Flux in ($erg\,cm^{2}\,s^{-1}$), line-of-sight velocity, and velocity dispersion in (km$\,$s$^{-1}$). The regions are marked in Fig.\ref{fig:regions}.
}
\end{table}

\begin{table}[h]
\centering
\caption{\small{As Table \ref{table:bridge} but for region SF\_1-4.}}
    \label{table:SF_1}
\begin{tabular}{|lll|c|c|c|c|}
\hline \hline
                & Vel &  &  &   &   &   \\ 
                & comp &          & {Region SF\_1} & {SF\_2} & {SF\_3} & {SF\_4} \\ \hline
H$\beta$\_NLR         & high     & Flux     & 9.44E-18    $\pm$ 2.76E-18    & 3.77E-17  $\pm$ 9.68E-18 &           -        & 2.24E-17  $\pm$ 3.04E-18 \\
                &          & Vlos     & 6802      $\pm$ 26        & 6526    $\pm$ 74     &           -        & 6838    $\pm$ 14     \\
                &          & Vel.Disp & 113.6       $\pm$ 27.2        & 305.1     $\pm$ 64.3     &           -        & 120.2     $\pm$ 13.6     \\
                & low      & Flux     & 3.82E-17    $\pm$ 4.47E-18    & 2.64E-17  $\pm$ 6.25E-18 &           -        & 4.59E-17  $\pm$ 4.47E-18 \\
                &          & Vlos     & 6122      $\pm$ 28        & 6040    $\pm$ 14     &           -        & 6198    $\pm$ 20     \\
                &          & Vel.Disp & 305.1       $\pm$ 31.1        & 120.3     $\pm$ 18.5     &           -        & 260.7     $\pm$ 22.1     \\
\hline
\end{tabular}
\end{table}

\begin{table}[h]
\centering
\caption{\small{As Table \ref{table:bridge} but for region SF\_5-8.}}
    \label{table:SF_2}
\begin{tabular}{|lll|c|c|c|c|}
\hline \hline
                & Vel &  &  &   &   &   \\ 
                & comp &          & {SF\_5} & {SF\_6} & {SF\_7} & {SF\_8} \\ \hline
H$\beta$\_NLR         & high     & Flux     & 8.48E-18  $\pm$ 2.31E-18 & 1.49E-17  $\pm$ 3.16E-18 & 1.05E-17  $\pm$ 3.28E-18 & 2.82E-17  $\pm$ 5.79E-18 \\
                &          & Vlos     & 6700    $\pm$ 30     & 6810    $\pm$ 13     & 6827    $\pm$ 26     & 6679    $\pm$ 33     \\
                &          & Vel.Disp & 131.8     $\pm$ 30.0     & 84.0      $\pm$ 14.3     & 108.2     $\pm$ 26.9     & 170.5     $\pm$ 28.4     \\
                & low      & Flux     & 2.93E-17  $\pm$ 3.25E-18 & 5.01E-17  $\pm$ 5.36E-18 & 5.14E-17  $\pm$ 5.21E-18 & 1.53E-17  $\pm$ 4.76E-18 \\
                &          & Vlos     & 6044    $\pm$ 22     & 6270    $\pm$ 23     & 6244    $\pm$ 22     & 6277    $\pm$ 38     \\
                &          & Vel.Disp & 255.4     $\pm$ 24.7     & 275.9     $\pm$ 25.7     & 275.0     $\pm$ 24.2     & 127.1     $\pm$ 29.0     \\
\hline
\end{tabular}
\end{table}

\begin{table}[h]
\centering
\caption{\small{As Table \ref{table:bridge} but for region SF\_9-12.}}
    \label{table:SF_3}
\begin{tabular}{|lll|c|c|c|c|}
\hline \hline
                & Vel &  &  &   &   &   \\ 
                & comp  &          & {SF\_9} & {SF\_10} & {SF\_11} & {SF\_12} \\ \hline
H$\beta$\_NLR         & high     & Flux     & 3.72E-17  $\pm$ 4.98E-18 &           -         &           -         &           -         \\
                &          & Vlos     & 6461    $\pm$ 30     &           -         &           -         &           -         \\
                &          & Vel.Disp & 283.1     $\pm$ 32.9     &           -         &           -         &           -         \\
                & low      & Flux     & 9.37E-18  $\pm$ 2.91E-18 & 1.34E-17  $\pm$ 6.81E-18  & 1.12E-17  $\pm$ 2.92E-18  & 1.84E-17  $\pm$ 1.59E-18  \\
                &          & Vlos     & 5906    $\pm$ 19     & 6647    $\pm$ 100      & 5811    $\pm$ 47      & 5674    $\pm$ 7       \\
                &          & Vel.Disp & 85.6      $\pm$ 21.4     & 203.6     $\pm$ 41.1      & 210.7     $\pm$ 47.0      & 88.0      $\pm$ 6.6       \\
\hline
\end{tabular}
\end{table}

\begin{table}[h]
\centering
\caption{\small{As  Table \ref{table:bridge} but for region SF\_13-15.}}
    \label{table:SF_4}
\begin{tabular}{|lll|c|c|c|}
\hline \hline
                & Vel &  &  &   &     \\ 
                & comp  &          & {SF\_13} & {SF\_14} & {SF\_15} \\ \hline
H$\beta$\_NLR         & high     & Flux     &           -         &           -         &           -         \\
                &          & Vlos     &           -         &           -         &           -         \\
                &          & Vel.Disp &           -         &           -         &           -         \\
                & low      & Flux     & 1.84E-17  $\pm$ 1.59E-18  & 1.84E-17  $\pm$ 1.59E-18  & 1.84E-17  $\pm$ 1.59E-18  \\
                &          & Vlos     & 5674    $\pm$ 7       & 5674    $\pm$ 7       & 5674    $\pm$ 7       \\
                &          & Vel.Disp & 90.3      $\pm$ 10.3      & 84.1      $\pm$ 1.8       & 94.4      $\pm$ 4.8       \\
\hline
\end{tabular}
\end{table}

\begin{table}[h!]
\centering
\caption{\label{table:CO_SF} \small{ $^{12}\text{CO~}(1- 0)$ and $^{12}\text{CO~}(2- 1)$ line-of-sight velocity, velocity dispersion, integrated flux, and H$_2$ molecular gas mass for the regions in the SF ridge and the bridge. }}
\begin{tabular}{|cc|c|c|c||c|c|c|c|c|} \hline \hline
        &  & \multicolumn{3}{l}{$^{12}\text{CO~}(1- 0)$}                                                                                                                       &    &  & \multicolumn{2}{l}{$^{12}\text{CO~}(2- 1)$}                                                                                                             \\
Region  &  & {Vlos        } & {Vel.Disp     } & {Flux          } & {M(H$_2$)      } & & {Vlos        } & {Vel.Disp     } & {Flux          }\\ 
  &  & {(km$\,$s$^{-1}$)} & {(km$\,$s$^{-1}$)} & {(K km$\,$s$^{-1}$)} & {      ($10^8M_{\odot}$)}& & {(km$\,$s$^{-1}$)} & {(km$\,$s$^{-1}$)} & {(K km$\,$s$^{-1}$)} \\ \hline 
b\_i    &  & 6458          $\pm$ 18         & 66.4           $\pm$ 18.3          & 1.01            $\pm$ 0.22           & 3.25          $\pm$ 0.70          &  & 5800          $\pm$ 22         & 50.5           $\pm$ 22.3          & 0.66            $\pm$ 0.30           \\
        &  & 6612          $\pm$ 10          & 39.9           $\pm$ 10.5          & 0.81            $\pm$ 0.18           & 2.61          $\pm$ 0.57          &  & 6669          $\pm$ 33         & 69.1           $\pm$ 33.4          &             $<0.70$           \\
        &  & 6788          $\pm$ 20         & 61.2           $\pm$ 22.1          & 0.57            $\pm$ 0.18           & 1.84          $\pm$ 0.59          &  & -              & -               &                 -             \\
        &  & 7120          $\pm$ 13         & 32.8           $\pm$ 11.5          & 0.31            $\pm$ 0.12           & 1.00          $\pm$ 0.37          &  & -              & -               & -                           \\ \hline
b\_ii   &  & -              & -               & -               &      -       &                                       & 5774          $\pm$ 24         & 70.0           $\pm$ 25.5          & 1.06            $\pm$ 0.38           \\
\hline
\end{tabular}
\tablefoot{
The regions are marked in Fig. \ref{fig:radio_regions}. 
The $^{13}\text{CO~}(1- 0)$ emission is absent in these regions.
}
\end{table}
\clearpage

\subsection{Regions in or near the NGC7318 pair and NGC7317}
\label{app:tables_7318}
\begin{table}[h!]
\caption{\small{Ionised gas and stellar flux, line-of-sight velocity, and velocity dispersion in the inner $1.5''$ of NGC7317, NGC7318A and B.}}
    \label{table:7318}
\begin{tabular}{|ll|c|c|c|} \hline \hline \noalign{\smallskip}
               &          & {NGC7318A} & {NGC7318B} & {NGC7317}  \\ \hline
H$\alpha$\_NLR        & Flux     & 9.91E-16   $\pm$ 1.14E-16  & 7.41E-16   $\pm$ 6.12E-17  &             -      \\
               & Vlos     & 6094     $\pm$ 53      & 6188     $\pm$ 34      &             -      \\
               & Vel.Disp & 408.5      $\pm$ 23.0      & 355.6      $\pm$ 23.1      &             -      \\ \hline
{[}NII{]} 6550 & Flux     & 1.26E-16   $\pm$ 2.07E-17  & 2.07E-16   $\pm$ 2.69E-17  &             -      \\
               & Vlos     & 5962     $\pm$ 164    & 6094     $\pm$ 613     &             -      \\
               & Vel.Disp & 456.0      $\pm$ 29.1      & 353.5      $\pm$ 27.0      &             -      \\ \hline
{[}NII{]} 6585 & Flux     & 1.08E-15   $\pm$ 1.25E-16  & 5.18E-16   $\pm$ 5.50E-17  &             -      \\
               & Vlos     & 5965     $\pm$ 61      & 6096     $\pm$ 38      &             -      \\
               & Vel.Disp & 453.6      $\pm$ 28.9      & 351.6      $\pm$ 26.8      &             -      \\ \hline
{[}OII{]} 3727 & Flux     & 2.81E-17   $\pm$ 4.86E-18  &              -         &             -      \\
               & Vlos     & 5949     $\pm$ 20      &              -         &             -      \\
               & Vel.Disp & 133.3      $\pm$ 20.1      &              -         &             -      \\ \hline
stellar        & Vlos     & 6787     $\pm$ 5       & 5973     $\pm$ 4       & 6734      $\pm$ 5    \\
               & Vel.Disp & 263.6      $\pm$ 10.7      & 168.9      $\pm$ 4.0       & 262.7       $\pm$ 6.9    \\  \hline
\end{tabular}
\tablefoot{
Flux in ($erg\,cm^{2}\,s^{-1}$), line-of-sight velocity, and velocity dispersion in (km$\,$s$^{-1}$). The regions are marked in Fig.\ref{fig:regions}.
}
\end{table}

\begin{table}[h!]
\caption{\label{table:CO_18} \small{ $^{12}\text{CO~}(1- 0)$, $^{12}\text{CO~}(2- 1)$ and $^{13}\text{CO~}(1- 0)$ line-of-sight velocity, velocity dispersion, integrated flux, and H$_2$ molecular gas mass in the regions in or near the NGC7318 pair and NGC7317. }}
$^{12}\text{CO~}(1- 0)$ \\
\begin{tabular}{|cc|c|c|c|c|c|} 
\hline\hline
Region  &  & {Vlos        } & {Vel.Disp     } & {Flux          } & {M(H$_2$)      } \\ 
  &  & {(km$\,$s$^{-1}$)} & {(km$\,$s$^{-1}$)} & {(K km$\,$s$^{-1}$)} & {      ($10^8M_{\odot}$)} \\\hline 
18a     &  &      -        &      -       &       -     &        -    \\ \hline
18b     &  & 6439 $\pm$ 14 & 37.3  $\pm$ 13.5 & 0.33 $\pm$ 0.12 & 1.05 $\pm$ 0.40 \\ 
        &  & 6631 $\pm$ 17 & 27.7  $\pm$ 13.6 & $<0.20$ & $<0.64$ \\\hline
18\_i   &  & 5882 $\pm$ 12 & 25.8  $\pm$ 11.7 & 0.25 $\pm$ 0.11 & 0.81 $\pm$ 0.37 \\
        &  & 6434 $\pm$ 11 & 34.3  $\pm$ 10.1 & 0.38 $\pm$ 0.12 & 1.21 $\pm$ 0.38 \\\hline
18\_iii        &  & 6639.0 $\pm$ 51 & 107.9 $\pm$ 51.0 & $<0.48$  & $<1.55$  \\\hline
17\_    &  &       -       &     -        &     -       &     -      \\ \hline
\end{tabular}
\vspace{0.5cm}
\\
$^{12}\text{CO~}(2- 1)$\\
\begin{tabular}{|cc|c|c|c|c|}
\hline\hline
Region  &  & {Vlos        } & {Vel.Disp     } & {Flux          }  \\ 
  &  & {(km$\,$s$^{-1}$)} & {(km$\,$s$^{-1}$)} & {(K km$\,$s$^{-1}$)}  \\\hline 
18a     &  & 5815 $\pm$ 16 & 70.0 $\pm$ 16.9 & 1.45 $\pm$ 0.34 \\
        &  & 6430 $\pm$ 17 & 25.3 $\pm$ 13.2 &  $<0.40$ \\ \hline
18b     &  & 5800 $\pm$ 25 & 70.0 $\pm$ 22.3 & 1.79 $\pm$ 0.44 \\
        &  & 6001 $\pm$ 41 & 70.0 $\pm$ 36.9 & 1.08 $\pm$ 0.44 \\
        &  & 6565 $\pm$ 29 & 70.0 $\pm$ 28.6 & 0.82 $\pm$ 0.35 \\ \hline
18\_i   &  & 5837 $\pm$ 22 & 70.0 $\pm$ 21.2 & 1.98 $\pm$ 0.48 \\
        &  & 6050 $\pm$ 35 & 70.0 $\pm$ 34.7 & 1.21 $\pm$ 0.48 \\
        &  & 6443 $\pm$ 29 & 70.0 $\pm$ 28.9 & 0.95 $\pm$ 0.40 \\ \hline
18\_iii &  & 5789 $\pm$ 22 & 70.0 $\pm$ 22.1 & 1.02 $\pm$ 0.33 \\ \hline
17\_    &  & 5621 $\pm$ 31 & 70.0 $\pm$ 34.2 & 1.30 $\pm$ 0.60 \\
        &  & 5800 $\pm$ 10  & 26.0 $\pm$ 9.9  & 1.11 $\pm$ 0.39 \\
        &  & 6052 $\pm$ 24 & 70.0 $\pm$ 23.8 & 1.58 $\pm$ 0.56 \\  \hline
\end{tabular}
\tablefoot{
The regions are marked in Fig. \ref{fig:radio_regions}. 
The $^{13}\text{CO~}(1- 0)$ emission is absent in these regions.
}
\end{table}
\clearpage

\end{appendix}
\end{document}